\documentclass[prd,10pt,tightenlines,aps,letterpaper,amsmath,amssymb,preprintnumbers,
               showpacs,floatfix,longbibliography,nofootinbib,superscriptaddress,twocolumn]{revtex4-2}
\usepackage{subfiles}
\usepackage{siunitx}
\usepackage{graphicx}
\graphicspath{ {./figures/} }
\usepackage[export]{adjustbox}
\usepackage[caption=false]{subfig}
\usepackage{bm}  
\usepackage{makecell} 
\allowdisplaybreaks 
\usepackage[hypertexnames=true]{hyperref}

\usepackage[capitalize]{cleveref}
\crefname{equation}{Eq.}{Eqs.}
\crefname{figure}{Fig.}{Figs.}
\crefname{table}{Tab.}{Tabs.}
\crefname{section}{Sec.}{Secs.}
\crefname{appendix}{App.}{Apps.}
\Crefname{table}{Table}{Tables}
\Crefname{figure}{Figure}{Figures}

\hypersetup{
    colorlinks=true,       
    linkcolor=blue,          
    citecolor=blue,        
    filecolor=blue,      
    urlcolor=blue           
}

\usepackage{slashed}
\usepackage{mathtools}
\usepackage{listings}
\usepackage{pgf}
\usepackage{fancyvrb}
\usepackage{longtable}
\usepackage{siunitx}
\usepackage{braket}
\usepackage{duckuments}
\usepackage{soul}
\usepackage{comment}

\usepackage{enumitem} 

\usepackage[normalem]{ulem} 
\usepackage{dsfont} 

\usepackage{tikz}
\usetikzlibrary{shapes,arrows,chains}
\usetikzlibrary{calc}
\usetikzlibrary{patterns}
\usetikzlibrary{patterns}
\usetikzlibrary{positioning,decorations.pathmorphing,decorations.markings}

\tikzset{>=latex}
\tikzset{every picture/.style={line width=0.75pt}}

\usepackage{array}
\newcolumntype{P}[1]{>{\centering\arraybackslash}p{#1}}

\usepackage{soul}

\DeclareMathAlphabet{\mathbbb}{U}{bbold}{m}{n}

\definecolor{listinggreen}{rgb}{0,0.6,0}
\definecolor{listinggray}{rgb}{0.5,0.5,0.5}
\definecolor{listingmauve}{rgb}{0.58,0,0.82}
\definecolor{listingkeywordcolor}{rgb}{1.0,0.4,0.0}
\definecolor{listinglightgray}{rgb}{0.8863,0.8863,0.8863}

\newcommand{\Id}[0]{\mathds{1}}

\newcommand{\mr}{rm}

\usepackage{colortbl}

\newcommand{\Egap}{\hat{E}_{\text{miss}}}
\newcommand{\lamgap}{\hat{\lambda}_{\text{gap}}}
\newcommand{\nn}{I=1}
\newcommand{\deut}{I=0}
\newcommand{\nprop}{$2.12\times 10^7$}

\begin{document}

\preprint{FERMILAB-PUB-25-0667-T}
\preprint{MIT-CTP/5985}

\title{Excited-state uncertainties in lattice-QCD calculations of multi-hadron systems}
\author{William Detmold}
\affiliation{Center for Theoretical Physics - A Leinweber Institute, Massachusetts Institute of Technology, Cambridge, MA 02139, USA}
\author{Anthony V. Grebe}
\affiliation{Department of Physics and Maryland Center for Fundamental Physics, University of Maryland, College Park, MD 20742, USA}
\affiliation{Fermi National Accelerator Laboratory, Batavia, IL 60510, USA}
 \author{Daniel C. Hackett}
\affiliation{Fermi National Accelerator Laboratory, Batavia, IL 60510, USA}
\author{Marc~Illa}
\affiliation{InQubator for Quantum Simulation (IQuS), Department of Physics, University of Washington, Seattle, WA 98195, USA}
\affiliation{Physical Sciences Division, Pacific Northwest National Laboratory, Richland, WA 99354, USA}
\author{Robert J. Perry}
\affiliation{Center for Theoretical Physics - A Leinweber Institute, Massachusetts Institute of Technology, Cambridge, MA 02139, USA}
\author{Phiala E. Shanahan}
\affiliation{Center for Theoretical Physics - A Leinweber Institute, Massachusetts Institute of Technology, Cambridge, MA 02139, USA}
\author{Michael L. Wagman}
\affiliation{Fermi National Accelerator Laboratory, Batavia, IL 60510, USA}
\collaboration{NPLQCD Collaboration}
\begin{figure}
  \vskip -1.0cm
  \leftline{\includegraphics[width=0.15\textwidth]{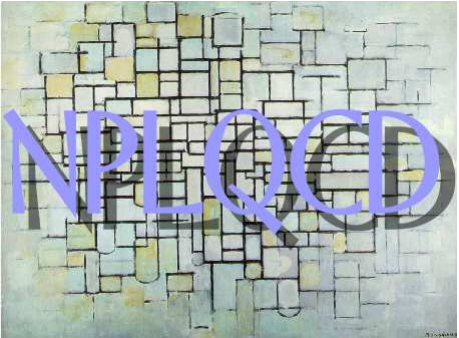}}
\end{figure}

\begin{abstract}
Excited-state effects lead to hard-to-quantify systematic uncertainties in lattice quantum chromodynamics (LQCD) spectroscopy calculations when computationally accessible imaginary times are smaller than inverse excitation gaps, as often arises for multi-hadron systems with signal-to-noise problems. Lanczos residual bounds address this by providing two-sided constraints on energies that do not require assumptions beyond Hermiticity, but often give very conservative systematic uncertainty estimates. Here, a more-constraining set of gap bounds is introduced for hadron spectroscopy. These bounds provide tighter constraints whose validity requires an explicit assumption about an energy gap. Exactly solvable lattice field theory correlators are used to test the utility of residual and gap bounds at finite and infinite statistics. Two-sided bounds and other analysis methods are then applied to a high-statistics LQCD calculation of nucleon-nucleon scattering at $m_\pi \sim 800$ MeV. Generalized eigenvalue problem (GEVP) and Lanczos energy estimators are compatible when applied to the same correlator data, but analyses including different interpolating operators show statistically significant inconsistencies. However, two-sided bounds from all operators are consistent. Under the assumption that the number of energy levels below $N\Delta$ and $\Delta\Delta$ thresholds is the same as for non-interacting nucleons, gap bounds are sufficient to constrain nucleon-nucleon scattering amplitudes at phenomenologically relevant precision. 
Lanczos methods further reveal that energy-eigenstate estimates from previously studied asymmetric correlators have not converged over accessible imaginary times. Nevertheless, data-driven examples demonstrate why assumptions are required to draw conclusions about the natures of two-nucleon ground states at these masses.
\end{abstract}

\maketitle

\tableofcontents

\section{Introduction}

Lattice quantum chromodynamics (LQCD) provides a powerful framework for predicting nonperturbative properties of hadrons and nuclei, although it has limitations. Along with the necessary extrapolations to the continuum and infinite-volume limits, in the context of hadron spectroscopy, physically motivated assumptions are often implicitly required to constrain formally infinite sets of LQCD energies from a finite number of correlation functions (correlators) computed stochastically for a finite, discrete set of operator separations. This can  make systematic uncertainty quantification challenging. The goal of this work is to make explicit the assumptions needed to constrain LQCD energy spectra and explore what constraints are possible with weaker assumptions than are usually used.

LQCD spectroscopy is particularly challenging for multi-hadron systems. The small energy gaps that exist in such systems mean that large  Euclidean times are required to isolate ground states. Further, the signal-to-noise ratios (SNRs) of  typical multi-hadron correlators decrease exponentially in the same limit of large imaginary time needed to ensure excited-state effects are suppressed~\cite{Parisi:1983ae,Lepage:1989hd}. Proof-of-principle studies of multi-hadron systems in LQCD have therefore generally been performed with larger-than-physical quark masses, where the signal-to-noise problem is less severe. 
Multi-baryon systems are particularly challenging, and to date no calculations of two-nucleon systems have been performed at the physical values of the quark masses.
These calculations can be separated into three different computational strategies: analysis of asymmetric (sets of) correlation functions constructed from different interpolating operators at the source and sink~\cite{Fukugita:1994na,Fukugita:1994ve,Beane:2006mx,Beane:2006gf,Beane:2009py,Yamazaki:2009ua,NPLQCD:2011naw,NPLQCD:2012mex,NPLQCD:2013bqy,Wagman:2017tmp,Orginos:2015aya,NPLQCD:2020ozd,NPLQCD:2020lxg,Yamazaki:2012hi,Yamazaki:2015asa,Berkowitz:2015eaa}, the HALQCD method\footnote{The theoretical strategy implemented by the HALQCD method~\cite{Ishii:2006ec} is not discussed in this work.}~\cite{Ishii:2006ec,Murano:2011nz,Aoki:2011gt,Ishii:2012ssm,HALQCD:2019wsz}, and
symmetric correlator matrix analyses~\cite{Francis:2018qch,Amarasinghe:2021lqa,Horz:2020zvv,Green:2021qol,Detmold:2024iwz,BaSc:2025yhy}.
While each of these methods should lead to equivalent information about the spectrum, historically they have led to quantitatively different low-energy spectra 
for two-nucleon systems,
and qualitatively different conclusions about the existence of bound states in the $I=0$, $S=1$ (``deuteron'') and the $I=1$, $S=0$ (``dineutron'') channels at unphysically large quark masses, where $I$ denotes the isospin, and $S$ the spin. In particular, studies employing asymmetric correlators favor the presence of bound states, while studies employing either the HALQCD method or correlator matrix analysis methods have not found evidence of such a bound state. The discrepancies between these different computational strategies 
imply the underestimation of systematic uncertainties in one or more of these calculations. 

Recently, a new class of approaches based on the Lanczos algorithm has been proposed to analyze LQCD correlators~\cite{Wagman:2024rid,Hackett:2024xnx,Ostmeyer:2024qgu,Chakraborty:2024scw,Hackett:2024nbe,Abbott:2025yhm}. 
One clear advantage of the Lanczos method, relative to other techniques that have previously been employed, is that it permits the calculation of a set of so-called \textit{residual bounds} which are two-sided bounds on energy eigenvalues that are guaranteed to hold regardless of the size of excited-state effects~\cite{Wagman:2024rid}. These residual bounds can be computed directly from the numerical correlation function data, and require no additional assumptions beyond Hermiticity of the transfer matrix. The existence of two-sided bounds enables the conversion of constraints on finite-volume (FV) energy eigenvalues for two-hadron systems into a set of scattering phase-shift values through L\"uscher's quantization condition \cite{Luscher:1986pf,Luscher:1990ux} or effective field theory~\cite{Barnea:2013uqa,Eliyahu:2019nkz,Detmold:2021oro,Sun:2022frr,Detmold:2023lwn}. However, the level of constraint provided by these bounds with the statistical precision obtainable using current-era computing has not been investigated. A major goal of this work is to quantify the constraining strength of these two-sided bounds. 

In this work, a second, more constraining set of bounds, termed \textit{gap bounds}~\cite{Davis:1970,Parlett,Parlett:1995,Demmel:1997,Haas:2025} are introduced to LQCD correlator analysis. These bounds rely on assumptions about the spectrum. If gaps to unresolved energy levels can be accurately estimated, for example using physics priors, these gap bounds provide a tighter set of two-sided bounds on energy eigenvalues than are available from residual bounds. These gap bounds constitute a compromise between the previously described residual bounds and more model-dependent estimates of systematic uncertainties, such as assuming ground-state saturation at statistically resolvable Euclidean time separations. 

It is natural to ask whether the Lanczos method and the associated residual and gap bounds can provide additional insight into systematic uncertainties in the two-nucleon spectrum at unphysically heavy quark mass.
To investigate, this study employs a high-statistics dataset constructed using quark propagators from \nprop\ sources, an order of magnitude more measurements than previous studies of spectroscopy in two-baryon systems. To benchmark and cross-check Lanczos methods, this dataset is also examined as a generalized eigenvalue problem (GEVP) for comparison.
The results are used to constrain nucleon-nucleon scattering amplitudes under various assumptions and study what can be concluded about the two-nucleon system near threshold.

This work is organized as follows. \Cref{sec:strategies} outlines the assumptions that are made in performing LQCD spectroscopy using various estimators and bounds. The spectroscopy methods employed in this work are briefly reviewed in \cref{sec:theory}. Theoretical properties of bounds on energy eigenvalues are discussed in \cref{sec:bounds}. 
The validity of residual bounds and gap bounds at finite statistics, and the degree of calibration of various systematic uncertainty quantification schemes, are investigated in \cref{sec:SHO} by studying correlators in a solvable scalar field theory with significant excited-state contamination and a signal-to-noise problem similar to that in LQCD.

These methods are then applied to numerical data obtained from a LQCD calculation summarized in \cref{sec:lqcd}. 
Physically motivated assumptions for the gap parameters required to compute gap bounds are discussed in \cref{sec:mind_the_gap}.
Energy estimators, residual bounds, and gap bounds computed from symmetric correlators and correlation matrices in the pion, nucleon, and two-nucleon $\nn$ and $\deut$ channels are presented in \cref{sec:lqcd-lanczos}.
Constraints on scattering phase shifts arising from gap bounds as well as statistical uncertainties on energy estimators are presented in \cref{sec:phase-shift}.
Asymmetric correlators that have historically been in tension with correlator matrix results are analyzed in \cref{sec:asymm}.
Possible scenarios for energy spectra consistent with LQCD results are explored in \cref{sec:adversary}.
Conclusions, and the assumptions required to reach them, are presented in \cref{sec:discussion}. 

\section{Assumptions and systematic effects in multi-hadron spectroscopy}
\label{sec:strategies}

A central challenge in hadron spectroscopy is robust uncertainty quantification in the estimation of energy eigenvalues. This work explores new strategies for systematic uncertainty quantification using two-sided bounds on energy eigenvalues that can be rigorously obtained by applying Lanczos methods to correlators with finite Euclidean-time extent. These energy eigenvalues also serve as important inputs which can be used to constrain multi-hadron interactions via the application of quantization conditions \cite{Luscher:1986pf,Luscher:1990ux} or effective field theory\ \cite{Barnea:2013uqa,Eliyahu:2019nkz,Detmold:2021oro,Sun:2022frr,Detmold:2023lwn}.

In the single-hadron sector, the key feature underlying the accuracy of LQCD spectroscopy is that correlation functions can be computed at imaginary times $t$ that are much larger than the inverse of the excitation energy $\delta_1$,
thereby providing a small parameter $e^{-\delta_1 t}$ that can be used to simplify the analysis.\footnote{The discussion in this work assumes that  QED is absent and that the light quark masses take their physical values or are larger. 
In the presence of QED, or if $m_q\to0$, additional pathologies appear as soft-photon and soft-pion excitations produce an accumulation of scattering poles even in a fixed finite volume. See Refs.~\cite{Beane:2014qha,NPLQCD:2020ozd} for a discussion of two or more charged hadrons in the presence of QED.}  Here, $\delta_n \equiv E_n - E_0$, where $E_0$ and $E_n$ are the ground-state and $n$th excited-state energies, respectively.\footnote{Here and below, units in which the lattice spacing is set to unity, $a=1$, are used.} Excited-state contributions to correlators are suppressed by $e^{-\delta_1 t}$ relative to ground-state contributions, and so when $e^{-\delta_1 t} \ll 1$, it is often possible to model correlation functions accurately as ground-state contributions plus small corrections.
Systematic uncertainties arising from excited-state effects are parametrically suppressed in this case.
Such systematic uncertainties can be further studied by fitting correlator results to truncated spectral expansions using different ranges of $t$ and  different numbers of excited states. 
This situation is analogous to the treatment of other LQCD systematic uncertainties such as FV effects that are parametrically suppressed by $e^{-m_\pi L}$ and can be further studied using explicit calculations with varying $L$. 

For generic multi-hadron systems, spectroscopy is fundamentally more challenging because of increasingly rapid SNR degradation with increasing $t$ and the presence of small energy gaps. Consequently, statistical control of correlators at $t \gg 1/\delta_1$ is not achievable with available computational resources and therefore $e^{-\delta_1 t}$ is not a small parameter. In this case, spectroscopic information can either be obtained from methods that are robust to these effects or be based on assumptions about these effects.

For two hadron systems, the robust bounding methods studied herein have not previously been applied.
To date, the determination of energy levels from (sets of) correlators 
has relied on an implicit assumption:
\begin{itemize}[leftmargin=*]
\item \textit{$N$-state-saturation assumption}: There exists a region of Euclidean time in which the correlator (matrix) is accurately described by a truncated spectral decomposition with $M \geq N$ states, where $N$ states approximate genuine energy eigenstates up to statistical noise (e.g., the $N$ lowest-energy states).
\end{itemize}
It is not possible to prove that $N$-state saturation holds at a target level of accuracy without knowledge of both the true energy spectrum and the overlaps of interpolating operators with energy eigenstates.
In particular, to be valid, the $N$-state-saturation assumption implicitly relies on the identification of interpolating operators that have sufficiently small overlaps with all states except for a particular (set of) $N$ state(s) of interest and sufficiently large overlaps with all the state(s) of interest. In some circumstances, the $N=1$ case is referred to as ground-state saturation.

For stable single-hadron states away from the chiral limit, it is expected that $\delta_1 \sim \Lambda_{\rm QCD}$.
In this case, $t \gg 1/\delta_1$ can often be statistically resolved in practical LQCD calculations, and all excited-state effects are suppressed by $e^{-\delta_1 t} \ll 1$.
$N$-state saturation may be expected to hold in this case, although it is not guaranteed to do so as its validity depends on the choice of interpolating operators that are used.
For finite-volume two-nucleon scattering states, where $\delta_1 \approx 4\pi^2 / (M_N L^2)$ with $M_N$ the nucleon mass and $L$ the spatial lattice extent, or nuclear bound states, where $\delta_1$ is expected to be on the order of a few MeV, it is impractical to compute statistically controlled correlation functions for $\delta_1 t \gg 1$. For computationally accessible results, where $\delta_1 t\alt1$, the saturation of correlation functions by a single state, or a fixed set of states, only follows from the  assumption that interpolating operators have been found that have much larger overlap with a given state than with all other states in the LQCD spectrum.
Similar issues will also affect even single-hadron states near the chiral limit, where effects from multi-hadron excited states including additional pions can be significant. 

Early works \cite{Fukugita:1994na,Fukugita:1994ve,Beane:2006mx,Beane:2006gf,Beane:2009py,Yamazaki:2009ua,NPLQCD:2011naw,NPLQCD:2012mex,NPLQCD:2013bqy,Wagman:2017tmp,Orginos:2015aya,NPLQCD:2020ozd,NPLQCD:2020lxg,Yamazaki:2012hi,Yamazaki:2015asa,Berkowitz:2015eaa} on two-baryon spectroscopy assumed $N$-state saturation of asymmetric correlators in order to obtain ground-state energy estimates. Physical arguments were used to justify the smallness of excited-state overlaps in correlation functions built from local sources (hexaquark operators) expected to overlap with bound states and plane-wave product sinks (dibaryon operators) expected to suppress overlap with non-zero relative momentum scattering states, as well as to select zero total momentum. 
Consistency checks based on stability to interpolating operator variation, stability of results obtained on multiple volumes, and agreement with pionless effective field theory (EFT$(\slashed{\pi}$)) provided further support for these $N$-state-saturation assumptions~\cite{Beane:2017edf}.

Some more recent works on two-baryon spectroscopy assumed $N$-state saturation of symmetric correlator matrices in order to obtain ground-state energy estimates~\cite{Francis:2018qch,Amarasinghe:2021lqa,Horz:2020zvv,Green:2021qol,Detmold:2024iwz,BaSc:2025yhy,Detmold:2024iwz}. Physical arguments were used to justify the smallness of excited-state overlaps in GEVP-optimized correlators where non-zero relative momentum dibaryon operators are used to identify and remove particular excited-state contributions. 
Consistency checks based on stability to interpolating operator variation provided support for these $N$-state-saturation assumptions. 
At asymptotically large $t$, GEVP analysis of a rank-$r$ correlator matrix has the advantage that excited-state effects are parametrically reduced from $O(e^{-\delta_1 t})$ to $O(e^{-\delta_r t})$; however, for finite $t$, this additional suppression is only achieved under the assumption that one or more of the interpolating operators that are used have significant overlap with each of the $r$ lowest energy states.
An $N$-state-saturation assumption is required to take energy estimates at face value from correlator matrices as well as from individual (symmetric or asymmetric) correlators.

The results of these previous studies clearly demonstrate that determinations of ground-state energies based on assuming $N$-state saturation for asymmetric correlators and those based on assuming $N$-state saturation for symmetric correlator matrices are inconsistent. Analysis of the high-statistics dataset studied here leads to the same conclusion. This implies that, despite the self-consistency checks passed for both symmetric and asymmetric correlators, $N$-state-saturation assumptions lead to results that are incorrect (statistically discrepant) in one or both approaches. 

These discrepancies highlight the importance of constraints on the two-baryon spectrum that require weaker assumptions.
Symmetric (matrices of) correlators, defined by the use of identical (sets of) interpolating operators at the source and sink, provide one-sided \emph{variational bounds} on energy eigenvalues that only assume energy eigenvalues are real and thermal effects are negligible.
However, these bounds are limited in their applicability, and, for example, cannot be used to meaningfully constrain scattering phase shifts in two-hadron systems. 
The residual bounds available in the Lanczos approach provide two-sided bounds on energy eigenvalues that do not require any assumptions beyond Hermiticity~\cite{Wagman:2024rid}.
The practical challenge to using residual bounds as systematic uncertainty estimates in spectroscopy is that residual bounds often provide exceedingly conservative error estimates in linear algebra applications~\cite{Parlett}.
Below, residual bounds are seen to be valid but conservative in applications to a solvable lattice quantum field theory.
In applications to high-statistics LQCD results below, residual bounds lead to percent-level constraints on two-nucleon energies but are insufficiently precise to determine scattering phase shifts. 

Gap bounds, introduced below, provide tighter two-sided bounds on energy eigenvalues than residual bounds in practice, but require a different explicit 
assumption:
\begin{itemize}[leftmargin=*]
\item \textit{No-missing-states assumption}: There are exactly $N$ states with energy less than a specified threshold $\Egap$ that make non-zero contributions to a correlator (matrix), neglecting thermal effects.
\end{itemize}
The validity of gap bounds therefore requires an explicit assumption about the energy spectrum, but no assumptions must be made about the sizes of interpolating operator overlaps with the $N$ states below $\Egap$.
This is particularly pertinent when undertaking spectroscopy very close to the physical values of the quark masses and near to the continuum, infinite-volume limit, as experimental information can be used to estimate the energy spectrum more easily than the overlaps of various interpolating operators.
Unlike $N$-state-saturation assumptions, which are always violated in the infinite-statistics limit\footnote{Single-state saturation would hold when an interpolating operator produces an exact energy eigenstate, but constructing an exact eigenstate in LQCD is not possible in practice.}
because an infinite number of states appear in the exact spectral representation, it is possible for a given no-missing-states assumption  to be valid at infinite statistics (and zero temperature).

Examples are shown below, both in solvable models and through tests of varying interpolating operator sets in LQCD, in which the no-missing-states assumption  appears valid for appropriate choices of $\Egap$ but $N$-state saturation is clearly violated and statistical fluctuations underestimate uncertainties in energy estimators. 

In cases where no-missing-states assumptions are valid, gap bounds may provide useful systematic-uncertainty estimates. 
This includes cases where $e^{-\delta_1 t}$ is not a small parameter, as well as other cases of finite-statistics bias in energy estimators resulting from excited-state effects. 
In particular, residual and gap bounds constrain the size of excited-state effects based on the smallness of a different parameter---the residual-norm-square $B$---that can be calculated explicitly and can be small even in cases where $e^{-\delta_1 t}$ is not small. 
However, it is important to note that it is also possible for a no-missing-states assumption to be violated in practice, especially if $\Egap$ is chosen based on the observed spectrum from interpolating operator sets that have small overlap with one or more low-energy states.

As a starting point for estimating spectral gaps in calculations of multi-hadron systems similar to those in nature, reasonable assumptions may amount to the use of the non-interacting hadron spectrum or hadronic effective-theory results.
Interactions lead to shifts in the spectrum that encode important physics, but in many systems in nature, these shifts are small fractions of the energies. If this is the case, lack of knowledge of the interacting energies would lead to parametrically small errors on gap-bound uncertainty estimates. 
In applying lattice quantum field theory methods to theories that are significantly different than nature (for example, when working at unphysical quark masses, or when the gauge group is not $SU(3)$), it is less clear as to how strong an assumption is being made when a given set of gap bounds is used.

\section{Methods for hadron spectroscopy in LQCD}
\label{sec:theory}

\subsection{Definitions and conventions}
\label{sec:theory-intro}

Lattice field theory approaches which discretize the temporal direction do not possess a continuous time-translation symmetry that can be used to define a Hamiltonian. Instead, one can define an operator $H = -\ln T$, where $T$ is the transfer matrix, which generates discrete Euclidean-time evolution and is related to the LQCD action. 
The Hamiltonian of the theory is the continuum limit of the operator $H$. 

The spectrum of finite-volume energy eigenvalues $E_n$ can be defined from the transfer matrix eigenstates $\{\ket{n}\ |\ n = 0,1,2\dots \}$ and corresponding eigenvalues $\lambda_n$ which satisfy
\begin{equation}
T\ket{n} = \lambda_n \ket{n},
\end{equation}
where
\begin{equation}
E_n = -\ln\lambda_n.
\end{equation}
There is a one-to-one correspondence between 
$\lambda_n$ and $E_n$ due to the monotonicity of the logarithm.
Thus, extracting information about the transfer-matrix eigenvalues allows information about the eigenvalues of the FV Hamiltonian to be determined (up to discretization effects).

A set of two-point correlation functions in thermal field theory is defined as
\begin{equation}
\widetilde{{C}}_{ab}(t)= \frac{1}{Z}{\rm Tr}[T^{(L_t-t)}\mathcal{O}_{a}T^{t}\mathcal{O}_b^\dagger],
\end{equation}
where $Z=\text{Tr}(T^{L_t})$, $L_t$ is the temporal extent of the lattice, and $a,b\in\{1,\dots N\}$ label the interpolating operators
${\cal O}_a$ that are constructed from temporally localized elementary fields and chosen to possess the quantum numbers of interest.\footnote{The dependence of correlation functions and quantities derived from them on $L_t$, the spatial volume, and action parameters is suppressed.}
The correlator matrix\footnote{Throughout this work, boldface quantities such as $\bm{C}$ denote matrices in the unwritten indices.} 
$\widetilde{\bm{{C}}}(t)$ can be constructed as
\begin{equation}
[\widetilde{\bm{{C}}}(t)]_{ab}=\widetilde{{C}}_{ab}(t).
\end{equation}

Each matrix element may be expressed as a spectral decomposition
\begin{equation}
\begin{split}
  \widetilde{{C}}_{ab}(t) &=
  \sum_{n,n'}
  [\mathcal{Z}_a^{n,n'}]^* \mathcal{Z}_b^{n,n'} e^{-\mathcal{E}_n t}e^{-\mathcal{E}_{n'} (L_t-t)} \\
&\equiv \sum_{n=-\infty}^{\infty} Z_{na}^* Z_{nb} e^{-E_n t},
    \label{eq:spectral_decomposition_thermal}
    \end{split}
\end{equation}
where $\mathcal{E}_n$ and $\mathcal{E}_{n'}$ are (vacuum-subtracted) energies of $T$ eigenstates for which the overlaps $\mathcal{Z}_a^{n,n'}=\langle n | {\cal O}_a^\dagger | n' \rangle$ are non-zero.\footnote{Note that the energy eigenstates $\ket{n}$ are normalized such that $\langle n|l \rangle=\delta_{nl}$ (labeling can be extended to cover cases with degeneracies but this is not made explicit here).} 
In the second line, thermal states are represented by terms with $n < 0$ and $E_n < 0$, and, for these terms, the overlaps $Z_{n,a}$ are exponentially suppressed by energies  multiplied by $L_t$. 
For non-thermal states, $Z_{n,a}$ is equal to $\mathcal{Z}_a^{n,0}$ and $E_n$ is equal to $\mathcal{E}_n$.

While the matrix $\widetilde{\bm{C}}(t)$  is real\footnote{Purely real/imaginary overlap factors can always be achieved for interpolating operators that are eigenstates of parity, as discussed in Appendix D of Ref.~\cite{Hackett:2024nbe}, which is sufficient for all interpolating operators considered below.} and symmetric in the infinite-statistics limit, $\widetilde C_{ab}(t)\ne \widetilde C_{ba}(t)$ at finite statistics due to fluctuations. The real, symmetric combination  can be constructed as 
\begin{equation}
\label{eq:Cabp}
C_{ab}^\prime(t) = \frac{1}{2} \text{Re}[\widetilde C_{ab}(t)+\widetilde C_{ba}(t)].
\end{equation}

In certain cases relevant for the operators studied in this work, contact terms associated with differences between time-ordered and normal-ordered operators arise when relating $\bm{C'}(t=0)$ to the tensor contractions of quark-propagator products explicitly computed in LQCD calculations. The latter can exhibit pathological features at $t=0$ that cannot be described by spectral representations~\cite{Zhang:2025hyo}. To avoid complications from such contact terms, correlation function matrices are normalized as: 
\begin{equation}
\label{eq:Cab}
C_{ab}(t) = \frac{{C}'_{ab}(t)}{\sqrt{{C}'_{aa}(2m_0){C}'_{bb}(2m_0)}},
\end{equation}
where $m_0 = 1$
and only results with $t \geq 2$ are included in analyses.  
This can be interpreted as replacing the operators $\mathcal{O}_a$ and $\mathcal{O}_b^\dagger$ appearing in \cref{eq:spectral_decomposition_thermal} with $\mathcal{O}_a T$ and $T \mathcal{O}_b^\dagger$, respectively, and therefore preserves the Hermiticity of correlation function matrices.

\subsection{Asymmetric correlation functions} 

Due to the computational cost, early two-nucleon spectroscopy calculations studied asymmetric correlators. These correlators are so named because they employ a different operator at the source than at the sink, and therefore have $a\neq b$. From Eqs.~\eqref{eq:spectral_decomposition_thermal}--\eqref{eq:Cab}, it is clear that the effective mass function 
\begin{equation}
E_\text{eff}^{ab}(t) = \ln\bigg[\frac{C_{ab}(t)}{C_{ab}(t+1)}\bigg]
\end{equation}
approaches the ground state energy as
\begin{equation}
\lim_{t\to\infty} E_\text{eff}^{ab}(t) = E_0,
\end{equation}
provided that neither operator is orthogonal to the ground state.
However, the lack of a symmetric correlator implies that the products of overlap factors $Z_{na}^{*} Z_{nb}$ are not positive definite. Therefore,  the correlator is not convex and the effective mass function  is not constrained to approach the ground state from above. As a result, it is possible for the products of overlap factors for the two or more states to combine in such a way as to produce an effective mass function that is consistent with a constant $E\ne E_0$ for times comparable to $m_{\pi}^{-1}\sim \Lambda_{\text{QCD}}^{-1}$ but still much smaller than $\delta_1^{-1}$. In fact, since the Hilbert space of QCD is infinite-dimensional~\cite{Kogut:1974ag}, there exist an infinite number of asymmetric correlation functions which produce an effective mass function with  an arbitrarily long, but  finite, region of time over which it assumes any constant value. This statement holds even in the infinite-statistics limit. Therefore,  a constant effective mass function is not necessarily a reliable indicator of the presence of a true FV energy eigenvalue, even at infinite statistical precision.

Nonetheless, it can be numerically less costly to achieve a given level of statistical precision with asymmetric correlation functions than with symmetric correlation functions, particularly for multi-hadron systems~\cite{Detmold:2012eu}.
In addition, as discussed in Ref.~\cite{Amarasinghe:2021lqa}, if the set of excited states with large overlap with the source interpolating operator is nearly orthogonal to those with large overlap with the  sink interpolating operator, then an asymmetric correlation function may converge more quickly to the ground state than either diagonal counterpart; see \cref{app:epsilon} for further discussion.
On the other hand, asymmetric correlators do not provide one-sided variational bounds and, as discussed in \cref{sec:asymm} below, do not lead to two-sided gap bounds centered around their effective masses.
Residual bounds, which in the asymmetric case of $C_{ab}$ for $a\ne b$ also require input from at least one of the diagonal correlators $C_{aa}$ or $C_{bb}$, are the only way to extract one- or two-sided bounds centered about $E^{ab}_\mathrm{eff}(t)$.

\subsection{Generalized eigenvalue problem  methods}
\label{sec:theory_gevp}

Due to algorithmic and hardware developments, it is now feasible to compute matrices of correlation functions (correlator matrices) for a set of $O(100)$ operators in two-nucleon systems~\cite{Horz:2020zvv,Amarasinghe:2021lqa,Detmold:2024iwz,BaSc:2025yhy}. The effective masses associated with each diagonal element of the correlator matrix (which therefore arise from symmetric correlation functions) provide upper bounds on the ground state energy, as will be discussed below in \cref{sec:bounds}.  An often-used analysis technique is to attempt to constrain the spectrum by solving the generalized eigenvalue problem (GEVP)~\cite{Fernandez:1987ph,Luscher:1990ck,Blossier:2009kd},
\begin{equation}
\sum_b C_{ab}(t) v_{bk}(t,t_0) = \Lambda_k(t,t_0) \sum_b C_{ab}(t_0) v_{bk}(t,t_0),
\end{equation}
where repeated indices are summed over. Here, $v_{bk}(t,t_0)$ and $\Lambda_k(t,t_0)$ are the eigenvectors and eigenvalues, respectively, for the $k$th state. The GEVP eigenvalues, also referred to as principal correlators, can be written as
\begin{equation}
\Lambda_k(t,t_0) = e^{-(t-t_0)E_k(t)} 
\label{eq:princ_corr} 
\end{equation}
and
obey the interlacing theorem~\cite{Cauchy:1821,Parlett}, which guarantees the existence of at least $k$ FV energy eigenvalues at or below $E_k(t)$ for any $t$~\cite{Fleming:2023zml}. On the other hand, direct analysis of the time dependence of the eigenvalues obtained from this procedure has the drawback that they cannot be rigorously related to a spectral decomposition.

It is possible to choose a fixed reference time $t_\text{ref}$ and interpret the eigenvectors obtained from a single pair of $(t_\text{ref},t_0)$ 
as weights to define a set of overlap-optimized interpolating operators that provide a set of $N$ variational bounds on the lowest energy eigenvalues. These are given explicitly by the diagonal elements of the matrix built from elements
\begin{equation}\label{eq:GEVP_corr}
  \hat{C}_{kl}(t) = \sum_{ab} v_{ak}^{*}(t_\text{ref},t_0) C_{ab}(t) v_{bl}(t_\text{ref},t_0).
\end{equation}
This approach has the advantage that the $\hat{C}_{kl}(t)$ obey an analogous spectral decomposition to the $C_{ab}(t)$, and therefore may be analyzed by performing convex multi-state fits to the diagonal correlators~\cite{Bulava:2016mks}. In Ref.~\cite{Detmold:2024iwz}, an algorithm is defined which determines a pair of $t_0$ and $t_\text{ref}$ that preserves both the interlacing theorem interpretation and the spectral decomposition.

Principal correlators have several useful formal properties, although finite-statistics effects can complicate their interpretation in practice.
At finite $t$, the effective mass constructed from a principal correlator in Eq.~\eqref{eq:princ_corr} is  constrained to approach the ground state from above. Further, at infinite statistics a principal correlator effective mass that is exactly constant between two or more timeslices indicates a true eigenstate has been identified. However, practical LQCD calculations which employ Monte Carlo methods produce stochastic estimators of correlators and their corresponding effective masses. In the presence of noise, it is possible for the overlap factors onto the ground state to be small enough that the effective mass function is statistically consistent with a constant other than the ground state energy over any finite range of $t$.\footnote{For example, principal correlators designed for optimal overlap with excited states are tuned to produce effective masses that do not approach the ground-state energy until very large $t$. More generally, it is straightforward to construct examples of energies and overlaps where principal correlators for multi-hadron systems have effective masses that are approximately constant but differ from all energy eigenvalues for times comparable to $m_{\pi}^{-1}$ or $\Lambda_{\text{QCD}}^{-1}$ but much smaller than the inverse gap $\delta_1^{-1}$.}
At finite statistical precision, symmetric correlators have the clear advantage of providing stochastically valid upper bounds on energy eigenvalues, but, just as with asymmetric correlators, an approximately constant principal effective mass function is not necessarily a reliable indicator of the presence of a true energy eigenvalue.

\subsection{Lanczos methods}
\label{sec:theory-lanczos}

While the variational method produces upper bounds on FV energy eigenvalues, it does not localize the energy eigenvalues to a finite range. 
Recently, a Lanczos (or filtered Rayleigh-Ritz) method has been shown to give two-sided bounds on FV energy levels~\cite{Wagman:2024rid,Hackett:2024xnx,Ostmeyer:2024qgu,Chakraborty:2024scw,Hackett:2024nbe,Abbott:2025yhm}. 
In particular, applying $m$ iterations of the Lanczos algorithm~\cite{Lanczos:1950zz} to a rank-$r$ correlator matrix provides Ritz values 
$\lambda_k^{(m)}$ for $k \in \{0,\ldots,\mr-1\}$, which are algebraic estimators of transfer-matrix eigenvalues with appropriate quantum numbers~\cite{Parlett}.
The Ritz values are related to energy estimators as $E^{(m)}_k \equiv - \log \lambda^{(m)}_k$. 
The algorithm also produces right and left Ritz vectors $\ket{y_k^{R/L(m)}}$ of the $m$th-iteration Krylov-space approximation of $T$,
\begin{equation}
    T^{(m)} = \sum_k \ket{y^{R(m)}_k} \lambda^{(m)}_k \bra{y^{L(m)}_k} ~ .
\end{equation}
The right and left Ritz vectors are constructed as superpositions of the right and left Krylov vectors 
$\{\ket{k_{tb}^R}\} = \{T^t \ket{ \psi_b } ~|~ \forall\ b,t\}$ and $\{\bra{k_{ta}^L}\} = \{\bra{\psi_a} T^t ~ |~ \forall\ a,t\}$
respectively, where at zero temperature $\ket{\psi_b}$ denotes the state obtained by acting on the vacuum with the interpolator $\mathcal{O}_b^\dagger$.
The operator $T^{(m)}$ is Hermitian for symmetric correlators at infinite statistics.
At finite statistics, it is non-Hermitian in general; however, for symmetric correlators, it admits a ``Hermitian subspace'' of states 
where $\ket{y^{R(m)}_k} = \ket{y^{L(m)}_k}$ and the corresponding $\lambda^{(m)}_k$ are real up to arithmetic precision~\cite{Hackett:2024xnx}.

Recent works~\cite{Fischer:2020bgv,Wagman:2024rid,Ostmeyer:2024qgu,Chakraborty:2024scw} have explored equivalences between (oblique) Lanczos, 
Prony's method~\cite{Prony}, first applied to LQCD in Refs.~\cite{Fleming:2004hs,Lin:2007iq,Beane:2009kya,Beane:2009gs,Fleming:2009wb}, matrix Prony~\cite{Beane:2009kya,Beane:2009gs}, and the generalized pencil-of-functions (GPOF) method~\cite{Hua:1989,Sarkar:1995} first applied to LQCD in Refs.~\cite{Aubin:2010jc,Aubin:2011zz}.
Further work~\cite{Hackett:2024nbe,Abbott:2025yhm} has elucidated the relation between oblique block Lanczos~\cite{Golub:1973,Cullum:1974,Golub:1977,Kim:1988,Kim:1990,Bai:1999,Abdel-Rehim:2014}, block Prony~\cite{Fleming:2023zml}, and GPOF for correlation function matrices. 

It has been shown that all of these methods, which compute an exact decomposition of any $r \times r$ correlator matrices of length $2m$ of the form 
\begin{equation}\label{eq:prony_ritz}
   \bm{C}(t) = \sum_{k=0}^{\mr-1} \bm{A}_k e^{-t E_k},
\end{equation}
constitute
the Prony-Ritz equivalence class~\cite{Abbott:2025yhm}. 
In the above equation, $\bm{A}_k$   
is a rank-one matrix, which can be interpreted as the product of overlap factors between Ritz vectors and  initial and final operators, that is $[\bm{A}_k]_{ab} \equiv [Z^R_{ka}]^* Z^L_{kb} $, where $Z^{R/L(m)}_{ka} = \braket{y^{R/L(m)}_k | \psi_a}$.
Ritz values and (un-normalized) right Ritz coefficients $\vec{P}^{R(m)}$  can be obtained from the solution of a GEVP: 
\begin{equation}\label{eq:hankel_gevp}
\bm{H}^{(1)} \vec{P}^{R(m)}_k = \bm{H}^{(0)} \vec{P}^{R(m)}_k \lambda_k^{(m)},
\end{equation}
where $\bm{H}^{(p)}$ is an $\mr\times \mr$ block Hankel matrix~\cite{Abbott:2025yhm} with components
\begin{equation}
[\bm{H}^{(p)}_{ij}]_{ab} = C_{ab}(i + j + p),
\end{equation}
and $\vec{P}^{R(m)}$ is a $\mr$-component vector with $[\vec{P}^{R(m)}_k]_{ta} = P^{R(m)}_{tak}$.
The corresponding left Ritz coefficients can be obtained as $\bm{P}^L = [\bm{H}^{(0)} \bm{P}^R]^{-1}$. The Ritz coefficients must be normalized in an additional step to obtain unit-normalized Ritz vectors~\cite{Hackett:2024xnx,Hackett:2024nbe,Abbott:2025yhm}.

It can be shown~\cite{Hackett:2024nbe} that block Lanczos with $m=1$ is equivalent to application of the GEVP correlator analysis method~\cite{Luscher:1990ck,Blossier:2009kd}. For $m>1$, more temporal information is directly analyzed, leading to faster convergence of eigenvalue estimates than the standard GEVP approach. In fact, near the continuum limit, where $\delta_r = E_r - E_0$ is small in lattice units, the Kaniel-Paige-Saad (KPS) bound~\cite{Kaniel:1966,Paige:1971,Saad:1980} can be used to show that excited state effects are asymptotically suppressed by
$e^{-2t \sqrt{\delta_r}}$ for $\delta_r \ll 1$~\cite{Hackett:2024nbe}. This improves on the standard GEVP method, which suppresses excited state contamination as $e^{- \delta_r t}$. In addition to this theoretical convergence bound, two-sided residual bounds that apply even at finite iteration counts (and therefore finite Euclidean times) can be explicitly computed in the Lanczos approach, as discussed in \cref{sec:bounds} below.

One difficulty in implementations of any of the algorithms in the Prony-Ritz equivalence class is the presence of \textit{spurious eigenvalues}. 
These eigenvalues arise due to finite precision from Monte Carlo noise in the correlators and/or round-off error when the Lanczos or Prony algorithm is executed with insufficient numerical precision.
Therefore, in order to extract robust estimates of transfer matrix eigenvalues from these algorithms, a scheme must be chosen to separate spurious and physical eigenvalues (i.e., Ritz values converging towards transfer-matrix eigenvalues at large $m$).
The Cullum-Willoughby test~\cite{Cullum:1981,Cullum:1985} provides a useful method for this identification of spurious eigenvalues in the scalar Lanczos method~\cite{Wagman:2024rid}, but it does not directly generalize to the block Lanczos case. Instead, the so-called ZCW test~\cite{Hackett:2024nbe} can be used, identifying  Ritz vectors as spurious states and corresponding Ritz values as spurious eigenvalues if 
\begin{equation}\label{eq:ZCW}
    \left| \sum_{ab} Z^{R(m)*}_{ka} [\bm{C}(2m_0)^{-1}]_{ab} Z^{L(m)}_{k b} \right| < \varepsilon_{\rm ZCW} ~ ,
\end{equation}
where $\varepsilon_{\rm ZCW}$ is a hyperparameter that must be chosen. The physical motivation for the ZCW test is that states whose overlaps are below a threshold set by statistical precision will make contributions to correlator spectral representations that cannot be resolved from noise. Therefore, Ritz vectors with overlaps much smaller than this threshold are more likely to be spurious states arising from noise than genuine indications that a small-overlap state has been resolved. 
Thermal modes, whose contributions to correlators grow with $t$, provide an exception to this generic expectation and can be identified through a thermal ZCW test; see Ref.~\cite{Hackett:2024nbe} for more details.
Note that at a fixed level of statistical precision, only a small number of the (infinitely many) states in the spectrum will have squared overlaps larger than $\varepsilon_{\rm ZCW}$. The lack of non-spurious states with energies in a given region should only be used to conclude that any such states have overlaps with the interpolating operators that are too small to be resolved from noise. 

Any choice of $\varepsilon_{\rm ZCW}$ leads to asymptotically unbiased estimators for all states whose overlaps are large enough to survive \cref{eq:ZCW} and be labeled as non-spurious.\footnote{The ZCW test with fixed $\varepsilon_{\rm ZCW}$ is applicable in the infinite-statistics limit; see \cref{sec:scalar-inf} for an example of the ZCW test applied to exact results for an analytically solvable correlator.} An adaptive choice in which $\varepsilon_{\rm ZCW} \rightarrow 0$ in the infinite-statistics limit 
is to identify the ``last symmetric iteration,''
defined here as the last iteration where Ritz values are all real and in the range $[0,1]$ and the corresponding Ritz vectors all have positive norms, and define $\varepsilon_{\rm ZCW}$ as the value of the left-hand-side of \cref{eq:ZCW} for that iteration divided by a hyperparameter $F_{\rm ZCW}$~\cite{Hackett:2024nbe}.
Approaching the infinite-statistics limit at fixed $F_{\rm ZCW}$ guarantees that $\varepsilon_{\rm ZCW} \rightarrow 0$ in the same limit because all Ritz values will satisfy the constraints used to define the last symmetric iteration by construction.\footnote{Modifications to account for genuine complex eigenvalues arising for some lattice actions
are possible but appear irrelevant at computationally accessible statistics, where all complex eigenvalues appear spurious by other definitions~\cite{Wagman:2024rid,Hackett:2024xnx,Ostmeyer:2024qgu,Chakraborty:2024scw,Hackett:2024nbe}. }

Filtering based on the Cullum-Willoughby or ZCW tests is sufficient to remove many 
obviously pathological Ritz values (for example, those that would violate QCD inequalities~\cite{Weingarten:1983uj}) by identifying them as spurious~\cite{Wagman:2024rid,Hackett:2024xnx,Hackett:2024nbe}.
However, for certain Lanczos iteration counts, the non-spurious Ritz values can include apparent extra energy levels or dislocations in the spectrum where an energy level appears to be missing.
These features can be associated with statistical noise, as evidenced by their high sensitivity to the subset of correlator data used in calculations.
They can lead to the appearance of outliers in distributions of energy estimators that, in some cases, can be understood as failures of \emph{state identification}, the process of assigning ordered labels $k \in \{0,1,\ldots\}$ to a set of (spurious and non-spurious) Ritz values $E_k^{(m)}$. 

Bootstrap resampling~\cite{Efron:1982,Davison:1997,Young:2014} can be used to provide outlier-robust estimators that are useful for mitigating the effects of imperfect state identification. 
As in Refs.~\cite{Wagman:2024rid,Hackett:2024nbe,Ostmeyer:2024qgu,Chakraborty:2024scw,Hackett:2024xnx}, bootstrap-median estimators are employed in this work with a nested bootstrap method used to compute the corresponding uncertainties.
For each of $N_{\rm boot}$ (one-level) bootstrap ensembles defined by resampling a given correlator (matrix), quantities such as Ritz values $\lambda_k^{(m,b)}$ can be computed for each ensemble $b \in \{1,\ldots,N_{\rm boot}\}$.
Bootstrap median estimators for a generic quantity $X_k^{(m)}$ are then defined as
\begin{equation}
  \overline{X}_k^{(m)} = \text{median}_{b'} \ X_k^{(m,b')} ~ .
\end{equation}
To compute uncertainties self-consistently, a nested bootstrap scheme is employed: $N_{\rm boot}$ bootstrap samples are drawn by resampling each of the $N_{\rm outer}$ one-level bootstrap ensembles of correlator (matrices).\footnote{Numerical results below use $N_{\rm outer} = N_{\rm boot}$ for simplicity, but this is not required.}
The $X_k^{(m,b,b')}$ are then computed for these two-level bootstrap samples.
Nested-bootstrap-median estimators are then defined by
\begin{equation}
  \overline{X}_k^{(m,b)} = \text{median}_{b'} \  X_k^{(m,b,b')},
\end{equation}
i.e., a median over the inner bootstraps.

Symmetric empirical bootstrap confidence intervals can then be used to define $1\sigma$ uncertainties for $\overline{X}_k^{(m)}$ as~\cite{Davison:1997}
\begin{equation}\label{eq:CI}
\begin{split}
  \delta \overline{X}_k^{(m)} &= \frac{1}{2}\left[ \text{Quantile}\left(\overline{X}_k^{(m,b)}, \frac{1 + \text{erf}(1/\sqrt{2})}{2}\right) \right. \\
    &\hspace{30pt} \left. - \text{Quantile}\left(\overline{X}_k^{(m,b)}, \frac{1 - \text{erf}(1/\sqrt{2})}{2}\right) \right],
    \end{split}
\end{equation}
and $n\sigma$ confidence intervals by multiplying the argument of the error function by $n$.
For a Gaussian distribution, these $n\sigma$ confidence intervals  approximate  $n \delta \overline{X}_k^{(m)}$ as the large statistics limit is taken.

The ZCW (or the Cullum-Willoughby) test in conjunction with bootstrap-median estimators is sufficient to remove most spurious eigenvalues in applications to individual correlators or to relatively small correlator matrices~\cite{Wagman:2024rid,Hackett:2024nbe,Ostmeyer:2024qgu,Chakraborty:2024scw,Hackett:2024xnx}.
This enables state identification to be performed using a simple \emph{filter-and-sort} strategy, in which the $k$th largest non-spurious Ritz value on each bootstrap ensemble is associated with the $k$th state. 
However, it is observed below that pathological results, such as multiple Ritz values appearing for some $m$ near where only a single Ritz value appears for other $m$, occur frequently enough in applications to larger correlator matrices to necessitate more sophisticated state identification strategies.
State identification based on an alternative \emph{symmetric Lanczos Ritz-vector labeling} (SLRVL) strategy analogous to eigenvector-based labeling in GEVP applications~\cite{Dudek:2010wm,Fischer:2020bgv} is introduced in \cref{sec:lqcd-lanczos} below and found to be sufficient to mitigate the obvious pathologies resulting from filter-and-sort state identification for large correlator matrices.

Other filtering and state identification approaches have been explored~\cite{Cushman:2019hfh,Chakraborty:2024scw,Ostmeyer:2025igc,Tsuji:2025zdn}.
Algorithms based on identifying peaks in Ritz-value distributions and sorting based on proximity to these peaks have been successfully applied in some cases~\cite{Cushman:2019hfh,Chakraborty:2024scw}, but for multi-hadron systems it is sometimes impossible to distinguish separate peaks associated with near-degenerate states (see~\cref{fig:histograms:deut} below for an example) without the additional eigenvector information used in Ritz-vector labeling.
Methods have also recently been proposed in which spurious eigenvalue filtering is performed in a basis other than the Ritz-vector basis by modifying negative eigenvalues of Hankel matrices~\cite{Ostmeyer:2025igc,Tsuji:2025zdn}.
This leads to bias in non-spurious Ritz values that must be removed through an extrapolation~\cite{Tsuji:2025zdn}.
The resulting unbiased estimators provide an interesting alternative to ZCW filtering and Ritz vector labeling whose application to multi-hadron spectroscopy is left to future work.

An issue that is common to all Prony-Ritz methods at finite statistics is \emph{stagnation},
which refers to undesirably slow convergence phenomena that have been observed in linear algebra applications of oblique Lanczos to finite-dimensional matrices~\cite{Leyk:1997,Gaaf:2016,Saad:2011} and discussed in the context of LQCD correlator analysis in Refs.~\cite{Hackett:2024xnx,Abbott:2025yhm}.
Stagnation in (noiseless) linear algebra contexts typically occurs when there is small overlap between vectors in the spans of the right and left Krylov vectors. 
Stagnation manifests in highly non-monotonic convergence patterns in which a set of Ritz values gives apparently stable results for many iterations but has one or more subsequent periods of relatively rapid convergence to true eigenvalues~\cite{Leyk:1997,Gaaf:2016,Saad:2011}. 
In the presence of statistical noise, all Lanczos applications must formally be viewed as oblique Lanczos and therefore stagnation is a concern~\cite{Hackett:2024xnx,Abbott:2025yhm}.
This implies that stability of Lanczos energy-estimators for many iterations does not necessarily signal that convergence has been achieved but may signal either that convergence has been achieved or the appearance of finite-statistics stagnation resulting from  poor overlap between the spans of right- and left-Krylov vectors due to noise at large $t$.
$N$-state-saturation assumptions are required in order to interpret statistical uncertainties as quantitative measures of the errors in Lanczos energy estimates, just as with analyses involving correlator fits and/or GEVP.

\section{Bounds on energy spectra: general formalism}\label{sec:bounds}

In LQCD studies of multi-hadron systems, $e^{-\delta_1 t}$ is often not a small parameter, leading to excited state effects that are not parametrically suppressed and difficulties in quantifying the resultant systematic uncertainties.
This motivates the study of two-sided bounds on the approximation errors of Lanczos energy estimates, which instead rely on the smallness of an explicitly calculable residual-norm-square parameter $B$. In this section, the formalism for variational, residual, and gap bounds at infinite statistics in Hermitian systems is introduced. Complications arising for finite-statistical precision estimates of correlation functions, where bounds are only valid stochastically, are deferred to \cref{sec:SHO}, below.

\subsection{Variational bounds}\label{sec:var}

{The variational principle of quantum mechanics has been applied to studies of spectroscopy for nearly a century~\cite{Hylleraas:1928,ABDELRAOUF1982163,suzuki2002stochastic} and has been used to prove the existence of atomic and nuclear bound states in various theories.
Variational methods in quantum field theory have also been applied to LQCD for many years~\cite{Wilson:1981,Fox:1981xz,Berg:1981zb,Falcioni:1981mm,Michael:1982gb,Luscher:1990ck,Fernandez:1987ph} and are often implemented by solving GEVPs as discussed in \cref{sec:theory_gevp}.
Variational bounds have also been used to constrain two-nucleon energy spectra in recent works~\cite{Amarasinghe:2021lqa,Detmold:2024iwz}.

The simplest variational bound
\begin{equation}
    E_0 \leq -\ln \frac{\braket{\psi|T|\psi}}{\braket{\psi|\psi}},
\end{equation}
can be proven straightforwardly for any state $\ket{\psi}$ by inserting a complete set of transfer matrix eigenstates $\ket{n}$ with eigenvalues $\lambda_n$.
The key point is that the ground-state energy $E_0 = - \ln \lambda_0$ is related to the largest eigenvalue of $T$, which is the maximum value of any normalized symmetric matrix element of $T$ (often called a Rayleigh quotient in linear algebra contexts)~\cite{Parlett,Abbott:2025yhm}.
Since $\lambda_0^{(m)}$ is a Rayleigh quotient of $T$ for the Ritz vector $\ket{y_0^{(m)}}$, in (infinite-statistics) Lanczos applications to symmetric correlation functions~\cite{Wagman:2024rid,Hackett:2024xnx} the Lanczos energy estimators $E_0^{(m)} = -\ln \lambda_0^{(m)}$ therefore satisfy
\begin{equation}
    E_0 \leq E_0^{(m)}.
\end{equation}
Effective masses for symmetric correlation functions provide variational bounds by similar arguments (or by taking $m=1$). 
Variational bounds are not applicable for asymmetric correlation functions because the spectral representation can include terms with negative coefficients or, equivalently, because their effective masses are not (symmetric) Rayleigh quotients; see \cref{sec:asymm}.
Further variational bounds
\begin{equation}
   E_k \leq E_k^{(m)} 
    \label{eq:variational_bound}
\end{equation}
follow from Cauchy's interlacing theorem~\cite{Cauchy:1821,Parlett}, implying that there are at least $k+1$ energy levels with energy less than or equal to $E_k^{(m)}$; see Refs.~\cite{Fleming:2023zml,Detmold:2024iwz}.
This interpretation as variational bounds applies to Lanczos energy estimators as well as to GEVP energy estimators.

An appealing feature of variational bounds is that they apply directly to the $k$th energy eigenvalue, rather than to an unspecified energy eigenvalue.
However, variational bounds are one-sided and cannot be used to provide evidence that the ground-state energy is larger than a certain value, as emphasized in Refs.~\cite{Amarasinghe:2021lqa,Detmold:2024iwz}.
When only variational bounds are available, it must be assumed that energy estimators have saturated these bounds in order to provide meaningful two-sided constraints on the LQCD spectrum.
This is equivalent to an $N$-state-saturation assumption, as introduced in \cref{sec:strategies} above.
Additionally, it must be noted that variational bounds hold only in the limit of zero temperature, as the transfer matrix energy spectrum is unbounded from below at any non-zero temperature.

\subsection{Residual bounds}\label{sec:residual}

Two-sided \emph{residual bounds} that are valid for finite Euclidean times can be derived in the Lanczos/Rayleigh-Ritz framework~\cite{Wagman:2024rid,Hackett:2024xnx,Hackett:2024nbe,Abbott:2025yhm} and provide an almost assumption-free approach to LQCD spectroscopy, requiring only Hermiticity of the transfer matrix.
Residual-bound estimators for a unit-normalized state $\ket{y}$ can be explicitly obtained through calculation of the norm-square $\left\| [ T - T^{(m)} ] \ket{y} \right\|^2 $, denoted $B$ below (often including sub/superscripts needed to label the associated state and Lanczos iteration). This quantity describes the finite-$m$ truncation error of replacing $T$ with its $m$-iteration Lanczos approximation $T^{(m)}$ when acting on this state.

The residual-norm-square $B$ decreases monotonically for the ground state in symmetric Lanczos applications and vanishes for Ritz vectors if and only if they are identical to eigenstates of $T$.
In many numerical linear algebra applications, it is common to bound the errors of using Ritz values as approximations to eigenvalues based on the size of $B$~\cite{Parlett}. 
In this way, $B$ can serve as a small parameter that can be used to bound the size of excited-state effects on finite-$t$ energy estimates even when $e^{-\delta_1 t}$ is not small.

Concretely, the residual bounds are inequalities guaranteeing that at least one genuine transfer-matrix eigenvalue $\lambda_n$ lies within a two-sided window of each Ritz value~\cite{Parlett,Wagman:2024rid},
\begin{equation}
    \min_{\lambda \in \{\lambda_n\}} \left| \lambda^{(m)}_k - \lambda \right| \leq  \sqrt{ B^{R/L(m)}_k }.
    \label{eq:residual_bound}
\end{equation}
Note that the eigenvalues associated with residual bounds for each Ritz value are not necessarily distinct in cases where multiple bounds overlap.
The fundamental quantities required to compute the residual bounds are the residual-norm-squares
\begin{equation} 
    B^{R/L(m)}_k \equiv  
     \frac{ \braket{y^{R/L(m)}_k | [T - T^{(m)}]^\dagger [T-T^{(m)}] | y^{R/L(m)}_k} }{ \braket{y^{R/L(m)}_k | y^{R/L(m)}_k} }.
     \label{eq:Bdef}
\end{equation}
Note that \cref{eq:residual_bound} applies simultaneously with either $B^{R(m)}_k$ or $B^{L(m)}_k$ on the right-hand side.\footnote{For states in the Hermitian subspace, $B_k^{R(m)}$ and $B_k^{L(m)}$ are identical (up to effects of numerical precision), even at finite statistics. 
For asymmetric correlators and Ritz vectors outside the Hermitian subspace in symmetric correlator analyses,
one may
define $B_k^{(m)}$ as 
the element of $\{ B_k^{R(m)}, B_k^{L(m)} \}$ with the minimum absolute value, at the bootstrap level.}
Estimators for $B^{R/L(m)}_k$ can be obtained in practice from the quantities appearing in the Lanczos recursions and norm-trick test~\cite{Wagman:2024rid,Hackett:2024xnx}, or they can be derived from Hankel matrices in the Rayleigh-Ritz perspective~\cite{Abbott:2025yhm} as
\begin{widetext}
\begin{equation}
    B^{L(m)}_k     = 
        \frac{\left[ \bm{P}^{L} \bm{H}_{LL}^{(2)} \bm{P}^{L\dagger} \right]_{kk}
        - 2 \mathrm{Re}[\lambda^{(m)}_k] \left[ \bm{P}^{L} \bm{H}_{LL}^{(1)} \bm{P}^{L\dagger} \right]_{kk}
        + |\lambda^{(m)}_k|^2 \left[ \bm{P}^{L} \bm{H}_{LL}^{(0)} \bm{P}^{L\dagger} \right]_{kk}}{ \left[ \bm{P}^{L} \bm{H}_{LL}^{(0)} \bm{P}^{L\dagger} \right]_{kk} }
     .
     \label{eq:Bexplicit}
\end{equation}
\end{widetext}
An analogous formula holds for $B_k^{R(m)}$~\cite{Abbott:2025yhm}.
Assuming $T=T^\dagger$, the Hankel matrices $\bm{H}_{LL}^{(p)}$ in \cref{eq:Bexplicit} can be identified with $\bm{H}^{(p)}$ for block Lanczos applications to symmetric correlator matrices.
In this case, for states in the Hermitian subspace $\bm{P}^L_k = [\bm{P}^R_k]^\dagger$ and $B^{R(m)}_k = B^{L(m)}_k$.
In applications to asymmetric correlators $C_{LR}(t) = \braket{ \psi^L | T^t | \psi^R}$ where $\ket{\psi^L} \neq \ket{\psi^R}$, under the same $T = T^\dagger$ assumption, $\bm{H}^{(p)}_{LL}$ can be identified with the Hankel matrix constructed from the corresponding symmetric correlator $C_{LL}(t) = \braket{\psi^L | T^t | \psi^L}$, and similarly for $\bm{H}^{(p)}_{RR}$.
The residual norms can be computed for all but the last two timeslices where correlator data is available. In particular, if $\ket{y}$ is a vector in the (oblique, block) Krylov space associated with correlator data from $[t_0, t]$, then $B$ can be computed if and only if data from $[t_0, t+2]$ are available and thus $\braket{y | T^2 | y}$ is computable.

Determinations of $B^{R/L(m)}_k$ provide  the upper and lower edges of two-sided windows that are guaranteed to contain at least one transfer-matrix eigenvalue $\lambda_n$ by \cref{eq:residual_bound},
\begin{equation}
  \lambda_n \in \left[ \lambda_k^{(m)} - \sqrt{B_k^{R/L(m)}},\  \lambda_k^{(m)} + \sqrt{B_k^{R/L(m)}} \right].
\end{equation}
Taking logarithms of the window edges provide analogous two-sided windows that are guaranteed to contain at least one genuine energy eigenvalue $E_n = -\ln \lambda_n$,
\begin{equation}
\begin{split}
   E_n &\in \left[ -\ln\left(\lambda_k^{(m)} + \sqrt{B_k^{R/L(m)}}\right), \right. \\
   &\hspace{20pt} \left. \  -\ln\left(\lambda_k^{(m)} - \sqrt{B_k^{R/L(m)}}\right) \right].
   \label{eq:E_window}
   \end{split}
\end{equation}
However, it is important to }note that the residual bounds can only constrain where \emph{one or more} eigenvalue(s) must exist. They do not guarantee that an interval contains any \emph{particular} eigenvalue such as~the ground state $E_0$. 
They also do not provide regions in which eigenvalues are guaranteed not to exist. 
Unlike variational bounds, residual bounds remain valid at non-zero temperature. 

Asymmetric and symmetric correlation functions both satisfy residual bounds provided that $T$ is Hermitian~\cite{Wagman:2024rid}.
In the asymmetric case, one or both of the symmetric correlation functions involving the source and/or sink operator are required to compute $B_k^{R/L(m)}$ in practice.

\subsection{Gap bounds}\label{sec:gap}

In numerical linear algebra applications, residual bounds often give conservative overestimates of the actual differences between Ritz values and target matrix eigenvalues~\cite{Parlett}.
In particular, in regimes where $B \ll 1$, errors in Lanczos-estimated eigenvalues tend to scale as $O(B)$ while residual bounds are parametrically larger, scaling as $O(\sqrt{B})$, and therefore underestimate how rapidly Lanczos/Rayleigh-Ritz approximations converge to the true eigenvalues.
This has motivated the development of rigorous $O(B)$ bounds on Ritz value approximation errors, a topic that has a long history~\cite{Parlett} dating back to the Davis-Kahan theorem \cite{Davis:1970} in 1970.
This theorem shows that Ritz-value approximation errors are bounded by $B$ times a coefficient that depends on the spectral gap between the closest and next-to-closest eigenvalues of the target matrix.

Two-sided bounds provided by the Davis-Kahan theorem and its refinements, collectively called \emph{gap bounds} below, require an explicit assumption about the size of this spectral gap. 
Although this means that gap bounds are more assumption-dependent than variational and residual bounds, only an explicit assumption about the spectrum of $T$ is required. 
There are no assumptions about interpolating-operator overlaps with eigenstates required for the validity of gap bounds.
Further, gap bounds are valid even when $e^{-\delta_1 t}$ is not small and/or when the spectrum includes near or exact degeneracies.

The utility of gap bounds is that they can provide accurate estimates of the sizes of Ritz-value approximation errors that are provably tight in some cases~\cite{Davis:1970,Parlett,Demmel:1997,Haas:2025}.
All gap-bound estimates are $O(B)$, and the $O(1)$ prefactors that enter the definitions of the various gap bounds require priors on the size of spectral gaps. Therefore, in the context of lattice field theories where the spectrum is not known, gap bounds only serve as  estimates of systematic uncertainties rather than assumption-free bounds.

The Davis-Kahan theorem~\cite{Davis:1970} and its refinements lead to gap bounds of the form~\cite{Parlett,Demmel:1997}
\begin{equation}
   \min_{\lambda \in \{\lambda_n\}} \left| \lambda^{(m)}_k - \lambda \right| \leq  \frac{ B^{R/L(m)}_k   }{ g_{k}^{(m)} }.
    \label{eq:gap_bound}
\end{equation}
Here, the gap parameter $g_k^{(m)}$ may be defined as~\cite{Parlett,Demmel:1997}
\begin{equation}\label{eq:gkdef}
   g_k^{(m)} \equiv \min_{\lambda_{n'} \in \{\lambda_n \neq \lambda\}} \left|  \lambda_{n'} - \lambda_k^{(m)} \right|,
\end{equation}
where again $\lambda$ is the closest eigenvalue to $\lambda_k^{(m)}$.
In this definition, $g_k^{(m)}$ is the gap between the Ritz value $\lambda^{(m)}_k$ and the second-closest eigenvalue of $T$ to $\lambda^{(m)}_k$.
Other variations of gap bounds that differ in the precise definition of the gap parameter $g_k^{(m)}$ can also be used~\cite{Demmel:1997}.
Provided  $g_k^{(m)}$ is known, gap bounds provide two-sided windows guaranteed to contain a genuine energy eigenvalue analogous to \cref{eq:E_window},
\begin{equation}
\begin{split}
   E_n &\in \left[ -\ln\left(\lambda_k^{(m)} + \frac{B_k^{R/L(m)}}{g_k^{(m)}} \right), \right. \\
   &\hspace{20pt} \left. \  -\ln\left(\lambda_k^{(m)} - \frac{B_k^{R/L(m)}}{g_k^{(m)}} \right) \right].
   \label{eq:E_window_gap}
\end{split}
\end{equation}

If it can be safely assumed that all $\lambda_{n'}$ in \cref{eq:gkdef} are 
farther from $\lambda_k^{(m)}$ than some value $\lamgap$, then the gap parameter estimate
\begin{equation}\label{eq:gkEgap}
  \hat{g}^{(m)}_k \equiv \left| \lamgap - \lambda_k^{(m)} \right| ~ ,
\end{equation}
will lead to valid two-sided bounds
\begin{equation}
\begin{split}
  E_n &\in \left[ -\ln\left(\lambda_k^{(m)} + \frac{B_k^{R/L(m)}}{\hat{g}_k^{(m)}} \right), \right. \\
   &\hspace{20pt} \left. \  -\ln\left(\lambda_k^{(m)} - \frac{B_k^{R/L(m)}}{\hat{g}_k^{(m)}} \right) \right].
   \label{eq:E_window_gap_est}
   \end{split}
\end{equation}
The assumption that there is no other $T$ eigenvalue closer to $\lambda_k^{(m)}$ than $\lamgap$ can be viewed as an instance of the no-missing-states assumption described in \cref{sec:strategies}.
As with residual bounds, gap bounds do not guarantee which eigenvalue is in a given bounding window.

One option that has been used to estimate gap bounds in a variety of contexts~\cite{Parlett,Demmel:1997} is to set $\lamgap$ equal to whichever of $\lambda_{k+1}^{(m)}$ or $\lambda_{k-1}^{(m)}$ is closer to $\lambda_k^{(m)}$.
However, it must be emphasized that this does not guarantee that  the corresponding gap bounds are valid, because it is always possible for additional unexpected energies to be present in an \emph{a priori} unknown energy spectrum.\footnote{Ref.~\cite{Parlett} further suggests a somewhat more conservative definition
\begin{equation}
\begin{split}
  \hat{g}_k^{(m)} =& \min_{\lambda_{l'}^{(m)} \in \{\lambda_l^{(m)} \neq \lambda_k^{(m)}\}} \left| |\lambda_{l'}^{(m)}  - \sqrt{B_{l'}^{R/L(m)}}|  \right. \\
   &\hspace{20pt}\left. - |\lambda_l^{(m)} + \sqrt{B_{l}^{R/L(m)}}| \right|,
   \label{eq:gap_approx_P}
   \end{split}
\end{equation}
but this definition is no more rigorous than the one described in the main text, which corresponds to \cref{eq:gap_approx_P} with $O(\sqrt{B})$ terms neglected.
Including $O(\sqrt{B})$ terms in the definition of $\lamgap$ in LQCD analyses has the unappealing feature that the relatively large variance of $B_l^{R/L(m)}$ leads to much larger variance in $\hat{g}_k^{(m)}$.}
This is particularly relevant for LQCD calculations of multi-particle systems, where  it is common for simple interpolating operators to have large overlap with some low-energy state but very small overlaps with others.
Practical strategies for estimating gap parameters in LQCD are discussed below in \cref{sec:mind_the_gap}.

Gap bounds apply to symmetric matrix elements $\braket{y | T | y} / \braket{y|y}$ computed using any state $\bigl| y \bigr>$.
This includes, but is not limited to, Ritz values computed from the symmetric Lanczos algorithm.
However, gap bounds do not constrain Ritz values computed from asymmetric correlators.
Instead, in this case, gap bounds apply only for symmetric Rayleigh quotients of the distinct $R/L$ Ritz vectors, which can each be computed with auxiliary results from the corresponding symmetric correlation functions,
as discussed in \cref{sec:asymm}.

\subsection{Haas-Nakatsukasa gap bounds}\label{sec:improved_gap}

Intuitively, one may expect that the effects of a nearby energy level on a given Ritz value can be approximately incorporated by performing a block Lanczos analysis using an additional interpolating operator that strongly overlaps with that nearby level.
In this case, the given Ritz vector will be automatically orthogonalized against an approximation to the nearby energy eigenstate, and any residual excited-state effects associated with that level should be small.
This intuition is formalized in Ref.~\cite{Haas:2025}, which extends the Davis-Kahan gap bounds discussed above in order to provide tighter gap bounds for Rayleigh-Ritz applications.

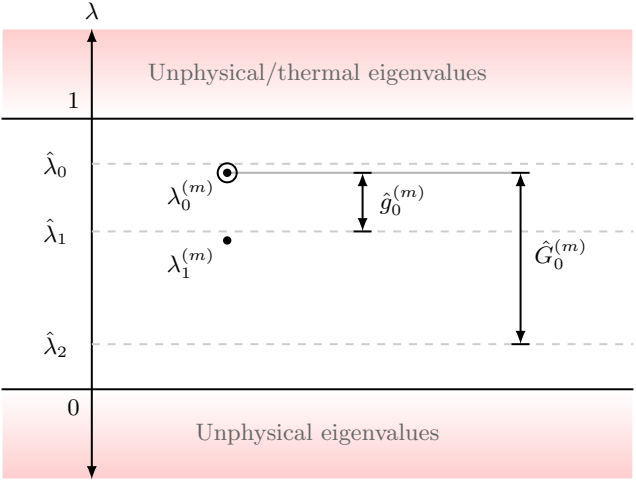
\begin{figure}
\centering
\begin{tikzpicture}[scale=1.2]
\filldraw[white, shading = axis, left color=red!20, right color=white, shading angle=180] (-1,-1) rectangle (6,0);
\filldraw[white, shading = axis, left color=red!20, right color=white, shading angle=0] (-1,3) rectangle (6,4);

\draw[<->] (0,-1) -- (0,4);
\draw[-] (-1,3) -- (6,3);
\draw[-] (-1,0) -- (6,0);

\node at (-0.2,3.2) {1};
\node at (-0.2,-0.2) {0};
\node at (0,4.2) {$\lambda$};

\node at (-0.4, 2.5)  {$\hat{\lambda}_0$};
\node at (-0.4, 1.75) {$\hat{\lambda}_1$};
\node at (-0.4, 0.5)  {$\hat{\lambda}_2$};

\draw[black!20,dashed] (0,2.5) -- (6,2.5);
\draw[black!20,dashed] (0,1.75) -- (6,1.75);
\draw[black!20,dashed] (0,0.5) -- (6,0.5);

\draw[black!30] (1.5,2.4) -- (4.75,2.4);

\node at (1.1, 2.15) {$\lambda_0^{(m)}$};
\node at (1.1, 1.40) {$\lambda_1^{(m)}$};

\filldraw[black] (1.5, 2.4) circle (1pt);
\filldraw[black] (1.5, 1.65) circle (1pt);
\draw[black]     (1.5, 2.4) circle (3pt);

\node at   (3.45,2.1) {$\hat{g}_0^{(m)}$};
\draw[<->] (3.0,1.75) -- (3.0,2.4);
\draw[-]   (2.9,1.75) -- (3.1,1.75);
\draw[-]   (2.9,2.4)  -- (3.1,2.4);

\node at   (5.20,1.5) {$\hat{G}_0^{(m)}$};
\draw[<->] (4.75,0.5) -- (4.75,2.4);
\draw[-]   (4.65,0.5) -- (4.85,0.5);
\draw[-]   (4.65,2.4) -- (4.85,2.4);

\node[black!60] at (2.5, 3.5) {Unphysical/thermal eigenvalues};
\node[black!60] at (2.5, -0.5) {Unphysical eigenvalues};

\end{tikzpicture}
\caption{
Example Davis-Kahan and Haas-Nakatsukasa gap parameter estimates $\hat{g}_0^{(m)}$ and $\hat{G}_0^{(m)}$ for $\lambda_0^{(m)}$ (circled point) with estimated $T$ eigenvalues shown as dashed lines and Ritz values shown as points. Since $\hat{G}_1^{(m)}> \hat{g}_1^{(m)}$, Haas-Nakatsukasa gap bounds will be tighter than the Davis-Kahan gap bounds in this case. Note that if the estimated $\lambda$ spectrum depicted was in fact incomplete and additional unexpected eigenvalues were present in the true spectrum, then gap bounds computed using either of these approaches could give incorrect bounds. In the case of Davis-Kahan gap bounds, this will occur if a second true eigenvalue is closer to $\lambda_0^{(m)}$ than $\hat{\lambda}_1$, while Haas-Nakatsukasa gap bounds will additionally fail if a third true eigenvalue is closer to $\lambda_0^{(m)}$ than $\hat{\lambda}_2$.
}
\label{fig:gap_parameter}
\end{figure}

The Haas-Nakatsukasa gap bounds derived in Ref.~\cite{Haas:2025} take the same basic form as the Davis-Kahan gap bounds for symmetric Rayleigh-Ritz:
\begin{equation}
   \min_{\lambda \in \{\lambda_n\}} \left| \lambda^{(m)}_k - \lambda \right| \leq  C_k^{(m)} \frac{ B^{R/L(m)}_k}{ G_{k}^{(m)} },
    \label{eq:Gap_bound}
\end{equation}
but differ in the prefactor $C_k^{(m)}$ and a different form of gap parameter $G_{k}^{(m)}$ in the denominator.
The benefit of Haas-Nakatsukasa gap bounds is that $G_k^{(m)}$ may be significantly larger than $g_k^{(m)}$ in \cref{eq:gap_bound}. 
The gap parameter $G_k^{(m)}$ is defined from the difference between $\lambda_k^{(m)}$ and the nearest $T$ eigenvalue that is not in one-to-one correspondence with another Ritz value. 
This difference is shown pictorially in Fig.~\ref{fig:gap_parameter}.
This can also be thought of as the smallest gap between a given Ritz value and the $T$ eigenvalues of Hilbert space states that have small overlap with all Ritz vectors present in a calculation.

Haas-Nakatsukasa gap bounds are complicated by the additional prefactor $C_k^{(m)}$ absent in \cref{eq:gap_bound}.
In practical applications, $B\ll 1$, and so $C^{(m)}_k$ can be expanded in $B$. 
Theorem~1 of Ref.~\cite{Haas:2025} shows that
$C_k^{(m)} = 1 + O(B)$ and therefore,
\begin{equation}
   \min_{\lambda \in \{\lambda_n\}} \left| \lambda^{(m)}_k - \lambda \right| \leq  \frac{ B^{R/L(m)}_k  }{ G_{k}^{(m)} }\left[ 1 + O\left( B^{R/L(m)}_k \right) \right].
    \label{eq:Gap_bound_exp}
\end{equation}
In the numerical results below, $B \ll 10^{-2}$ is always achieved after a few Lanczos iterations. 
Further, in all examples below (involving symmetric correlator matrices where gap bounds are computable) $B$ becomes consistent with zero within $1\sigma$ statistical uncertainties after several iterations.
In this situation, the statistical uncertainties on estimates of the right-hand-side of \cref{eq:Gap_bound_exp} are orders of magnitude larger than the $O(B^2)$ factors arising from sub-leading terms in this Taylor expansion. These terms are therefore neglected in the following analysis.
This leads to constraints on energy eigenvalues analogous to \cref{eq:E_window_gap},
\begin{equation}
\begin{split}
   E_n &\in \left[ -\ln\left(\lambda_k^{(m)} + \frac{B_k^{R/L(m)}}{G_k^{(m)}} \right), \right. \\
   &\hspace{20pt} \left. \  -\ln\left(\lambda_k^{(m)} - \frac{B_k^{R/L(m)}}{G_k^{(m)}} \right) \right].
   \label{eq:E_window_Gap}
   \end{split}
\end{equation}

As with Davis-Kahan gap bounds, estimating the gap parameter $G_k^{(m)}$ requires a no-missing-states assumption.
The essential difference is that, for Haas-Nakatsukasa gap bounds, if it is assumed that exactly $N$ energy levels are present below $\Egap$, then the gap parameter estimate
\begin{equation}\label{eq:GkEgap}
  \hat{G}_k^{(m)} \equiv \left| e^{-\Egap} - \lambda_k^{(m)} \right|,
\end{equation}
will lead to the two-sided bounds
\begin{equation}
\begin{split}
  E_n &\in \left[ -\ln\left(\lambda_k^{(m)} + \frac{B_k^{R/L(m)}}{\hat{G}_k^{(m)}} \right), \right. \\
   &\hspace{20pt} \left. \  -\ln\left(\lambda_k^{(m)} - \frac{B_k^{R/L(m)}}{\hat{G}_k^{(m)}} \right) \right]
   \label{eq:E_window_Gap_est}
   \end{split}
\end{equation}
for all Ritz values with $k \in \{0,\ldots,N-1\}$.
In particular, under no-missing-states assumptions, Haas-Nakatsukasa gap bounds for the ground state are sensitive to the gap between $E_0^{(m)}$ and $E_N$, while Davis-Kahan gap bounds involve the gap between $E_0^{(m)}$ and $E_1$.
Depending on the spectrum and details of the correlator matrix used, $\hat{G}_k^{(m)}\gg \hat{g}_k^{(m)}$ can be realized, making  Haas-Nakatsukasa gap bounds significantly stronger than Davis-Kahan gap bounds.

Practical estimation of $\Egap$ depends on the physics of the energy spectrum in question and is discussed for two-nucleon systems in LQCD in \cref{sec:mind_the_gap}.

\subsection{Combining gap and variational bounds}\label{sec:gap_var}

The no-missing-states assumption required to produce gap bounds is also sufficient to allow variational bounds to be usefully combined with gap bounds at zero temperature.
Under this assumption---that exactly $N$ states are present below $\Egap$---each Lanczos energy estimator satisfying $E_k^{(m)} < \Egap$ provides a variational  bound on the corresponding $E_k$.
It is straightforward to  test explicitly that exactly $N$ Ritz values are also present below $\Egap$, i.e., that the no-missing levels-assumption is valid for the spectrum estimated from the Ritz values.\footnote{If this is not the case, it signals that additional interpolating operators are required to accurately determine the spectrum and/or that the $\Egap$ assumption needs to be modified.} 
If this is the case, gap bounds for $E_k^{(m)}$ can be interpreted as constraints on particular energies $E_k$ rather than on an unspecified $E_n$.

Under this no-missing-states assumption, the intersection of Davis-Kahan gap and variational bounds on $E_k$ is itself a valid bound and constrains $E_k$ to lie in the two-sided window 
\begin{equation}
   E_k\in\left[ -\ln\left(\lambda_k^{(m)} + \frac{B_k^{R/L(m)}}{g_k^{(m)}} \right), \  E_k^{(m)} \right].
   \label{eq:E_window_gap_var}
\end{equation}
Identical logic applies to combinations of Haas-Nakatsukasa gap and variational bounds, which therefore lead to two-sided constraints on $E_k$ of the form
\begin{equation}
   E_k\in\left[ -\ln\left(\lambda_k^{(m)} + \frac{B_k^{R/L(m)}}{G_k^{(m)}} \right), \  E_k^{(m)} \right],
   \label{eq:E_window_Gap_var}
\end{equation}
where $O(B^2)$ corrections to the lower bound are neglected as described in \cref{sec:improved_gap}.

\section{Bounds on energy spectra: noisy data analysis}
\label{sec:SHO}

Noise has non-trivial effects on the quantities entering the computation of the two-sided bounds on energy eigenvalues.
In particular, $B_k^{R/L(m)}$ is positive for a Hermitian $T^{(m)}$, but stochastic estimators lead to estimates of $B_k^{R/L(m)}$ which are frequently negative.\footnote{This is possible due to the assumption of $T = T^\dagger$ in the derivation of $B$~\cite{Wagman:2024rid,Abbott:2025yhm}, which is violated by the  non-Hermiticity of $T$ arising from noise in the Lanczos/Rayleigh-Ritz treatment.}
While stochastic estimates of $B_k^{R/L(m)}$ are correct in expectation, these non-positive fluctuations complicate the interpretation of bootstrap-by-bootstrap bounds. 
Therefore, in order to interpret the bound estimator in a stochastic sense, it is necessary to define a prescription to handle negative residual-norm-squares.

To explore such a prescription, stochastic estimates of bounds are computed in an exactly solvable toy model and compared to exact results. In this study, residual bounds are found to be very conservative both at finite and infinite statistics, while gap bound constraints at a given statistical confidence level are found to range from well-calibrated to conservative depending on the size of excited-state effects and statistical precision.
This behavior is contrasted with that of energy estimators---both from multi-state fits and from Ritz values---which give constraints under 
$N$-state-saturation assumptions that systematically underestimate true uncertainties in examples where excited-state effects are large.

\subsection{Toy model: complex harmonic oscillator}

The lattice-regularized complex harmonic oscillator is an exactly solvable $0+1$-dimensional lattice field theory defined by the action
\begin{equation}
\begin{split}
    S_\varphi &= \sum_{t=0}^{L_t - 1} \left\lbrace \varphi(t)^* \left[ -\varphi(t + 1) + 2\varphi(t)- \varphi(t - 1) \right] \right. \\
    &\hspace{50pt} \left. + M^2 |\varphi(t)|^2  \right\rbrace,
    \end{split}
\end{equation}
where $\varphi$ is a complex scalar field with periodic boundary conditions where $\varphi(t) \equiv \varphi(t\text{ mod } L_t)$.
The family of interpolating operators
\begin{equation}\label{eq:Ophi}
  \mathcal{O}_{\varphi}^k(t) = e^{k\varphi(t)} - 1
\end{equation}
leads to correlator matrices
\begin{equation}\label{eq:Cphi}
  C^{kk'}_{\mathcal{O}}(t) = \left< \mathcal{O}_{\varphi}^k(t) \mathcal{O}_{\varphi}^{k'}(0)^* \right>,
\end{equation}
which overlap with the single-particle ground-state with energy $E_0 = 2 \, \text{arcsinh}(M/2)$, as well as with single-particle states with energies $E_n=(2n+1) E_0$ and multi-particle states with energies $E_n = (n+1)E_0$ that are identical to a harmonic oscillator spectrum apart from $n$-dependent multiplicities that are unimportant to the discussion below.
These correlator matrices feature significant excited-state contamination, as well as exponential signal-to-noise problems analogous to those of baryon correlators in LQCD~\cite{Wagman:2016bam,Detmold:2018eqd,Detmold:2025mhw}.
For the numerical demonstrations below, $L_t = 64$ and $M = 0.25$.

This toy model has several practically useful features.
First, for the particular family of interpolating operators in \cref{eq:Ophi}, it is possible to compute correlator matrix elements at finite $t$ exactly as
\begin{equation}\label{eq:Cphi_exact}
  C_{\mathcal{O}}^{kk'}(t) = \exp\left(kk'D^{-1}_{t0}\right) - 1,
\end{equation}
where 
\begin{equation}
D_{tt'} = (2+M) \delta_{tt'} - \delta_{t(t'+1)} - \delta_{(t+1)t'},
\end{equation}
is the $L_t \times L_t$ matrix defining the quadratic action $S_{\varphi}$, with the Kroenecker deltas defined mod $L_t$ to implement periodic boundary conditions.
This enables noiseless calculations of Ritz values and two-sided bounds for fixed iteration counts that can be compared to those coming from finite-statistics analyses.
Second, the action corresponds to a multivariate complex Gaussian 
distribution with covariance matrix $(2D)^{-1}$ and zero means, so it is trivial to sample exactly.
Correlators are not Gaussian distributed in this case~\cite{Detmold:2018eqd,Detmold:2025mhw}.
Use of interpolating operators that are nonlinear in the fields leads to further non-Gaussianity.

Note that in these examples, relatively precise correlators are obtained for $\delta_1 t \lesssim 3$--4, which makes $e^{-\delta_1 t}$ significantly smaller than in the LQCD studies below.
Reliable quantification of excited-state effects is likely to be even more challenging for multi-hadron systems in LQCD than for these examples, and indeed distinct failure modes of energy estimators are observed in studies of multi-baryon correlators where $\delta_1 t \lesssim 1$ below.
The fact that underestimated errors are nevertheless seen in these examples demonstrates that uncertainties on energy estimators do not always adequately account for excited-state effects even when 
ground-state overlaps and excitation gaps are not extremely small.

\subsection{$B$ at finite statistics}\label{sec:B}

Examination of a typical finite-statistics distribution of residual-norm-squares $B$  reveals the aforementioned negative fluctuations.
Residual bounds involve $\sqrt{B}$, and it is therefore necessary to define a prescription for dealing with these negative residual-norm-squares to obtain real-valued residual-bound estimators. 
Gap bounds are linear in $B$. Therefore, gap bounds constructed with either $B$ or $|B|$ give rise to real-valued, asymptotically unbiased estimators.
The implications of these different choices are discussed further below.

This section studies the statistics of these quantities under (nested) bootstrapping in the harmonic oscillator model.
The analysis herein employs Hermitian-subspace and ZCW filtering.
State identification is performed using a SLRVL scheme\footnote{
The simple filter-and-sort state identification used in Refs.~\cite{Wagman:2024rid,Hackett:2024xnx,Ostmeyer:2024qgu,Hackett:2024nbe} also leads to reasonable results. 
However, for the large correlator matrices studied in LQCD applications below, filter-and-sort state identification leads to obvious pathologies discussed in \cref{sec:stateID}, which motivates the introduction of the SLRVL scheme therein.
SLRVL is therefore employed here to test a single methodology consistently throughout this work.} defined by assigning the ordering of the $\lambda_k^{(m_{\rm last})}$ based on maximizing the overlap between the corresponding Ritz vectors and a set of reference Ritz vectors obtained from the largest iteration where no spurious eigenvalues appear, denoted $m_{\rm last}$.
All the $\lambda_k^{(m_{\rm last})}$ are real and positive by construction and there is therefore no ambiguity about assigning labels $k \in \{0,\ldots,m_{\rm last}-1\}$ based on the ordering $\lambda_0^{(m_{\rm last})} > \ldots > \lambda_{m_{\rm last}-1}^{(m_{\rm last})}$.
Having all Ritz vectors pass the Hermitian-subspace test further implies that all off-diagonal elements of $T^{(m_{\rm last})}$ in the Lanczos basis are positive-definite. Consequently, the subsequent Lanczos correlator analysis can be interpreted as symmetric Lanczos without the need to specify separate definitions for left- and right-
vectors. 

Broad tails and clear evidence of non-Gaussianity are observed in the $B_0^{(m,b,b')}$ distributions in both the harmonic oscillator model and the LQCD results below.
These tails, in conjunction with the non-positive-definiteness of $B_0$ estimators, mean that the precise way that outlier-robust estimators are constructed can lead to $B_0$ estimators with significantly different uncertainties.
The strategy used in this work is to perform bootstrap (inner) medians to define $\overline{B}_0^{(m)}$ and subsequently form two-sided bounds using $|\overline{B}_0^{(m)}|$; see \cref{app:lanczos} for further discussion.

The large-$m$ correlation structure of Lanczos energy estimators shows that the useful information that can be extracted from a correlator (matrix) saturates once noise becomes too large.  
As shown in \cref{app:lanczos}, large correlations are visible in $\overline{E}_0^{(m)}$ for large $m$, as previously observed in LQCD studies~\cite{Wagman:2024rid,Hackett:2024xnx,Ostmeyer:2024qgu,Chakraborty:2024scw,Hackett:2024nbe}.
Interestingly, the correlation structure of $\overline{B}_0^{(m)}$ is quite different. 
Only few-percent-level long-range correlations appear.
Significant nearest-neighbor anti-correlations are visible at intermediate $m$, indicating a tendency for $\overline{B}_0^{(m)}$ to flip signs between iterations in the oblique Lanczos regime.
This behavior suggests that a median of $\overline{B}_k^{(m)}$ across the last several iterations will provide a large-$m$  estimate 
\begin{equation}\label{eq:window}
\begin{split}
\overline{B}_k &\equiv \text{median}_m \{ \overline{B}_k^{(m_{\rm max}-m_{\rm window})}, \ldots, \overline{B}_k^{(m_{\rm max})} \},   
 \end{split}
\end{equation}
that has smaller uncertainty than the estimators at individual $m$.
Here, $m_{\rm max}$ denotes the largest Lanczos iteration 
included in a correlator analysis, and $m_{\rm window}$ denotes a hyperparameter specifying how many $m$ to include in the large-$m$ median.

Analogous large-$m$ median results for generic bootstrap-median estimators $\overline{X}_k^{(m)}$ are denoted by $\overline{X}_k$ below.
The corresponding large-$m$ medians of two-level bootstrap inner medians
\begin{equation}
     \overline{X}_k^{(b)} \equiv \text{median}_m \{ \overline{X}_k^{(m_{\rm max}-m_{\rm window},b)}, \ldots, \overline{X}_k^{(m_{\rm max},b)} \},
\end{equation}
are used to compute uncertainties $\delta \overline{X}_k$ using the same procedure as in \cref{eq:CI}.
The outer-bootstrap residual-norm-square estimators $\overline{B}_k^{(b)}$ used to compute $\delta \overline{B}_k$ can also be used to define outlier-robust estimators for two-sided bound windows straightforwardly.
Residual-bound window edges can be defined for each outer bootstrap sample by\footnote{
In all cases of physical interest below, $\overline{\lambda}_k^{(b)} \mp \sqrt{ | \overline{B}_k^{(b)} | }$ will be in the physical eigenvalue range $[0,1]$ and the logarithm is real valued. In generic cases such as when a Hermitian subspace does not exist, residual- and gap-bound estimators can be defined from the real parts of the corresponding logarithms.
}
\begin{equation}\label{eq:Rboot}
  \mathcal{R}_k^{(b)\pm} = - \ln\left( \overline{\lambda}_k^{(b)} \mp \sqrt{ \left| \overline{B}_k^{(b)} \right| } \right).
\end{equation}
In particular, the lower and upper edges of the residual-bound window at $68\%$ confidence are defined as the 16th percentile of $\mathcal{R}_k^{(m,b)-}$ and the 84th percentile of $\mathcal{R}_k^{(m,b)+}$, respectively.
Analogous gap-bound window edges can be defined from empirical bootstrap confidence intervals of
\begin{equation}\label{eq:Gboot}
  \mathcal{G}_k^{(b)\pm}(g) = -\ln\left( \overline{\lambda}_k^{(b)} \mp \frac{\left|\overline{B}_k^{(b)}\right|}{g} \right),
\end{equation}
where $g$ denotes the gap. The appropriate value for $g$ depends on the interpolator set and whether Davis-Kahan or Haas-Nakatsukasa gap bounds are used.
In the complex harmonic oscillator examples in this section, exact results for the spectrum are used to compute gap parameters since they are available. 

\subsection{Infinite vs. finite statistics}\label{sec:scalar-inf}

\begin{figure}[t!]
    \includegraphics[width=0.48\textwidth]{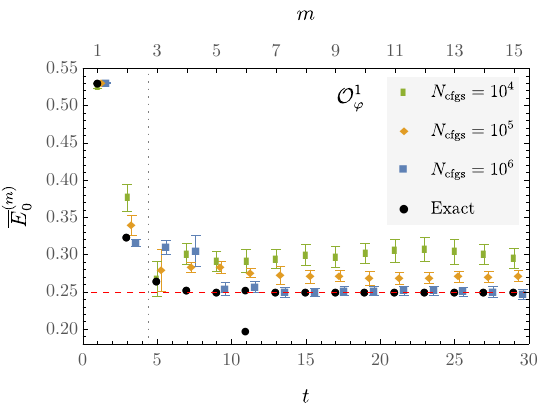}
    \caption{
    Finite-statistics ($N_{\rm cfgs}\in\{10^4,10^5,10^6\}$) and exact results for Lanczos energy estimators, with $E_0$ shown as a dashed red line for comparison. 
      The dotted line is placed before the median values of the smallest $m$ where spurious eigenvalues are identified for $N_{\rm cfgs}=10^6$ by the Hermitian subspace and ZCW tests with $F_{ZCW} = 10$; the previous iteration (at the inner bootstrap level) is used for SLRVL state identification, as described in the main text. 
      Infinite-statistics points are filtered using the ZCW test with $\varepsilon_{\rm ZCW} = 10\, e^{-E_0 L_t}$ in order to remove thermal states; the largest-magnitude non-spurious Ritz value is then used to define $E_0^{(m)}$. For $t=11$ at infinite statistics, this ZCW cut leaves two non-spurious energy estimators $E_0^{(m)}$ and $E_1^{(m)}$ that are close to $E_0$; both are shown. 
    \label{fig:complex-scalar-exact1}
    }
\end{figure}

\begin{figure}[t!]
  \includegraphics[width=0.48\textwidth]{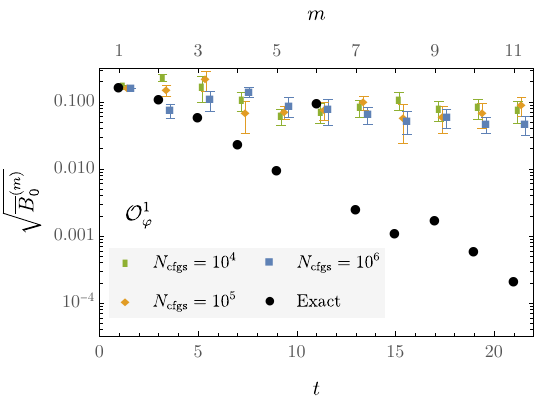}
    \caption{
    Finite-statistics ($N_{\rm cfgs}\in\{10^4,10^5,10^6\}$) and exact results for Lanczos residual-norm-square estimators.
    }
    \label{fig:complex-scalar-exact2}
\end{figure}

As shown in \cref{fig:complex-scalar-exact1,fig:complex-scalar-exact2}, exact computations of $C_{\varphi}^{11}(t)$ using \cref{eq:Cphi_exact} enable finite-statistics results for $\overline{E}_0^{(m)}$ and $\overline{B}_0^{(m)}$ to be directly compared with exact results for $E_0^{(m)}$ and $B_0^{(m)}$  at a given $m$.
Finite-statistics bias is visible in $\overline{E}_0^{(m)}$ for $N_{\rm cfgs} \in \{10^4, 10^5\}$ as discussed in \cref{sec:scalar-calibration} below.\footnote{The first iteration where spurious eigenvalues appear at finite statistics, $m_{\rm last}+1$, tends to have particularly large deviations from $E_0$.
Since $m_{\rm last}$ increases (slowly) with $N_{\rm cfgs}$, fixed-$m$ results have non-monotonic uncertainty scaling over the range of $N_{\rm cfgs}$ where a given $m$ goes from typically having one to zero spurious eigenvalues.
Similar pathologies have been observed in previous studies~\cite{Wagman:2024rid,Hackett:2024xnx,Hackett:2024nbe}, although the precise mechanism leading to them is not fully understood.}
Finite-statistics results for $\overline{B}_0^{(m)}$ are significantly larger than exact results for large $m$, which makes finite-statistics bounds more conservative than the corresponding exact bounds.

\begin{figure}[t!]
  \includegraphics[width=0.48\textwidth]{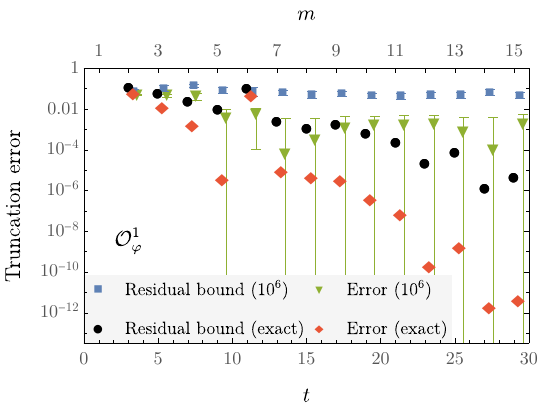}\vspace{15pt}
    \includegraphics[width=0.48\textwidth]{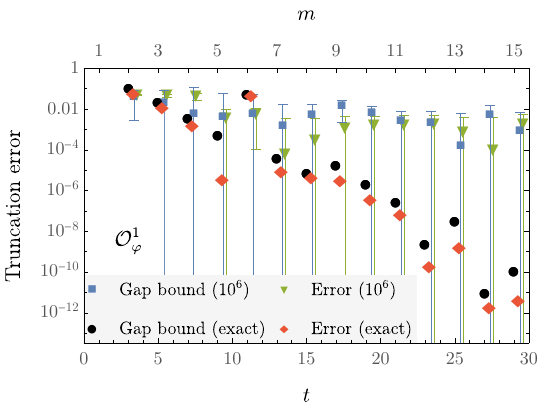}
    \caption{
      Comparisons of finite-$m$ truncation errors with the widths of residual bounds (top) and gap bounds (bottom), for both exact and finite-statistics results.
    The values at $m=1$ are not shown because $\lambda_0^{(1)}$ is closer to $\lambda_1$ than $\lambda_0$ and bounds therefore involve the first excited state rather than the ground state.   
    Non-monotonicity of finite-$m$ truncation errors and the non-monotonicity in $m$ of bound widths both arise from thermal effects discussed in the main text.
    }
    \label{fig:complex-scalar-exact3}
\end{figure}

The fact that the energy spectrum is exactly calculable (at non-zero lattice spacing) further enables residual and gap bounds to be compared with the exact finite-$m$ truncation errors, i.e., differences between $E_0^{(m)}$ and $E_0$.
Residual bounds are seen to provide valid, but conservative, bounds on true uncertainties at both finite and infinite statistics, as seen in \cref{fig:complex-scalar-exact3}.
Gap bounds are similarly verified to be valid for all $m$ at infinite statistics, as well as valid stochastically for all $m$ at finite statistics.\footnote{For both finite-statistics and exact calculations, gap bounds are computed using the exact spectrum to compute $g_0^{(m)}$. For the scalar correlators considered in this section, Haas-Nakatsukasa and Davis-Kahan gap bounds coincide; see \cref{sec:scalar-matrices} for analogous studies of correlator matrices where non-trivial differences between the two types of gap bounds arise.}

At infinite statistics, \cref{fig:complex-scalar-exact3} shows that finite-$m$ truncation errors decrease faster than residual bounds and roughly at the same rate as gap bounds.
This is consistent with general linear algebra expectations~\cite{Parlett,Demmel:1997} and $O(\sqrt{B})$ vs $O(B)$ scaling arguments above.
In particular, residual bounds are around $10$ times larger than finite-$m$ truncation errors at small $m$ while they are $10^6$--$10^7$ times larger for $m \in [14,16]$.
Exact gap bounds and finite-$m$ truncation errors both  scale in an empirically similar way with $m$ and approximately follow  $O(B) \sim O(e^{-2t\sqrt{\delta_1}})$ scaling. This matches the convergence rate predicted by the KPS bound.
Exact residual bounds decay more slowly and follow $O(\sqrt{B}) \sim O(e^{-t \sqrt{\delta_1}})$ scaling.

At finite statistics, finite-$m$ truncation errors stop decreasing with $m$ once the statistical uncertainties are comparable to the errors themselves.
Finite-statistics residual and gap bounds both stop decreasing significantly with $m$ at a similar point to where the finite-$m$ truncation errors saturate.
The residual bounds are significantly looser than the finite-$m$ truncation errors also at finite statistics, and they are clearly resolved from zero even at large $m$.
The gap bounds are again on the same order as the actual finite-$m$ truncation errors in the noisy case.
Gap bounds are consistent with zero for $m \gtrsim 3$.
The results of these comparisons show that, at least in this harmonic oscillator model,  gap bounds are valid for all $m$, both in the exact case and (stochastically) at finite statistics.
Gap bounds are nearly saturated in the finite-statistics case, suggesting that finite-statistics gap bounds can provide useful systematic-uncertainty estimates.
More-detailed calibration tests of finite-statistics gap bounds are explored below by repeating this analysis with many independent Monte Carlo ensembles.

\begin{figure}[t!]
    \includegraphics[width=0.48\textwidth]{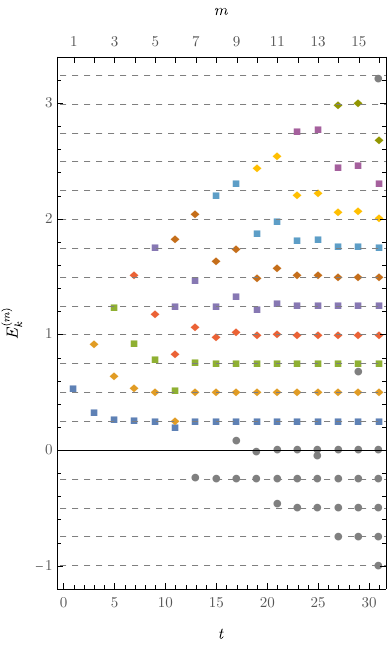}
    \caption{
    Infinite-statistics results for Lanczos energy-estimators as a function of iteration count are compared to the exact spectrum (gray dashed lines), which, including thermal states, is given by $(n+1)E_0$ for all $n \in \mathbb{Z}$.
    Energies for states passing the ZCW test with $\varepsilon_{\rm ZCW} = e^{-E_0 L_t}$ are shown as colored squares and diamonds; those for states with overlaps below this ZCW cut are shown as gray circles.
    Non-monotonic features visible for $m \in \{6,9,15\}$ arise from thermal effects.
    \label{fig:complex-scalar-thermal}
    }
\end{figure}

Although thermal effects are not visible in the finite-statistics results shown in \cref{fig:complex-scalar-exact1}, they lead to significant effects on exact results that must be treated with care.
With $m=16$, there are Lanczos energy estimators within one part in $10^4$ of $E_n = (n+1)E_0$ for $n \in \{-5,\ldots,5\}$ as well as less accurate estimates for higher-energy states with $n \in \{6,\ldots,10\}$; see \cref{fig:complex-scalar-thermal}.
The last iteration where all infinite-statistics Ritz values are greater than $E_0$ is $m=5$. 
Iteration $m = 6$, which involves correlator data with $t \in [0,11]$, is particularly interesting in that there are two infinite-statistics Lanczos energy estimators with energies of $\approx 0.196$ and $\approx 0.252$ that lie relatively close to the exact ground-state energy $\approx 0.249$ (unusual features are also visible for $m=6$ in \cref{fig:complex-scalar-exact1,fig:complex-scalar-exact2,fig:complex-scalar-exact3}).
They have squared overlaps $[Z_0^{(m=6)}]^2$ of $5\times 10^{-3}$ and 0.32, respectively.
The small overlap in the former case suggests that the energy below $E_0$ may be associated with a thermal state. 
For $m > 6$ where the energies of thermal states are resolved more accurately, their overlaps approximately follow the pattern $Z_n^2 \sim e^{- |n| E_0 L_t}$ for thermal states with $n \leq -1$, which physically represent $|n|$-particle states with additional contributions from $|n|$-particle ``backwards-propagating'' thermal modes.\footnote{Corrections to this scaling from higher-multiplicity states are suppressed by further powers of $e^{-E_0 L_t}$. The leading $e^{-|n| E_0 L_t}$ scaling of various overlap factors has been explicitly verified by performing calculations with larger $L_t$.}
In particular, $n=-1$ corresponds to a zero-energy state with overlap $\approx e^{-E_0 L_t}$ arising from the product $e^{-E_0 t} e^{-E_0(L_t - t)} = e^{-E_0 L_t}$ of one ``forwards-propagating'' and one ``backwards-propagating'' mode.
See Ref.~\cite{Detmold:2011kw} for further discussion of multi-particle thermal states.

Due to these thermal states, the ground-state cannot be simply associated with the largest-magnitude Ritz value. 
Thermal states can be identified and removed in some instances by cutting states with $\lambda_k^{(m)} > 1$; however, the presence of $\lambda_{-1} = 1$ complicates this process by leading to Ritz values satisfying $\lambda_k^{(m)} < 1$ that correspond to $E_k^{(m)} \approx 0$.
Alternatively, the ZCW test provides a straightforward way to identify and remove thermal states from exact results.
Thermal states in the harmonic oscillator example have squared-overlaps suppressed by powers of $e^{-E_0 L_t}$, while some non-thermal states have significantly larger overlaps.
Therefore applying the ZCW test with $ e^{-E_0 L_t} \ll \varepsilon_{\rm ZCW} \ll 1$ should remove thermal modes while leaving some non-thermal modes unaffected.
Infinite-statistics results in \cref{fig:complex-scalar-exact1,fig:complex-scalar-exact2,fig:complex-scalar-exact3,fig:complex-scalar-thermal} use $\varepsilon_{\rm ZCW} = 10  e^{-E_0 L_t}$ for concreteness.

These infinite-statistics results with fixed $\varepsilon_{\rm ZCW}$ demonstrate how thermal states are removed and unbiased results for all energy estimators associated with a set of low-energy states ($n \leq 11$ in this example) are obtained by taking the infinite-statistics limit with fixed $\varepsilon_{\rm ZCW}$ in the range $ e^{-E_0 L_t} \ll \varepsilon_{\rm ZCW} \ll 1$; see \cref{fig:complex-scalar-thermal}.
The thermal modes themselves (and the $n \geq 12$ states in this example) could be extracted by selecting different ranges of squared-overlaps as in the thermal ZCW test discussed in Ref.~\cite{Hackett:2024nbe}.

\subsection{Calibration tests}\label{sec:scalar-calibration}

\begin{figure}[h!]
    \includegraphics[width=0.48\textwidth]{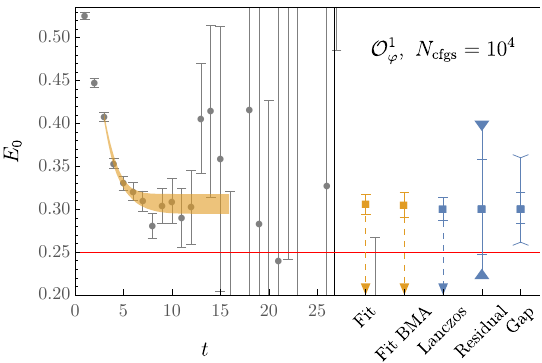}\vspace{15pt}
    \includegraphics[width=0.48\textwidth]{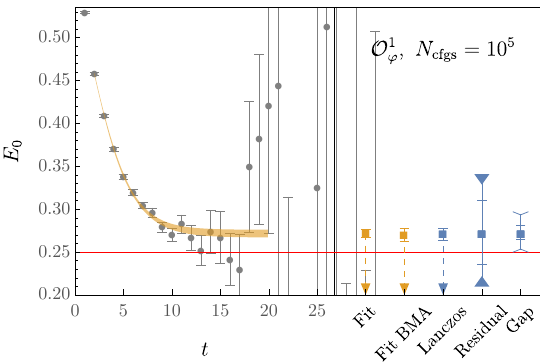}\vspace{15pt}
    \includegraphics[width=0.48\textwidth]{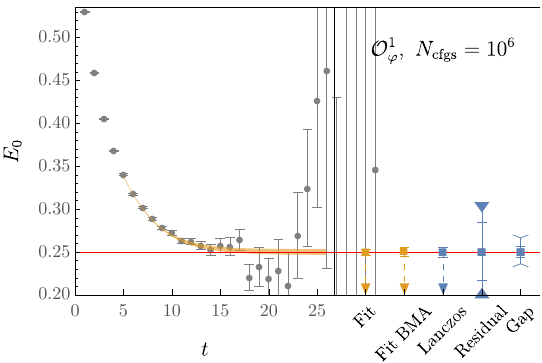}
    \caption{
      Effective masses compared with highest-weight fit (orange band, left, and first point, right) and Bayesian model-averaged~\cite{Jay:2020jkz} (BMA) multi-state-fit results (second orange point, right). Lanczos energy estimators, residual bounds, and gap bounds are shown as blue points, right. Arrows denote one-sided variational bounds. Horizontal lines denote bootstrap medians of two-sided bounds; triangles (chevrons) denote 68\% confidence intervals for residual (gap) bounds.
    \label{fig:complex-scalar}
    }
\end{figure}

\Cref{fig:complex-scalar} shows results for $\overline{E}_0$ defined analogously to \cref{eq:window} with $m_{\rm max} = 15$ and $m_{\rm window} = 3$ and analogous large-$m$ medians of residual- and gap-bound estimators for $N_{\rm cfgs} \in \{10^4, 10^5, 10^6\}$ with $F_{\rm ZCW} = 10$ and SLRVL state identification.
These estimators and bounds are compared with effective masses, as well as multi-state fits to correlator results with in the range $[t_{\rm min}, t_{\rm max}]$ for all minimum times $t_{\rm min}$ with $t_{\max}$ chosen as the last point where $E_{\rm eff}(t)$ has SNR greater than 0.25 and $N_{\rm states}$ selected using the Akaike Information Criterion (AIC)~\cite{AkaikeAIC}, as detailed in Ref.~\cite{NPLQCD:2020ozd}. 
Linear shrinkage~\cite{Ledoit:2004} is used to regularize covariance matrix estimation in the fits, as described in Ref.~\cite{Rinaldi:2019thf},
and a number of checks on optimizer convergence are performed, as detailed in Ref.~\cite{NPLQCD:2020ozd}.
Logarithms of energy gaps and overlap-factor products are used as fit variables to improve numerical stability as described in Ref.~\cite{Amarasinghe:2021lqa}.
Here, Bayesian model averages (BMA) over $t_{\rm min}$ are performed using the weights defined in Ref.~\cite{Jay:2020jkz} with flat priors after removing unconstraining fits with greater than ten times the minimum fit uncertainty.\footnote{Qualitatively similar results are obtained using the $p$-value / variance weights defined in Ref.~\cite{Rinaldi:2019thf} that were used in Refs.~\cite{Amarasinghe:2021lqa,Detmold:2024iwz}. 
The step of removing unconstraining fits also does not change the qualitative results of this section. }

For the ensemble with $N_\mathrm{cfgs} = 10^6$ shown in \cref{fig:complex-scalar}, all estimators converge to the true $E_0$ within statistical uncertainties.
However, for the lower-statistics results shown (which correspond to subsets of this particular $N_\mathrm{cfgs} = 10^6$ ensemble), all estimators assuming $N$-state saturation---Ritz values, effective mass ``plateaus,'' and model-averaged multi-state fit results---disagree with the true value of $E_0$ at 2-3$\sigma$.
Interpreted as variational bounds, they are valid: all are overestimates of $E_0$.
The residual bound clearly provides a conservative over-estimate of the finite-$m$ truncation error, while the $1\sigma$ gap bound provides a relatively accurate uncertainty estimate.
Gap bounds are satisfied at $\approx 1.3\sigma$ for the cases of $N_{\rm cfgs} \in \{ 10^4, 10^5\}$.

It is noteworthy that underestimated error estimates from statistical uncertainties with $N_{\rm cfgs} \in \{ 10^4, 10^5\}$ are not signaled by obvious curvature in effective masses or unacceptable $\chi^2 / \text{dof}$ in multi-state fits.
In particular, the highest-weight fits shown in \cref{fig:complex-scalar} have $\chi^2 / \text{dof}$ equal to 0.50, 0.68, and 0.82 for $N_{\rm cfgs}$ equal to $10^4$, $10^5$, and $10^6$ respectively.\footnote{The optimal shrinkage parameters used for each case range from $\approx 0.026$ for $N_\mathrm{cfgs} = 10^4$ to $\approx 0.007$ for $N_\mathrm{cfgs} = 10^6$. Shrinkage tends to lead to reductions in $\chi^2/\text{dof}$, and one may worry that this is the cause of acceptable $\chi^2/\text{dof}$ resulting from fits that disagree with the exact $E_0$. It can be explicitly verified that this is not the case: otherwise identical fits without shrinkage give a model-averaged fit result $E_0 = 0.302(16)$ for $N_\mathrm{cfgs} = 10^4$, which shows a $3\sigma$ deviation from the exact result $E_0 \approx 0.249$, and a highest-weight fit $\chi^2 / \text{dof} = 0.72$.

Increasing the minimum value of $t_{\rm min}$ used in fits or varying the cutoff on $t_{\rm max}$ does not change the conclusion that model-averaged $E_0$ show 2-3$\sigma$ discrepancies from the exact $E_0$ for $N_{\rm cfgs} \in \{ 10^4, 10^5\}$.
One-state fits with larger $t_{\rm min}$ give qualitatively similar results to the two-state fits preferred by BMA.
For example, 
a one-state fit for $N_{\rm cfgs} = 10^4$ with $t_{\rm min}$ of 7 gives $E_0 = 0.292(12)$, a $3\sigma$ deviation from the exact result $E_0 \approx 0.249$, with $\chi^2/\text{dof} = 0.46$, quantitatively verifying the lack of apparent curvature in the effective mass for $t \gtrsim 7$ in \cref{fig:complex-scalar}.}

To test the generality of these observations based on the single ensemble for each level of statistics shown in Figs.~\ref{fig:complex-scalar-exact1}--\ref{fig:complex-scalar}, calibration tests are performed using 200 independent ensembles for each of $N_{\rm cfgs} \in \{10^4, 10^5, 10^6\}$. 
Nested bootstrap Lanczos analyses, as well as bootstrapped fits with model averaging over fit ranges as described above, are performed independently for each ensemble.
Finite-$m$ truncation errors are compared with variational, residual, and gap bounds as well as statistical uncertainties on Ritz values and two different versions of multi-state fits: one using Bayesian model averaging, and the other by choosing the highest-weight fit from the same pool of fits.

To assess calibration of each species of uncertainty estimate, counts of the number of ensembles in which the true error exceeds $n\sigma$ of the uncertainty estimate are performed for varying $n$, and the resulting ratios to the expectation assuming Gaussian fluctuations are constructed.
\Cref{fig:complex-scalar-calibration} shows the results as a function of $n\sigma$.
A perfectly calibrated estimator will lie along the dashed line at unity---an $n\sigma$ uncertainty should be violated at a rate of $1-{\rm erf}(n/\sqrt{2})$. 
Overly conservative uncertainties lie below the line.
Insufficiently conservative uncertainty estimates lie above the line, where the error rate exceeds the $n\sigma$ expectation.

Several conclusions can be drawn from this exercise:

\begin{figure}[]
    \includegraphics[width=0.48\textwidth]{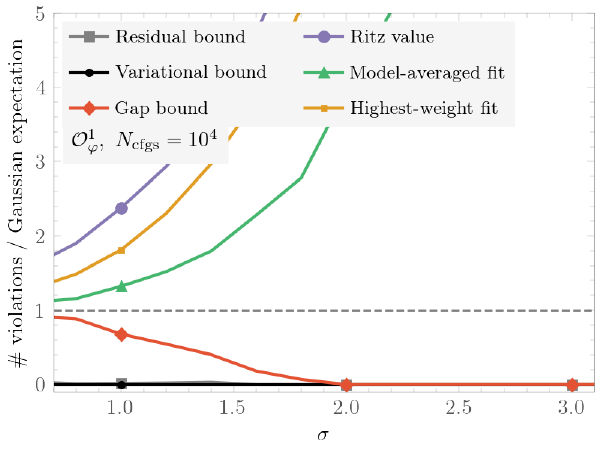}\vspace{15pt}
    \includegraphics[width=0.48\textwidth]{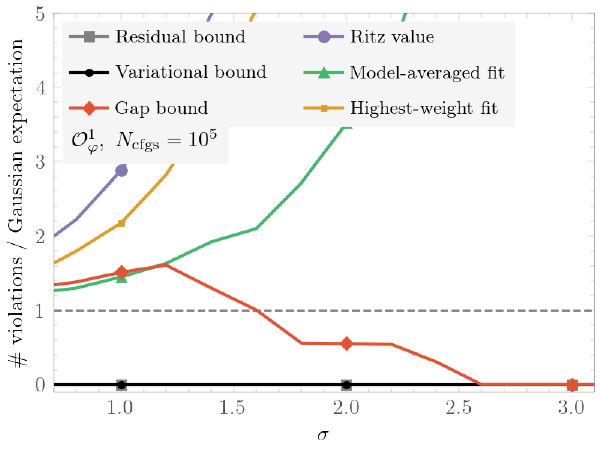}\vspace{15pt}
    \includegraphics[width=0.48\textwidth]{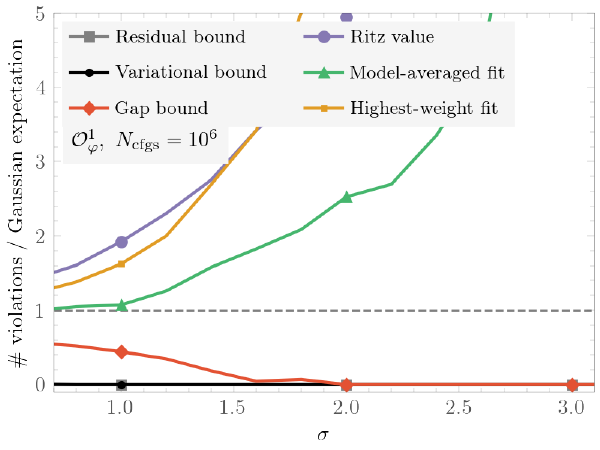}
    \caption{
        Calibration tests for various energy estimators  and bounds using 200 independent ensembles of harmonic-oscillator field configurations with the level of statistics shown. Each curve shows the number of ensembles where an estimator or bound on ground-state energy finite-$m$ truncation errors is violated, divided by the number expected to be violated for a Gaussian-distributed error estimate.
        Plot markers are placed at integer values of $n\sigma$ to guide the eye.
    }
    \label{fig:complex-scalar-calibration}
\end{figure}

\begin{itemize}[leftmargin=*]
    \item First, estimators assuming $N$-state saturation, i.e., Ritz values and multi-state fit results, are insufficiently conservative at all levels of statistics considered for this example.
Calibration slightly improves with increasing $N_{\text{cfgs}}$, but even at the highest level of statistics, $N_\mathrm{cfgs}=10^6$, both Ritz values and fit results are poorly calibrated.
Model-averaged fits are better calibrated than both Ritz values and individual fits and have reasonable $1\sigma$ calibration, especially at the highest level of statistics; however, they have many more $2\sigma$ and $3\sigma$ discrepancies than expected for a Gaussian error distribution at all levels of statistics for this example.

\item Second, variational and residual bounds are completely reliable but overly conservative.
Variational bounds are satisfied on all 200 ensembles generated at all three levels of statistics.
Consequently, the many observed discrepancies between energy estimators and $E_0$ always involve energy estimators that are larger than $E_0$.
This indicates that finite-statistics effects cause energy estimators to systematically over-estimate the true ground state energy, at all levels of statistics for this example.
Residual bounds are equally conservative, as  no violations of these bounds are seen in any of the  200 ensembles.

\item Third, gap bounds provide the intermediate behavior desired of a systematic uncertainty estimate: they are either well calibrated or slightly conservative in this toy model.
At $N_\mathrm{cfgs} = 10^4$ and $10^6$, gap bounds are violated at $1\sigma$ in a non-zero fraction of ensembles, but less than the Gaussian expectation. \
They are satisfied at $2\sigma$ in all 200 ensembles for these levels of statistics.
At intermediate statistics, $N_\mathrm{cfgs} = 10^5$,  where the interplay between excited-state and finite-statistics effects is particularly severe, gap bounds provide relatively well-calibrated uncertainty estimates.
They are violated by $1\sigma$ slightly more than expected for a Gaussian estimate and violated by $2\sigma$ slightly less than the corresponding expectation.
No $3\sigma$ violations are observed in all 200 ensembles.

\end{itemize}

\begin{figure}[]
    \includegraphics[width=0.48\textwidth]{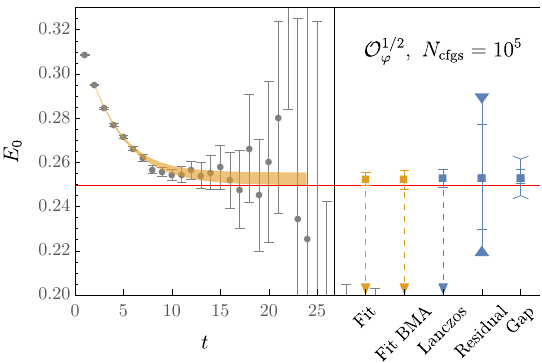}\vspace{15pt}
    \includegraphics[width=0.48\textwidth]{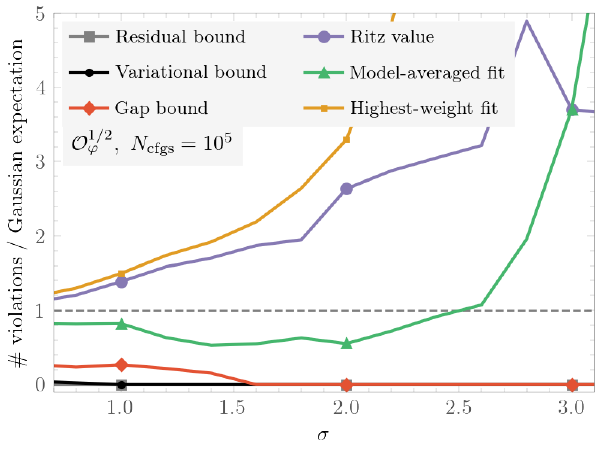}
    \caption{
      Effective masses, energy estimators, and two-sided bounds analogous to  \cref{fig:complex-scalar} but for the $\mathcal{O}_{\varphi}^{1/2}$ interpolating operator, top. Calibration tests analogous to \cref{fig:complex-scalar-calibration} for $\mathcal{O}_{\varphi}^{1/2}$, bottom.
    }
    \label{fig:complex-scalar-easy}
\end{figure}

Note that these calibration tests have focused on a particular interpolating operator in  a particular $0+1$-dimensional scalar field theory, and not all features will appear for other operators or theories.
It should be noted that the particular interpolating operator $\mathcal{O}_{\varphi}^1$ considered above has overlap with all $U(1)$ particle-number sectors and significantly worse excited-state contamination than other operators that have been investigated, for example in Ref.~\cite{Wagman:2024rid}. 
Operators $\mathcal{O}_{\varphi}^k$ with $k<1$ lead to more precise results with less excited-state contamination\footnote{Operators with $k \gtrsim 2$ are too noisy to achieve reliable fit results at the studied levels of statistics.} than $\mathcal{O}_{\varphi}^1$. 
Energy estimators, bounds, and calibration tests for $\mathcal{O}_{\varphi}^{1/2}$ are shown in \cref{fig:complex-scalar-easy}.
Ritz values and highest-weight fits are statistically better behaved than for $\mathcal{O}_{\varphi}^1$, although they still have slightly more $1\sigma$ violations and significantly more $2\sigma$ and $3\sigma$ violations than Gaussian expectations.
Model-averaged fits provide well-calibrated uncertainty estimates at $1\sigma$ and $2\sigma$, although they have significantly more $3\sigma$ violations than Gaussian expectations.
Gap bounds provide conservative uncertainty estimates that are again violated at $1\sigma$ in a non-zero fraction of ensembles but less than Gaussian expectations. Residual and variational bounds are never violated at $1\sigma$. 

\subsection{Correlation function matrices}\label{sec:scalar-matrices}

\begin{figure}[t!]
    \includegraphics[width=0.48\textwidth]{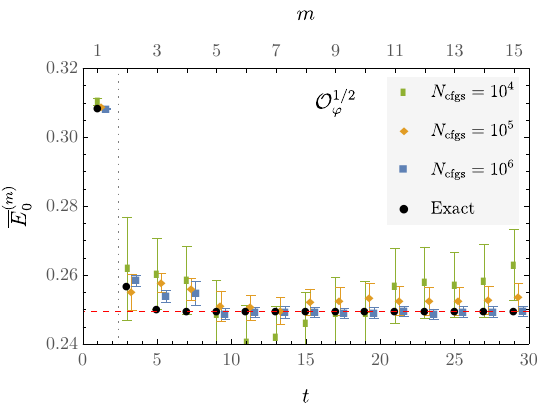}
    \includegraphics[width=0.48\textwidth]{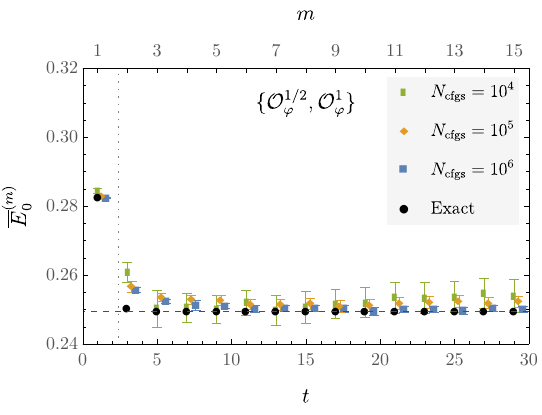}
    \caption{
    Comparisons of finite-statistics and exact results for Lanczos energy estimators as in \cref{fig:complex-scalar-exact1} (note, however, the much reduced vertical plot range) for $\mathcal{O}_{\varphi}^{1/2}$ interpolating operator, upper, and block Lanczos with $r=2$ correlator matrices constructed from $\{ \mathcal{O}_{\varphi}^{1/2}, \mathcal{O}_{\varphi}^{1} \}$, lower.
    \label{fig:complex-scalar-block-exact1}
    }
\end{figure}

\begin{figure}[t!]
  \includegraphics[width=0.48\textwidth]{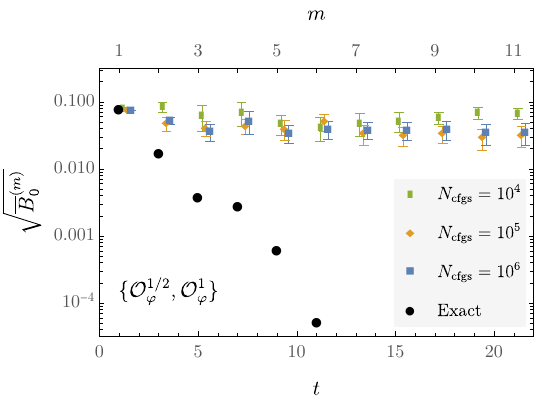}
    \caption{
      Block Lanczos results for residual-norm-square estimators analogous to \cref{fig:complex-scalar-exact2}.
    }
    \label{fig:complex-scalar-block-exact2}
\end{figure}

\begin{figure}[t!]
  \includegraphics[width=0.48\textwidth]{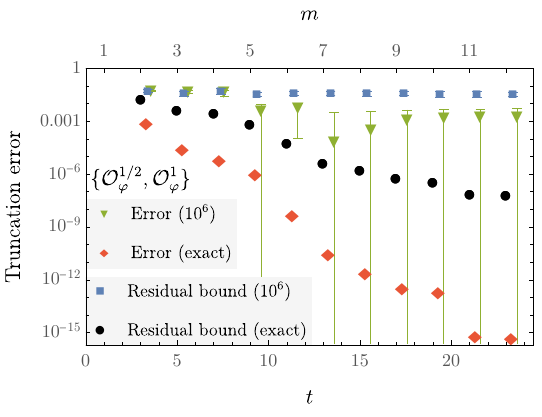}\vspace{15pt}
    \includegraphics[width=0.48\textwidth]{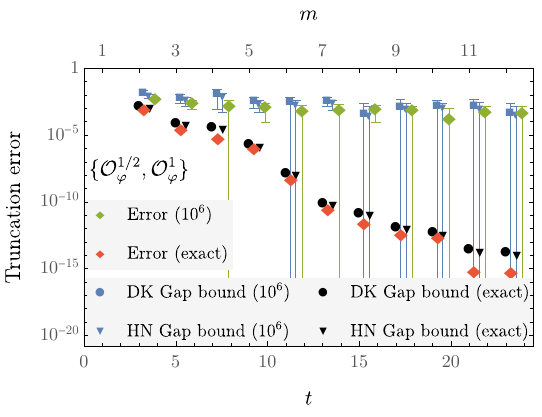}
    \caption{
      Comparisons of finite-$m$ truncation errors with the widths of residual bounds (top) and gap bounds (bottom) as in \cref{fig:complex-scalar-exact3} for the block Lanczos case.
    }
    \label{fig:complex-scalar-block-exact3}
\end{figure}

Correlator matrices constructed from the pair of interpolating operators $\{ \mathcal{O}_{\varphi}^{1/2}, \mathcal{O}_{\varphi}^1 \}$ considered above can be used to compare exact and finite-statistics block-Lanczos results.
As seen in \cref{fig:complex-scalar-block-exact1}, exact block-Lanczos results for $E_0^{(m)}$ converge to $E_0$ significantly faster than scalar results using either interpolating operator alone.
Somewhat faster convergence is seen at finite statistics, but in this case, the effect is milder.
Smaller uncertainties are achieved by block-Lanczos energy estimators than scalar Lanczos with either interpolator.

Exact and finite-statistics results show noteworthy differences when comparing $r=1$ (scalar) and $r=2$ results for $B$ and associated two-sided bounds, as shown in \cref{fig:complex-scalar-block-exact2,fig:complex-scalar-block-exact3}.
Exact block-Lanczos results for $B_0^{(m)}$ in \cref{fig:complex-scalar-block-exact2} are strictly smaller than exact results from scalar Lanczos with either interpolator in this example.
This is a general feature that can be expected based on optimality properties of Ritz values for symmetric Lanczos~\cite{Abbott:2025yhm}.
The same behavior is observed here in small-$m$ finite-statistics results; however, for large-$m$ where $\overline{B}_0^{(m)}$ is consistent with zero, its uncertainties $\delta \overline{B}_0^{(m)}$ are larger by a factor of 1--2 for $r=2$ block Lanczos than for $\mathcal{O}_{\varphi}^{1/2}$.
This counter-intuitive behavior has been previously observed in LQCD examples~\cite{Hackett:2024nbe}, and may be a generic feature of  results that are far from the infinite-statistics limit (i.e., results where several spurious eigenvalues are present and $T^{(m)}$ is non-Hermitian).

Finite-$m$ truncation errors are seen to decrease more rapidly for $r=2$ block Lanczos than for either scalar correlator. Their scaling with $m$ is consistent with the exponentially faster $O(e^{-2 t \sqrt{\delta_2}})$ scaling for the $r=2$ case predicted by the KPS bound.
Exact gap-bound widths, both Davis-Kahan and Haas-Nakatsukasa varieties, scale as $O(B) \sim O(e^{-2 t \sqrt{\delta_2}})$. 
Exact residual-bound widths decrease more slowly with $m$ as $O(\sqrt{B}) \sim O(e^{-t \sqrt{\delta_2}})$.

In this example, Haas-Nakatsukasa gap-bound widths are computed under the no-missing-states assumption that there are exactly two states\footnote{This discussion neglects the presence of thermal modes and to take these into account, the next-closest Ritz value is $\lambda_{-1} = e^{-E_{-1}} = 1$, which is closer to $\lambda_0$ than $\lambda_2$ and should be used to compute $G_0^{(m)}$.
The formally incorrect determination of $G_0^{(m)}$ used in the main text leads to valid bounds in practice due to the thermal suppression of the overlap of the $n=-1$ state. } with energies below $\Egap = E_2$, which can be verified to be self-consistent for all $m \geq 2$ in the sense that $E_0^{(m)} < E_1^{(m)} < E_2$.
The gap parameter for this case therefore corresponds to $G_0^{(m)} = \lambda_0^{(m)} - \lambda_2$.
On the other hand, Davis-Kahan gap-bound widths are computed using the gap parameter $g_0^{(m)} =  \lambda_0^{(m)} - \lambda_1$.
This leads to Haas-Nakatsukasa gap bounds being tighter than Davis-Kahan gap bounds by an approximately $m$-independent factor of $\approx 2$. 

At infinite statistics, both Haas-Nakatsukasa and Davis-Kahan gap bounds are observed to be valid for all $m$ using exact results.
For relatively large $m$, exact Haas-Nakatsukasa bounds are larger than exact finite-$m$ truncation errors by factors of 3--5 on many iterations. 
Davis-Kahan gap bounds are larger by an additional factor of $\approx 2$.
Both are tighter than residual bounds by orders of magnitude.
At finite statistics, the ratio between Davis-Kahan and Haas-Nakatsukasa gap bounds is similar; both appear stochastically valid for all $m$.

It is noteworthy that the effects on gap bounds of $\delta \overline{B}_0^{(m)}$ growing with increasing $r$ can be partially mitigated by using Haas-Nakatsukasa gap bounds.
In cases where the gap parameter increases with $r$, the $B / G$ ratios entering Haas-Nakatsukasa gap bounds may stay approximately constant or decrease even when $B$ increases.
In this example, block Lanczos has $\delta \overline{B}_0$ that is thirty percent larger than scalar analysis of $\mathcal{O}_{\varphi}^{1/2}$ and Davis-Kahan gap bounds that are corresponding looser.
However, the increased gap parameter for $r=2$ in the Haas-Nakatsukasa case leads to gap bounds that are twenty percent tighter for $r=2$ block Lanczos than for scalar analysis of $\mathcal{O}_{\varphi}^{1/2}$.

\section{LQCD data}
\label{sec:lqcd}

In order to examine systematic uncertainties from correlator analysis, and the assumptions going into them, in the QCD context, a very high statistics dataset is required. Here, the basic properties of the dataset studied in this work are introduced. 

\subsection{Ensemble specifications}
\label{sec:lqcd-data}

\begin{figure}
    \centering
    \includegraphics[width=\linewidth]{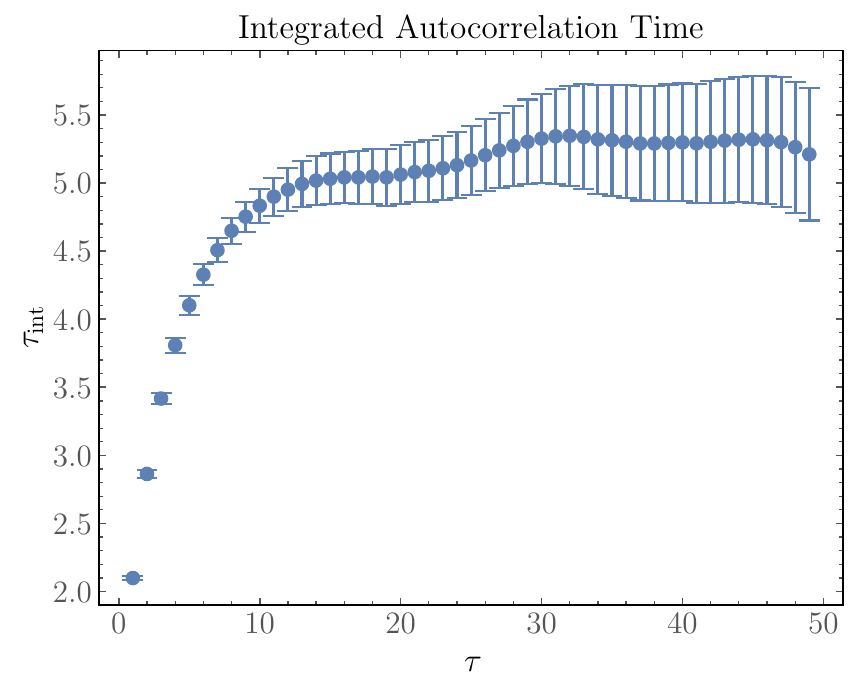}
    \caption{
    The integrated autocorrelation time of the nucleon two-point correlator at source-sink separation $t = 10$, in units of saved configurations (10 trajectories).  While the effect of autocorrelations cannot be neglected, a bin size of 50 saved configurations is sufficient to remove autocorrelations. 
    }
    \label{fig:autocorrelations}
\end{figure}

The ensemble of gauge field configurations used in this work was generated on a  $L^3\times L_t=24^3\times48$ lattice geometry using a tree-level-improved L\"uscher-Weisz gauge action \cite{Luscher:1984xn} with $\beta=6.1$ and $N_f=3$ degenerate flavors of clover fermions \cite{Sheikholeslami:1985ij} defined with links
that had a single level of  stout-smearing~\cite{Morningstar:2003gk} with $\rho=0.125$ applied and using the Sheikoleslami-Wohlert improvement parameter $c_{\rm SW}=1.2493$.\footnote{This action leads to a transfer matrix that is Hermitian in the low-energy sector of Hilbert space where $E\ll1/a$~\cite{Luscher:1976ms,Luscher:1984is}.
}
The corresponding lattice spacing is $a=0.145$ fm \cite{NPLQCD:2012mex} using $\Upsilon$ spectroscopy for scale setting\footnote{Wilson flow scales computed on a subset of the ensemble  are $t_0/a^2=0.9348(7)$ and $w_0/a=1.0138(5)$.} and the quark mass was chosen such that the pion mass is $m_\pi\sim800$ MeV. A subset of the gauge fields employed here were also
employed in previous studies of two-nucleon spectroscopy
in Refs.~\cite{NPLQCD:2012mex,NPLQCD:2013bqy,Berkowitz:2015eaa,Wagman:2017tmp,Detmold:2024iwz}.
The ensemble contains $N_{\rm cfgs}=16,368$ configurations generated in 17 independent streams, each with individual configurations separated by 10 rational hybrid Monte-Carlo trajectories \cite{clark2006rationalhybridmontecarlo} of length $\tau = 1.414$ after 500 thermalization trajectories.
Autocorrelations of various correlation functions were analyzed, and an example of the integrated autocorrelation time for the nucleon two-point correlator is shown in Fig.~\ref{fig:autocorrelations}.  The autocorrelation times of the $NN$ correlation functions are virtually identical to that of the single nucleon. The  remaining correlations in the data are handled with binning using a conservative bin size of 50 in units of the saved configurations for data-processing expediency.

Sparsening~\cite{Detmold:2019fbk} is employed to compute sparsened timeslice-to-all propagators.
On each configuration, a sparse grid of ${\cal S}^3$ propagators was generated with ${\cal S}=6$ on timeslices $t_0\in\{0,8,16,24,32,40\}$ and used to build the correlation functions. 
The quark fields at the source and sink were both Gaussian smeared~\cite{Gusken:1989ad,Gusken:1989qx} before sparsening  with a width of 2.1 using links subject to eight hits of three-dimensional stout smearing with $\rho = 0.16$.

\subsection{Operators}

The isodoublet nucleon field $N_\sigma = (p_\sigma, n_\sigma)^T$ is defined as 
\begin{equation}
\begin{split}
N_{\sigma,f}(x)  =& \epsilon_{abc}\frac{1}{\sqrt{2}} \big[ u^{a T}(x) C \gamma_5 P_+ d^b(x) 
\\
& - d^{a T}(x) C \gamma_5 P_+ u^b(x) \big] P_{\sigma} q_f^c(x) \ ,
\end{split}
\end{equation}
where $q^c = (u^c, d^c)^T$, $C = \gamma_2\gamma_4$ is the Euclidean charge conjugation matrix, $P_\pm = (1\pm\gamma_4)/2$ is the parity projector, and
\begin{equation}
  \begin{split}
    P_\sigma = \begin{cases} P_{+} \left(  \Id - (-1)^\sigma i \gamma_1 \gamma_2 \right)/2, & \sigma \in \{0,1\}\ 
    \\ 
    P_{-} \left( \Id - (-1)^\sigma i \gamma_1 \gamma_2 \right)/2, & \sigma \in \{2,3\}\ 
    \end{cases},
  \end{split}
\end{equation}
projects the quark spin to a specific row $\sigma \in \{0,\ldots,3\}$ of the spinor representation ($G_1^+ \oplus G_1^-$) of the lattice isometry group $O_h^D$.

The focus of this work is on the low-energy spectrum in the deuteron and dineutron channels. A set of operators which can be classified based on their spatial structure are considered. In particular, this work studies hexaquark ($H$) operators, constructed from products of six quark fields centered at the same point, and dibaryon ($D$) operators, constructed from products of plane-wave single-nucleon operators. 

Dibaryon operators are constructed from products of single nucleon operators that are individually projected to definite momentum.
Dibaryon operators with zero total three-momentum (the current study is restricted to this frame) are defined by
\begin{equation}
\label{eq:dibaryon}
\begin{split}
  D^{I}_{\rho}(\vec{n},t) =& \sum_{\vec{x}_1,\vec{x}_2 \in \Lambda_{\mathcal{S}}} 
  e^{i{\frac{2\pi}{L}} \vec{n}\cdot(\vec{x}_1-\vec{x}_2) } 
  \\
  &\times \sum_{\sigma,\sigma',u,u'} v_{\rho}^{\sigma\sigma'}  P_{uu'}^{(I)} N_{\sigma, u}(\vec{x}_1,t)  N_{\sigma', u'}(\vec{x}_2,t)\ ,
\end{split}
\end{equation} 
where $x_i=(\vec{x}_i,t)$ for $i\in\{1,2\}$ are lattice coordinates, 
$\Lambda_{\mathcal{S}}$ is a sparse sublattice with $L/\mathcal{S}$ sites in each dimension that is introduced to make the volume sums computationally tractable as described in Refs.~\cite{Detmold:2019fbk,Li:2020hbj,Amarasinghe:2021lqa}.
The index
$\rho \in \{0,\ldots,3\}$ labels spin-singlet ($\rho=0$) and spin-triplet ($\rho \in \{1,2,3\}$) dibaryon operators, and $v_{\rho}^{\sigma\sigma'}$ denotes Clebsch-Gordan coefficients (explicitly presented in Ref.~\cite{Amarasinghe:2021lqa}) projecting the product of two spin-1/2 operators into the particular dibaryon spin state. 
Two-nucleon operators including ``lower'' spin components $\sigma \in \{2,3\}$ were studied in Ref.~\cite{Detmold:2024iwz}, but effective energies computed using these operators suggest they predominantly overlap with higher-energy states than operators only involving the ``upper'' spin components $\sigma \in \{0,1\}$.
In this work, only upper spin-component fields are therefore considered.
Below, $\sigma \in \{0,1\}$ components are abbreviated as $\sigma \in \{ \uparrow, \downarrow \}$ for simplicity.
Projection to operators with definite isospin is accomplished using $P^{(0)} = i\tau_2$ and $P^{(1)} = i\tau_2 \tau_3$, where $I_z=0$ is chosen for simplicity.
Dibaryon operators with relative momenta $\vec{k} = (2\pi/L)\vec{n}$ are included with
\begin{equation}\label{eq:mtm}
\begin{split}
  \vec{n} &\in \big\{ (0,0,0),\ (0,0,1),\ (0,1,1),\ (1,1,1), \\
  & \hspace{20pt} \ (0,0,2), \ (0,1,2), \ (1,1,2), \,\ldots  \big\}\ ,
\end{split}
\end{equation}
where 
the ellipsis denotes all momenta related to the ones that are shown by cubic group transformations.

To enable direct comparisons with previous calculations, a local hexaquark operator is also considered. Previously, the low-energy spectrum has been studied with a complete basis of local hexaquark operators~\cite{Detmold:2024iwz}. This basis included operators which could not be written as a product of two color-singlet nucleon operators. However, at the statistical precision obtained in that work, effective energies for these ``hidden color'' operators suggested that they predominantly overlap with higher-energy states than hexaquark operators constructed from a product of two color singlets.
Therefore, only hexaquark operators which can be written as products of single-nucleon operators are considered in this work. 
Zero-momentum color-singlet-product hexaquark operators are defined as
\begin{equation}
  H^{I}_{\rho}(t) = \sum_{\vec{x} \in \Lambda_{\mathcal{S}}}  \sum_{\sigma,\sigma',u,u'} v_{\rho}^{\sigma\sigma'}  P_{uu'}^{(I)} N_{\sigma u}(\vec{x},t)  N_{\sigma' u'}(\vec{x},t)\ .
  \label{eq:hexaquark}
\end{equation} 
The $I_z=0$ case considered in this work involves
\begin{equation}
\begin{split}
H_0^{1} &= \sqrt{2} \sum_{\vec{x}\in\Lambda_\mathcal{S}}  p_{\downarrow}(\vec{x},t) n_{\uparrow}(\vec{x},t) - p_{\uparrow}(\vec{x},t) n_{\downarrow}(\vec{x},t), \\
H_1^{0} &= 2 \sum_{\vec{x}\in\Lambda_\mathcal{S}}  p_{\uparrow}(\vec{x},t) n_{\uparrow}(\vec{x},t), \\
H_2^{0} &= \sqrt{2} \sum_{\vec{x}\in\Lambda_\mathcal{S}}  p_{\downarrow}(\vec{x},t) n_{\uparrow}(\vec{x},t) + p_{\uparrow}(\vec{x},t) n_{\downarrow}(\vec{x},t), \\
H_3^{0} &= 2 \sum_{\vec{x}\in\Lambda_\mathcal{S}}  p_{\downarrow}(\vec{x},t) n_{\downarrow}(\vec{x},t),
\end{split}
\end{equation}
where the normalizations result from a product of spin and isospin Clebsch-Gordan coefficients.

Nucleon-nucleon scattering at the lowest energies predominantly occurs though the ${}^1S_0$ channel for $\nn$ and the coupled ${}^3S_1 - {}^3D_1$ channels for $\deut$.
When rotational invariance is broken to the cubic group, these subduce to the $A_1^+$ irreducible representation (irrep) for $\nn$ and the $T_1^+$ irrep for $\deut$.
The hexaquark operators defined above naturally transform in these irreps.

Dibaryon operators can be constructed to transform in a variety of irreps depending on the choice of linear combinations of plane-wave pairs related by cubic transformations.
It is convenient to construct dibaryon operators that transform in definite irreps since they do not mix under Euclidean time evolution and can be analyzed separately.
In practice, a correlator matrix involving all plane-wave operators with momenta in \cref{eq:mtm} is constructed and multiplied by change-of-basis matrices to convert to a basis of operators transforming in definite irreps, exactly as detailed in  Ref.~\cite{Amarasinghe:2021lqa}; see Refs.~\cite{Luu:2011ep,Morningstar:2013bda,Detmold:2024ifm} for more details.
This work is restricted to the $A_1^+$ irrep in the $\nn$ channel and the $T_1^+$ irrep in the $\deut$ channel, as relevant for $S$-wave nucleon-nucleon scattering. Higher partial waves can be studied using different irrep projections of the same operators, as in Refs.~\cite{Berkowitz:2015eaa,Amarasinghe:2021lqa,Detmold:2024iwz}. 

Shorthand notation is useful for describing the operator sets studied in this work. 
The $\deut$ and $\nn$ channels are analyzed separately below, and explicit isospin indices on interpolating operators are suppressed when clear from context.
Dibaryon operators are labeled by the magnitude of the integer triplet specifying the relative momentum $|\vec{n}|$ between the two nucleon operators. 
In situations where it is obvious, a simplified notation is employed to ensure readability of figures. The hexaquark operator is denoted $H$, and the dibaryon operators are denoted $D_{|\vec{n}|}$. 

Interpolating operator sets are further denoted through the shorthand $j D$ for sets of $j$ dibaryon operators and $jD+H$ for sets of $j$ dibaryon and one hexaquark operators, e.g., $6D$ or $7D+H$.
Typically, dibaryon operators with all relative momenta below a definite cutoff are included, e.g., $1D$, $2D$, $3D$, ... in the $\nn$ channel and $1D$, $3D$, $6D$, ... in the $\deut$ channel.
Sets with other operator orderings will be denoted as $jD'$ and their content will be described explicitly when relevant.

\subsection{Thermal effects}

The boundary conditions in the temporal direction that are imposed on the fermion and boson fields in this work correspond to a finite-temperature quantum field theory. This implies that two-point correlation functions contain contributions to the spectral decomposition that arise from thermal states in which color-singlet states propagate around the temporal boundary. Given the high statistical resolution of this calculation, it is important to quantify the significance of these effects. 

The single-nucleon interpolating operators used here contain a positive-parity projector and consequently thermal contributions to the nucleon two-point function are from negative-parity states.  From  analysis of the nucleon correlator in the region $t\in[L_t/2, L_t]$, the negative-parity nucleon energy is found to be $M_{N^*}=1.577(8)$ assuming $N$-state saturation. Interpreting the observed signal as the $N^*$ rather than a pion-nucleon scattering state $M_{N^*} \in [1.46, 1.74]$ at 68\% confidence  
via residual bounds.\footnote{Calculations using different negative-parity interpolating operators (involving negative-parity quark spinor components) on the same ensemble in Ref.~\cite{Detmold:2024iwz} obtained $M_{N^*} = 1.636(8)$ assuming $N$-state saturation.  Variational and residual bounds obtained in either case are consistent with one another, but it is  inconsistent to interpret both results as constraints on the negative-parity nucleon ground state under $N$-state saturation.} 
This should be compared with the lowest positive-parity nucleon energy $M_N = 1.2035(2)$ obtained assuming $N$-state saturation and $M_N \in [1.20,1.21]$ at 68\% confidence ($M_N \in [1.19, 1.21]$ at 95\% confidence) via residual bounds.

Assuming $N$-state saturation of both $N$ and $N^*$ correlators, ground-state overlaps are 0.790(1) and 0.17(1) for $N$ and $N^*$, respectively, and the ratio of $N^*$ to $N$ contributions at $t=L_t/2=24$ is suppressed by $3\times 10^{-5}$ at 68\% confidence.
Thermal effects are consequently negligible until statistical precision reaches a commensurate level.

In the two-nucleon sector, which is bosonic and therefore has the same eigenstates propagating both forward and backward in time, thermal effects are nonetheless suppressed. All interpolating operators that are used are built from single nucleon operators with positive-parity projectors, which, in the limit of no interactions, would source a combination of two negative parity nucleons propagating backward in time. It is expected that even with interactions, the overlap of this interpolating operator onto states similar to two positive parity nucleons  propagating backward in time will be small (this small overlap onto the ground state from ``wrong-parity'' dinucleon operators was seen in Ref.~\cite{Detmold:2024iwz}). 
Explicit calculations of dinucleon correlators with $t \in [L_t/2, L_t]$, computed for a subset of correlators with an order of magnitude less statistics than the full ensemble, validate this prediction and suggest that the dominant thermal states in the two-nucleon sector have excitation gaps larger than $2(M_{N^*} - M_N)$.
These results justify neglecting thermal effects in multi-state fits to one- and two-nucleon correlators over time ranges that are subsets of $[0,L_t/2]$.

Lanczos methods automatically incorporate thermal states as growing exponentials within spectral representations, which isolates non-thermal states from thermal effects.
In particular, the ZCW filtering used here removes thermal states as spurious because their overlaps are exponentially small in $L_t$.
A thermal ZCW test can be used to explicitly identify thermal states and determine their energies~\cite{Hackett:2024nbe}; however, results with $t \leq L_t/2$ do not provide any evidence for thermal states that are resolved from noise.

\section{Estimating gaps for LQCD spectroscopy}\label{sec:mind_the_gap}

Applying the gap bounds discussed in \cref{sec:gap,sec:improved_gap} to LQCD correlator data requires estimates of gap parameters $\hat{g}_k^{(m)}$ or $\hat{G}_k^{(m)}$.
Since the LQCD spectrum at unphysical quark masses and finite lattice spacing is not known \emph{a priori}, these gap parameters are not known exactly and must be estimated based on numerical data and physical expectations about the spectrum.

The first assumption required to estimate gap parameters for multi-hadron states is
that the ground-state energies of single-hadron states are approximately known.\footnote{Single-hadron energies are  more straightforward to determine as excited state effects are typically milder. This same assumption is required for the application of L{\"u}scher's quantization condition~\cite{Luscher:1986pf,Luscher:1990ux} and its generalizations.} 
Given a pair of single-hadron ground-state energies $M_1$ and $M_2$, the non-interacting two-hadron energy spectrum in the center-of-mass frame includes levels at 
\begin{equation}
\begin{split}
    \hat{E}_{n}(M_1, M_{2}) &= \sqrt{ M_1^2 + n \left( \frac{2\pi }{L} \right)^2 } \\
    &\hspace{20pt} + \sqrt{ M_{2}^2 + n \left( \frac{2\pi }{L} \right)^2 }.
    \end{split}
\end{equation}
While interactions shift these energies, small discrepancies between true and estimated energy gaps 
do not necessarily spoil the utility of gap bounds and their effects may be of similar order to other error-on-the-error effects. However, if the interactions are strong enough that the true spectrum contains bound states and resonances, the gap bounds will be unreliable. The second assumption required for the gap parameter estimates in this section is that this is not the case (estimates suitable for other scenarios may also be constructed; see e.g.~\cref{sec:srq}).

The lowest-energy states in the two-nucleon sector have non-interacting energies given by $\hat{E}_{n}(M_N, M_N)$, where the multiplicity of states with a given $n$ is non-trivial~\cite{Grosswald1985}.
Additional non-interacting energy levels arise from the presence of other hadrons.
For example, the two-nucleon sector at $m_\pi \approx 800$ MeV includes non-interacting $NN$ levels with energies $\hat{E}_n(M_N,M_N)$ as well as $\Delta \Delta$ levels with energies $\hat{E}_n(M_\Delta,M_\Delta)$ where $M_\Delta$ is the mass of the (stable) $I=3/2$ baryon ground state, which has been previously computed to be  $M_\Delta=1.3321(21)$~\cite{Chang:2015qxa} under $N$-state-saturation assumptions for the same action parameters.

\begin{table}[]
 \renewcommand{\arraystretch}{1.4}
    \centering
    \begin{ruledtabular}
    \begin{tabular}{c c c c}
   $\deut$ operators & $\nn$ operators & $\Egap$ & Energy
    \\
    \hline
         - & - & $\hat{E}_0(M_N, M_N)$ &  2.407 \\
         $1D$ & $1D$ & $\hat{E}_1(M_N, M_N)$ &  2.463 \\
         $3D$ & $2D$ & $\hat{E}_2(M_N, M_N)$ & 2.518 \\
         $6D$ & $3D$ & $\hat{E}_3(M_N, M_N)$ & 2.572 \\
         - & $\geq 4D$ & $\hat{E}_1(M_N, M_\Delta)$ & 2.589 \\
         $8D$ & - & $\hat{E}_4(M_N, M_N)$ & 2.625 \\
         - & - & $\hat{E}_2(M_N, M_\Delta)$ & 2.642 \\
         $\geq 10D$ & - & $\hat{E}_0(M_\Delta, M_\Delta)$ & 2.664 \\
         - & - &$\hat{E}_5(M_N, M_N)$ & 2.677  \\
    \end{tabular}
    \end{ruledtabular}
    \caption{Non-interacting energy levels relevant for low-energy $NN$ systems, ordered according to their total energy. Note that the 0-shell $N\Delta$ system has different quantum numbers and therefore does not mix with $\deut$ $NN$ states. In the $\nn$ channel, $\Egap$ for $jD+H$ interpolator sets is identified with $\Egap$ for the $(j+1)D$ interpolator set. In the $\deut$ channel, $\Egap$ for $jD+H$ is similarly taken to be $\Egap$ for the dibaryon interpolator set with one more relative momentum shell included.} 
    \label{tab:nonintE}
\end{table}

The $\deut$ channel involves $NN$ and $\Delta\Delta$ states.
If $r$ plane-wave $NN$ interpolators are present, then the first non-interacting ``missing level'' is either the $(r+1)$th $NN$ state or the first $\Delta\Delta$ state.
Concretely, in the $\deut$ channel, the energy used to compute Haas-Nakatsukasa gap-bound parameter estimates $\hat{G}_k^{(m)}$ via \cref{eq:GkEgap} with $\Egap$ defined as
\begin{equation}
    \Egap^{\deut} = \text{min} \{ \hat{E}_{r+1}(M_N, M_N),\ \hat{E}_0(M_{\Delta}, M_{\Delta}) \},
\end{equation}
where $r$ is the number of interpolators employed.
The non-interacting energies of this form relevant for the calculations below are reported in \cref{tab:nonintE}.

In the $\nn$ channel, additional $N\Delta$ states with energies $\hat{E}_n(M_N,M_{\Delta})$ are present.
Spin-singlet $N\Delta$ states must be in odd partial waves to have overall positive parity and therefore $\hat{E}_0(M_N,M_\Delta)$ is not relevant for $NN$ systems.
Therefore in the $\nn$ channel, 
the Haas-Nakatsukasa gap bound parameters $G_k^{(m)}$ are computed via \cref{eq:GkEgap} with $\Egap$ defined as
\begin{equation}
    \Egap^{\nn} = \text{min} \{ \hat{E}_{r+1}(M_N, M_N),\ \hat{E}_1(M_N, M_{\Delta}) \}
\end{equation}
as $\hat{E}_{1}(M_N, M_\Delta)<\hat{E}_{0}(M_\Delta, M_\Delta)$.
The numerical values of these energies are also shown in \cref{tab:nonintE}.

With $m_\pi \approx 800$ MeV, non-interacting $N\pi$ energies are significantly larger than inelastic nucleon excitation energies.
Therefore, in analyses of single-nucleon ground-state energies, the Davis-Kahan gap bound is used with gap parameter estimates $\hat{g}_0^{(m)}$ computed via  \cref{eq:gkEgap} with 
    $\lamgap = e^{-E_1^{(m)}}$.
An analogous gap parameter estimate is used  for single-pion correlators.

\section{Correlation function matrix analysis}
\label{sec:lqcd-lanczos}

Lanczos analyses of noisy systems with dense energy spectra using large correlator matrices share many similarities with previous analyses of single-hadron systems described in Refs.~\cite{Wagman:2024rid,Hackett:2024xnx,Ostmeyer:2024qgu,Chakraborty:2024scw,Hackett:2024nbe}.
Two important differences that arise for the present dataset are highlighted below:
\begin{itemize}[leftmargin=*]
  \item \emph{State identification}. Labeling the physical states that remain after spurious-state filtering is more challenging when more than a few excited states are of interest and there are close-to-degenerate eigenvalues expected. A new state-identification scheme based on overlaps with Ritz vectors from early Lanczos iterations where all states are non-spurious is therefore used below. 
  \item \emph{Stagnation}. 
  Apparent plateaus of Lanczos energy-estimators as a function of the iteration number 
    may signal that the information available at a given level of statistics is exhausted, rather than that convergence within statistical uncertainties has been achieved. 
    This behavior can be viewed as a finite-statistics analog of the stagnation phenomena arising from finite-precision effects in oblique Lanczos~\cite{Leyk:1997,Gaaf:2016,Saad:2011} discussed in Refs.~\cite{Hackett:2024xnx,Abbott:2025yhm}.
    Stagnation is a general concern for all finite-statistics LQCD analyses (using either Lanczos or multi-state fit techniques), but it appears to be especially severe for multi-hadron systems where gaps are small.
    Observations of severe stagnation in this LQCD two-nucleon dataset strongly motivate the importance of considering two-sided bounds on energies that do not assume convergence within statistical uncertainties.
\end{itemize}

\subsection{Symmetric correlation functions}\label{sec:diag}

\begin{figure}
  \includegraphics[width=0.48\textwidth]{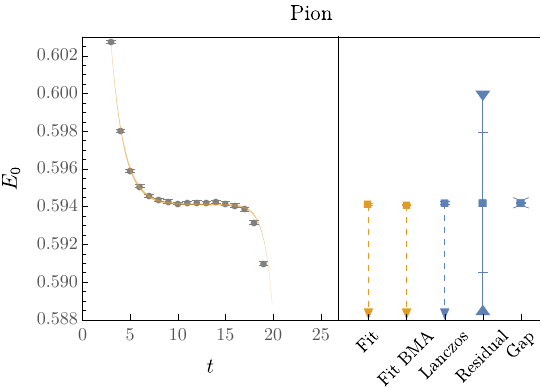}\vspace{15pt}
  \includegraphics[width=0.48\textwidth]{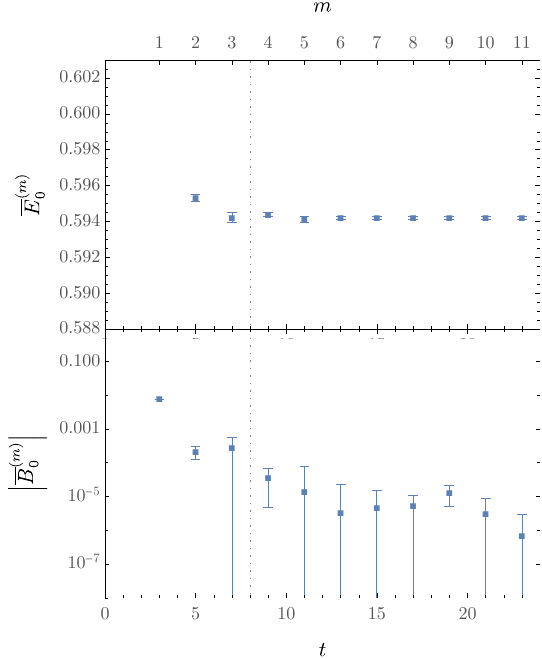}
    \caption{
      Upper: Pion effective masses and multi-state fit results (orange) along with Lanczos energy estimators and two-sided residual and gap bounds from a median of $m\in \{8,9,10\}$ (blue); details are as in \cref{fig:complex-scalar}. Lower: Iteration dependence of Lanczos energy estimators and residual-norm-square estimators. 
      A vertical dashed lines is placed after the median $m$ where spurious eigenvalues first appear; the previous iteration (at the inner bootstrap level) is used for SLRVL state identification; see \cref{sec:stateID}.
     Correlator results with $t \in \{0,1\}$ are omitted to avoid complications arising from contact terms. 
     \label{fig:pion-Lanczos-EMP}
     }
\end{figure}

\begin{figure}
    \includegraphics[width=0.48\textwidth]{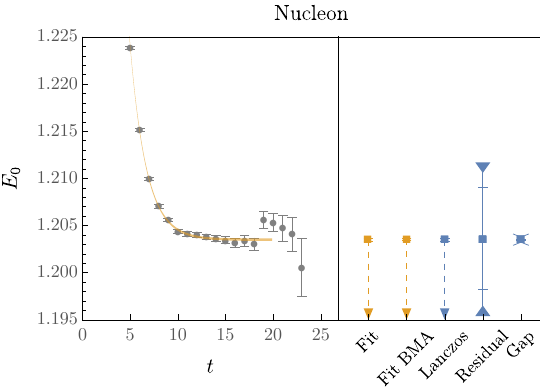}\vspace{15pt}
    \includegraphics[width=0.48\textwidth]{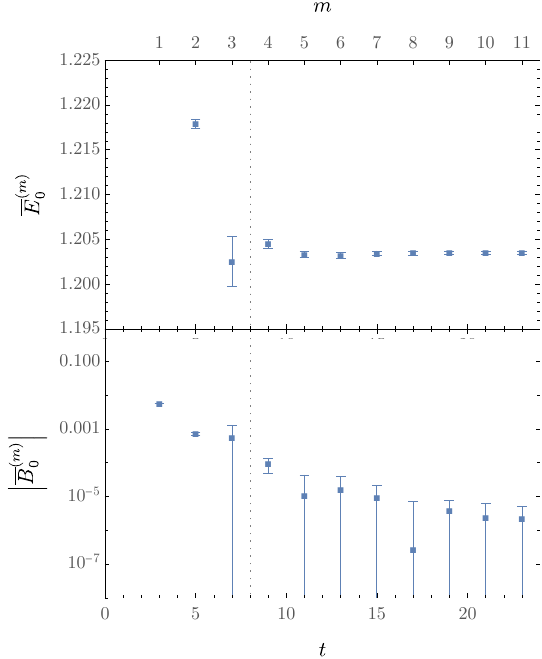}
    \caption{
      Upper: Nucleon effective masses, multi-state fit results, Lanczos energy estimators, and two-sided bounds analogous to \cref{fig:pion-Lanczos-EMP}. 
      Lower: Iteration dependence of Lanczos energy estimators and residual-norm-square estimators, with details as in \cref{fig:pion-Lanczos-EMP}.
      Note that the unusual behavior near $t=20$ appears to be a statistical fluctuation.
    }
    \label{fig:nuc-Lanczos-EMP}
\end{figure}

The pion and nucleon sectors are each studied here using a single interpolating operator, and their analysis is presented in Figs.~\ref{fig:pion-Lanczos-EMP} and \ref{fig:nuc-Lanczos-EMP} and \cref{tab:results}.
Sub-per-mille precision on both pion and nucleon effective masses is achieved for $t \alt 20$, almost to the midpoint of the temporal extent of the lattice. 
Lanczos energy estimators show rapid convergence until saturation at $t \sim 9$ in both cases.
This is also around when spurious eigenvalues begin to appear.\footnote{Here and below, Ritz values are filtered using the Hermitian-subspace and ZCW tests with $F_{\rm ZCW} = 10$. State identification is performed using SLRVL.
}
Statistical uncertainties in Lanczos energy estimators decrease until $t \sim 21$ and then saturate in both cases. 
Final results for Lanczos energy estimators $\overline{E}_k$ and corresponding two-sided bound estimators shown in \cref{tab:results} are taken from medians over iterations $m \in \{8,9,10\}$ (the largest iterations where residual-norm-squares can be computed from data with $t \in [2,L_t/2]$).
For iterations in this large-$m$ region,  correlations between energy estimators are $O(1)$ (see \cref{app:lanczos}).

\begin{figure}
    \includegraphics[width=0.48\textwidth]{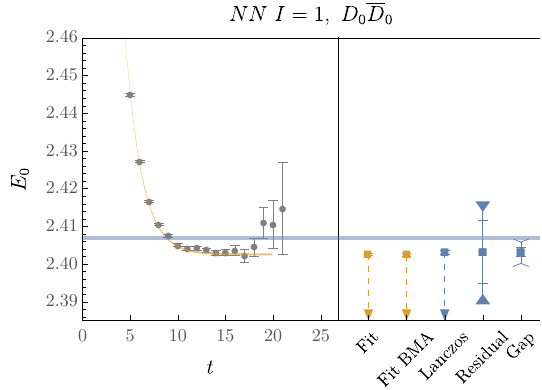}\vspace{10pt}
    \includegraphics[width=0.48\textwidth]{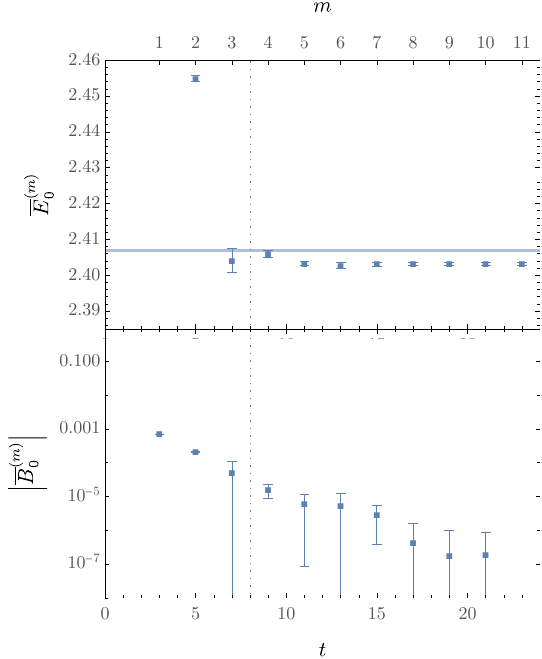}
    \caption{
      Upper: Dineutron effective masses, multi-state fit results, Lanczos energy estimators, and two-sided bounds using $D_0 \overline{D}_0$ correlators. Lower: Iteration dependence of Lanczos energy estimators and residual-norm-square estimators, with details as in \cref{fig:pion-Lanczos-EMP}.
      The blue horizontal band shows the threshold $2M_N$; its width corresponds to gap bounds for the lowest-energy nucleon state. 
    }
    \label{fig:nn-Lanczos-EMP}
\end{figure}

\begin{figure}
    \includegraphics[width=0.48\textwidth]{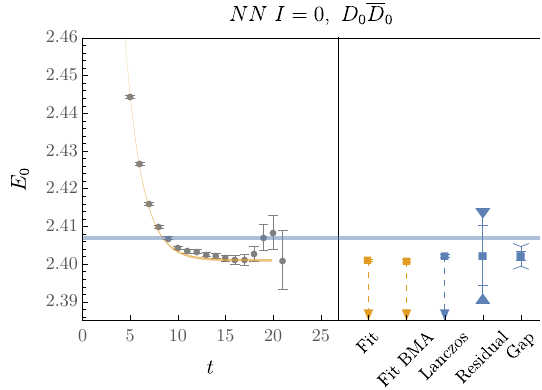}\vspace{10pt}
    \includegraphics[width=0.48\textwidth]{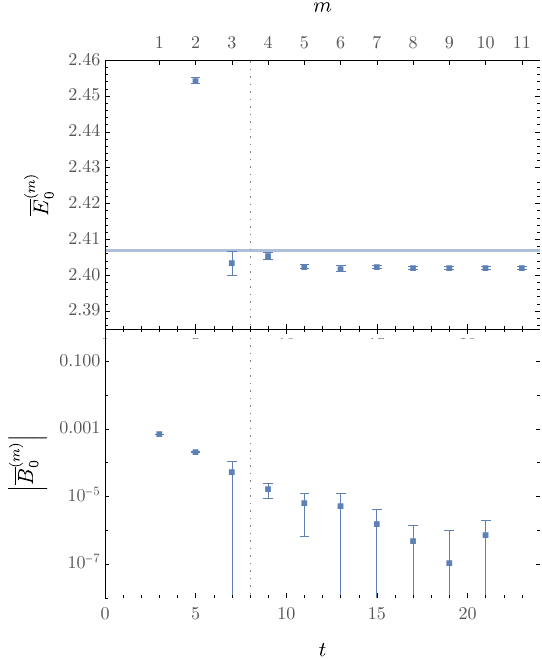}
    \caption{
      Upper: Deuteron effective masses, multi-state fit results, Lanczos energy estimators, and two-sided bounds using $D_0 \overline{D}_0$ correlators. Lower: Iteration dependence of Lanczos energy estimators and residual-norm-square estimators, with details as in \cref{fig:nn-Lanczos-EMP}.
     }
    \label{fig:deut-Lanczos-EMP}
\end{figure}

Qualitatively similar behavior is observed for  both the $\nn$ and $\deut$
 symmetric two-nucleon correlators built from $D_0$ operators, as shown in \cref{fig:nn-Lanczos-EMP,fig:deut-Lanczos-EMP}.
Noise effects enter slightly earlier than for the single hadron systems: spurious eigenvalues appear around $t \sim 7$, and uncertainties saturate, and large correlations arise, around $t \agt 17$.
Ground-state gap bounds are computed using Haas-Nakatsukasa gap bounds with $\Egap$ appropriate for the $1D$ interpolator set, as described in \cref{sec:mind_the_gap}.

For symmetric correlator matrices at infinite statistics with negligible thermal effects, the residual-norm-squares $\overline{B}_k^{(m)}$ must decrease monotonically with increasing $m$~\cite{Parlett}.
The behavior of $\overline{B}_k^{(m)}$ therefore provides another measure of convergence of the Lanczos approach.
For finite-statistics results where convergence is limited by noise (as in the examples of \cref{sec:SHO}), the $\overline{B}_k^{(m)}$ are expected to decrease monotonically for some number of iterations, become consistent with zero within $1\sigma$, and subsequently saturate. 
In these LQCD results, 
 $\overline{B}_0^{(m)}$ becomes consistent with zero at $1\sigma$ for $t \gtrsim 11$ for pion and nucleon correlators and $t \gtrsim 13$ for $NN$ correlators.
This suggests that saturation of residual-norm-squares in these examples is due to finite-statistics effects.

\begin{figure*}
    \includegraphics[width=0.48\textwidth]{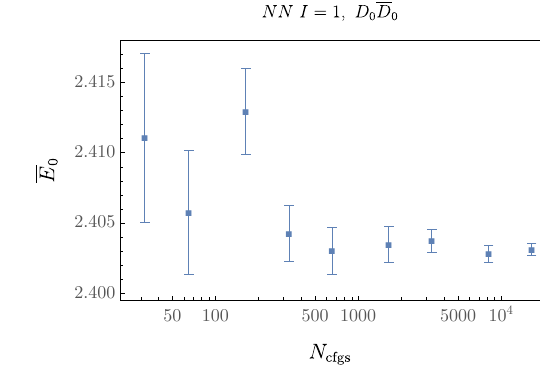}\qquad
    \includegraphics[width=0.48\textwidth]{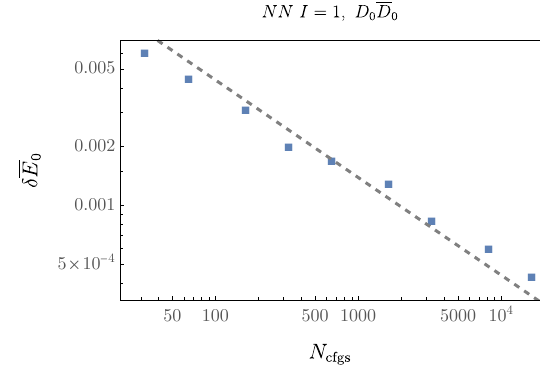}\vspace{15pt}
    \includegraphics[width=0.48\textwidth]{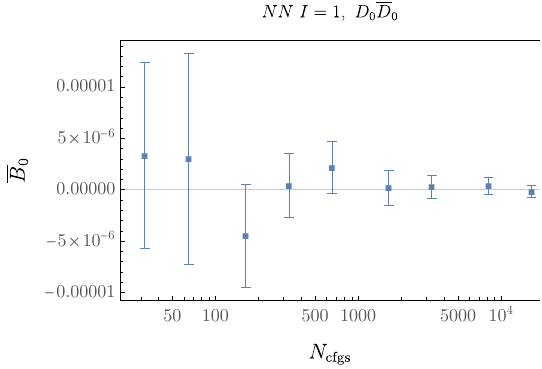}\qquad
    \includegraphics[width=0.48\textwidth]{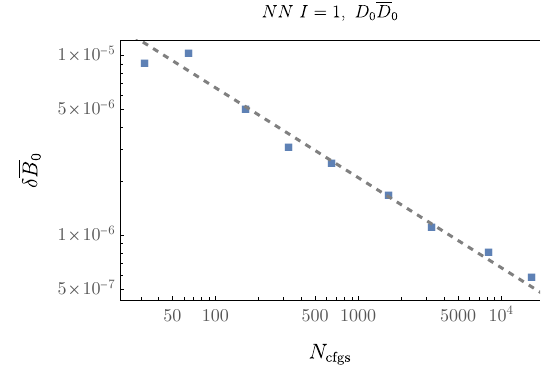}
    \caption{
      Convergence of Lanczos ground-state energy estimators, top, and residual-norm-square estimators, bottom, from a median of iterations $m \in \{8,9,10\}$ as a function of $N_{\rm cfgs}$.
     The left panels show the central values and the right panels show the behavior of the statistical uncertainties.
    }
    \label{fig:nn-stats}
\end{figure*}

Further evidence that saturation of residual-norm-squares arises from finite-statistics effects is given by studying the scaling of Lanczos results with $N_{\rm cfgs}$. 
The  $\nn$ $NN$ channel provides a representative example of this scaling behavior and is shown in \cref{fig:nn-stats}.
Clear $1/\sqrt{N_{\rm cfgs}}$ scaling is observed for the statistical uncertainties of the large-$m$ medians of both $\overline{E}_0$ and $\overline{B}_0$. 
Residual-norm-squares are consistent with zero at $1\sigma$ for all values of $N_{\rm cfgs}$ that are studied, and the central values of $\overline{B}_0$ decrease $\propto 1/\sqrt{N_{\rm cfgs}}$ along with the corresponding uncertainties.
This demonstrates that it is finite-statistics effects, rather than finite-$m$ truncation effects, that are responsible for setting the magnitude of $\overline{B}_0$.
Since $\overline{B}_0$ is still consistent with zero for the full statistical ensemble studied here, it is likely that even higher statistics calculations of the same correlators would further decrease $\overline{B}_0$ with $\propto 1/\sqrt{N_{\rm cfgs}}$ scaling and thus continue to improve the corresponding residual and gap bounds.

\begin{table*}[t]
    \centering
    \begin{ruledtabular}
    \begin{tabular}{cllll}
    State & $\overline{E}_0$  & $\delta \overline{E}_0$ & Gap & Residual
    \\
    \hline
      $\pi$              & 0.594204 & 0.000062 & 0.00012 & 0.0056 \\
      $N$               & 1.20349 & 0.00018 & 0.00027 & 0.0078 \\
      $N^*$             & 1.5772 & 0.0076 & 0.024 & 0.12 \\ \hline
      $NN\ \nn, \ 1D$         & 2.40311 & 0.00043 & 0.0029 & 0.013 \\
      $NN\ \deut, \ 1D$       & 2.40207 & 0.00041 & 0.0026 & 0.012 \\\hline

      $NN\ \nn, \ 3D$   & 2.40252 & 0.00041 & 0.0021 & 0.017 \\
      $NN\ \deut,\ 10D$ & 2.40094 & 0.00041 & 0.0025 & 0.026 \\
     \end{tabular}
    \end{ruledtabular}
    \caption{Lanczos ground-state energy estimator results. Columns from left to right show the state name, large-$m$ median Lanczos energy estimators, statistical uncertainties, gap lower bound widths at $68\%$ confidence, and residual lower bound widths at 68\% confidence. Scalar Lanczos results (top four rows) show Davis-Kahan gap bounds with $\hat{g}_0^{(m)}$ estimated as described in \cref{sec:mind_the_gap}, while block Lanczos results (bottom two rows) show Haas-Nakatsukasa gap bounds with $\hat{G}_0^{(m)}$ estimated as described in \cref{sec:mind_the_gap} for the interpolator sets indicated (which lead to the most constraining gap bounds).
    \label{tab:results}
    }
\end{table*}

\begin{figure}
    \includegraphics[width=\columnwidth]{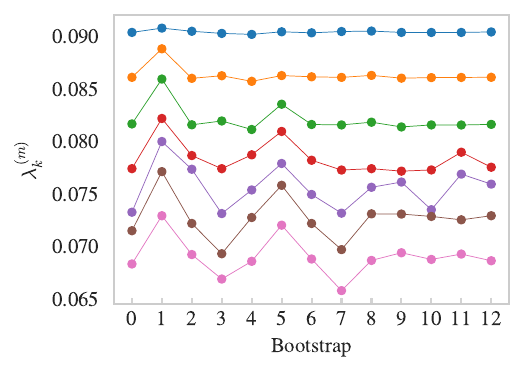} 
    \caption{
      Illustrating the issue of indexing dislocations in filter-and-sort state identification, block Lanczos estimates are shown for the highest seven eigenvalue (lowest seven energy) estimators for the $\nn$ channel with the $7D+H$ interpolating operator set for different (inner) bootstrap ensembles. 
      Different colors indicate the labels assigned to each state; same-color points (linked by lines to guide the eye) are identified as the same state for subsequent averaging between different bootstrap ensembles.
    }
    \label{fig:dineut-filter-and-sort-boots}
\end{figure}

\begin{figure*}
\includegraphics[width=0.98\textwidth]{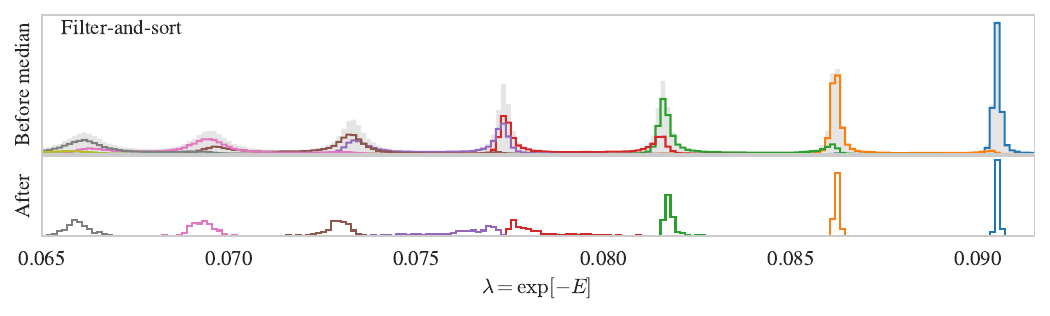}
\includegraphics[width=0.98\textwidth]{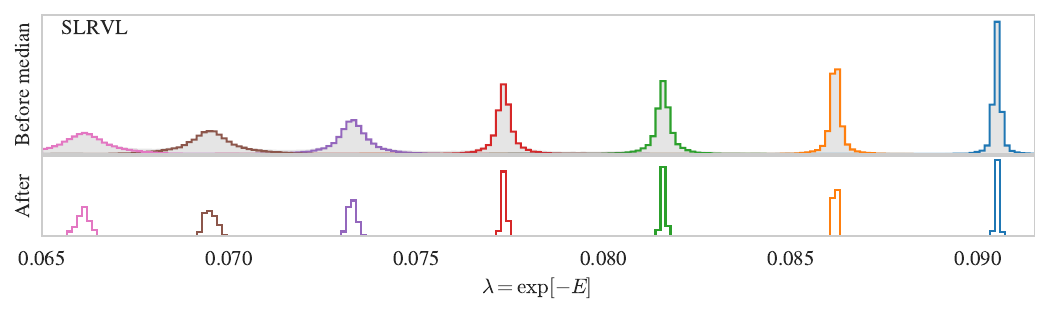}
    \caption{
      Ritz-value histograms for the $\nn$ channel using the $7D+H$ interpolating operator set from  nested bootstrap sampling with $N_{\rm boot} = 200$ (gray, both panels). Colored histogram outlines show the Ritz values labeled as particular states using filter-and-sort state identification after the Hermitian subspace and ZCW tests, top, compared with those obtained using SLRVL labeling as described in the main text, bottom. The upper and lower parts of the two sub-panels show the Ritz-value histograms before and after inner bootstrap medians are taken. In both cases the upper sub-panels include gray background histograms showing the full set of non-spurious Ritz values before state identification.
    }
    \label{fig:histograms:dineut}
\end{figure*}

\begin{figure*}
\includegraphics[width=0.98\textwidth]{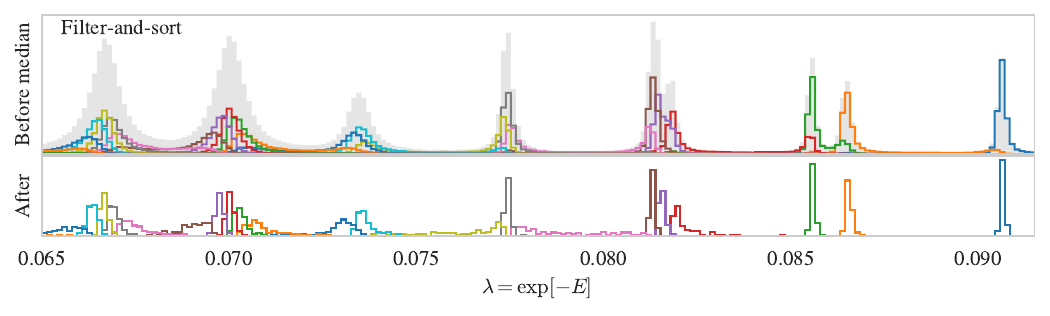}
\includegraphics[width=0.98\textwidth]{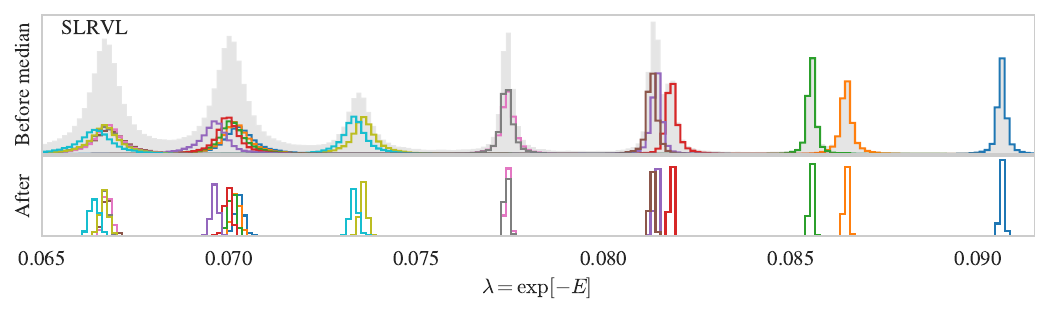}
    \caption{
      Ritz-value histograms for the $\deut$ channel with the $20D+H$ interpolating operator set, with details as in \cref{fig:histograms:dineut}.
    }
    \label{fig:histograms:deut}
\end{figure*}

\subsection{State identification}\label{sec:stateID}

\begin{figure*}
    \includegraphics[width=0.48\textwidth]{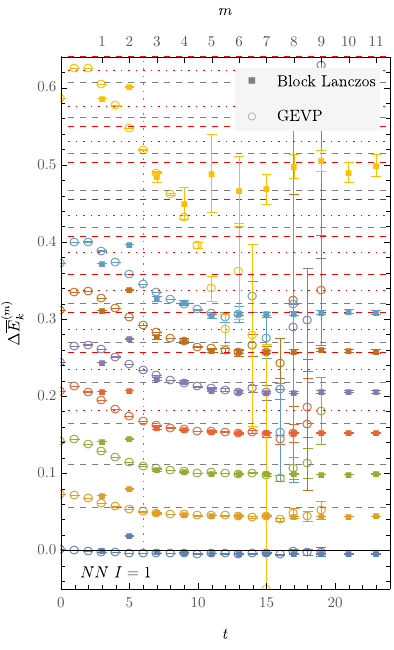} 
   \quad \includegraphics[width=0.48\textwidth]{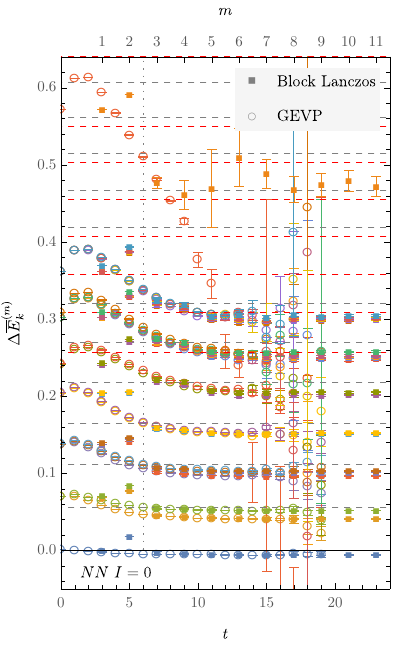} 
    \caption{
      Block-Lanczos correlated energy-shift estimators compared with effective-mass differences of GEVP-optimized correlators for both $NN$ channels. Different colors correspond to estimators for different states. Dashed gray, dashed red, and dotted red lines show non-interacting $NN$, $\Delta\Delta$, and $N\Delta$ scattering levels, respectively.
      GEVP results are shown for $t<20$ as subsequent results have uncertainties larger than the plot range.
    }
    \label{fig:deut-Lanczos-GEVP-comp}
\end{figure*}

Depending on the level of statistics and the correlator matrix rank, state identification in block Lanczos is more challenging than for scalar correlators and requires the introduction of new methodology. 
Previous works on the Lanczos approach~\cite{Wagman:2024rid,Ostmeyer:2024qgu,Hackett:2024nbe,Hackett:2024nbe} have explored a simple filter-and-sort scheme: after removing spurious states, the surviving states are numerically sorted based on their Ritz values, such that $\lambda^{(m)}_k\geq \lambda^{(m)}_{k+1}$.

As illustrated in \cref{fig:dineut-filter-and-sort-boots}, filter-and-sort state identification may fail for larger interpolator sets due to e.g.~``dislocations'' in the indexing resulting from imperfect state filtering.
This occurs when multiple Ritz values on a given bootstrap ensemble are nearly coincident (e.g.~within $1\sigma$) with some $\overline{\lambda}_k^{(m)}$.
One of these multiple eigenvalues will be labeled as $\overline{\lambda}_k^{(b,m)}$ with the other mislabeled as $\overline{\lambda}_{k+1}^{(b,m)}$.
This can in turn cause a Ritz value on this bootstrap ensemble nearly coincident with $\overline{\lambda}_{k+1}^{(m)}$ to be mislabeled as $\overline{\lambda}_{k+2}^{(b,m)}$.
In this way, the presence of a multiple eigenvalue for the $k$th state leads to the appearance of anomalously large discrepancies between $\overline{\lambda}_{l}^{(b,m)}$ and $\overline{\lambda}_{l}^{(m)}$ for all $l > k$.
These multiple eigenvalues appear in only some bootstrap ensembles and not others, resulting in inconsistent indexing of states and thus, downstream, the averaging-together of estimators for different states.
This is somewhat analogous to the issue of eigenvector tracking in GEVP correlator matrix analyses~\cite{Dudek:2010wm,Fischer:2020bgv}. 

The top sub-figures of \cref{fig:histograms:dineut,fig:histograms:deut} illustrate the consequences of dislocations when filter-and-sort indexing is applied to analyses of large correlator matrices.
For example, in the top panel of \cref{fig:histograms:dineut},  the unlabeled set of eigenvalues for the $\nn$ channel (in gray) form a clear peak structure.
However, when labeled by filter-and-sort (colored histograms), bleeding between peaks is visible.
For example, the first excited state (orange) includes some points more naturally associated with (i.e., closer to) the ground-state peak (blue).
The issue becomes more severe higher up the spectrum: the third and fourth excited states (red, purple) are split nearly evenly between two different peaks.
The result of taking a median over inner bootstraps \footnote{The inner median relies on state identification to determine which inner-bootstrap estimators to combine.} (bottom sub-panels) further provides clear indications of a problem.
For example, for the third and fourth excited states in the $\nn$ channel in \cref{fig:histograms:dineut}, the post-median distribution of estimates lies primarily \emph{between} peaks of the original histogram.
This means that the median will return estimates in  regions where there is near-zero density of actual Ritz value estimates.
The same issue is apparent in several states in the more complicated $\deut$ channel shown in \cref{fig:histograms:deut}.

Absent a perfect filtering prescription, a better approach to state labeling is required for large sets of interpolating operators.
In the $\nn$ channel, these can be corrected empirically or by adopting a cluster algorithm for state identification such as in Refs.~\cite{Cushman:2019tcv,Chakraborty:2024scw}.
However in the deuteron channel, the presence of nearly degenerate energy levels associated with admixtures of $S$- and $D$-wave  states makes histogram-based state identification more challenging.

In this work,  an approach which exploits the 
additional information about Ritz vectors made available by the Lanczos formalism is explored.
In particular, inner products between Ritz vectors after different numbers of Lanczos iterations $m$, $m'$ are straightforward to compute:
\begin{equation}
  \braket{y^{(m)}_k | y^{(m')}_l} = \sum_{s,t=1}^{m'} \sum_{a,b} P^{(m)*}_{kta} H^{(0)}_{ta,sb} P^{(m')}_{lsb},
\end{equation}
where $m' \leq m$ is taken for concreteness.
The resulting information is similar to that used in eigenvector-based labeling in GEVP analyses~\cite{Dudek:2010wm,Fischer:2020bgv}, but differs by the nontrivial metric $H^{(0)}$ in the inner product, and in its interpretation in terms of overlaps between states.

For the first few iterations ($m \lesssim 3$ in this dataset for the $NN$ systems), no spurious states appear
and the dislocation issues discussed above are absent. 
This implies that applying filter-and-sort state identification will lead to stable results that are consistent between different bootstrap ensembles for $\mr$ different states. 
Further, results for these iterations can be interpreted as arising from the symmetric, rather than oblique, Lanczos algorithm and therefore the Ritz values associated with these states can be viewed as optimal $T$ eigenvalue approximations within an $\mr$-dimensional block Krylov space.
Oblique Lanczos results associated with larger $m$ can provide better convergence, so simply stopping the Lanczos iteration when spurious states arise does not optimally extract information from correlator data.
However, the small-$m$ Ritz vectors provide a useful set of reference vectors to label states from larger-$m$ iterations.

The particular Ritz-vector based labeling scheme employed here is symmetric Lanczos Ritz vector labeling (SLRVL), as briefly introduced in the discussion of the complex harmonic oscillator.
The precise definition of SLRVL state identification is as follows:
\begin{itemize}[leftmargin=*]
  \item Identify the last symmetric Lanczos iteration $m_\mathrm{last}$, defined as the largest $m$ where there are zero spurious eigenvalues identified by the Hermitian-subspace and ZCW tests.
  \item Sort the states at $m_\mathrm{last}$ by the magnitudes of their Ritz values, $\lambda^{(m_{\rm last})}_k$. 
  \item For all $m > m_{\rm last}$, apply spurious-state filtering and then compute the inner products $\braket{y^{(m)}_l| y^{(m_\mathrm{last})}_l}$ between all non-spurious states and the ``label vectors'' $\ket{y^{(m_\mathrm{last})}_l}$.
    \item Reorder the non-spurious states $\ket{y^{(m)}_k}$: for label-vector index $l$, define $\ket{y^{(m)}_l}$ to be the state with greatest absolute value\footnote{Empirically, Ritz vectors between different iterations may be nearly anti-parallel, rather than parallel. This reflects only an unphysical sign ambiguity in the eigenvector normalization, and thus requires one to consider the absolute value of the Ritz-vector dot product.} of $\braket{y^{(m)}_l| y^{(m_\mathrm{last})}_l}$.
\end{itemize}
Within a nested median approach, SLRVL is applied independently for each inner bootstrap ensemble.
Note that in this scheme, the same $\ket{y^{(m)}_k}$ may be replicated as the highest-overlap vector with multiple different label vectors $\ket{y^{(m_\mathrm{last})}_l}$ in principle; however, the approximate orthogonality of the label vectors makes this rare, and pathological behavior arising from this possibility has not been observed in practice.
It is also possible for more non-spurious states to survive filtering for $m$ than for $m_{\rm last}$, in which case the Ritz values that do not have largest overlap with one of the $m_{\rm last}$ label vectors are discarded. 

Since Ritz vectors at iteration $m$ that have maximal overlap with non-spurious label vectors at $m_{\rm last}$ are unlikely to be spurious themselves, the SLRVL state-identification scheme reduces the importance of ZCW filtering at the target $m$. 
In this analysis, nearly identical results are obtained if Hermitian-subspace filtering and SLRVL state identification are applied without ZCW filtering for each target $m$.
However, it remains important to employ ZCW to define $m_\mathrm{last}$, as spurious states may appear at small $m$ that are not diagnosed by the Hermitian subspace test.

The results of applying SLRVL to block Lanczos analyses of $NN$ correlator matrices are shown in the bottom subfigures of \cref{fig:histograms:dineut,fig:histograms:deut}.
The improved labeling is apparent: all peaks in the Ritz value distribution are clearly associated with a single label and median estimators have densities closely corresponding to the peaks in the unlabeled distribution.
The issue of the inner median producing estimates in low-density regions between peaks is resolved.
In the $\nn$ channel, SLRVL resolves an apparent near-degenerate pair of states (red, purple at top of \cref{fig:histograms:dineut}) that arises as an artifact of taking an inner median with poor state labeling; appropriately labeled, the same eigenvalues become the single red peak in the lower panel of \cref{fig:histograms:dineut}.

Further improvements are apparent in the case of the $\deut$, where there are several near-degenerate levels.
Consider the cluster of three near-degenerate levels near the non-interacting $|\vec{n}|=2$ shell energy  with eigenvalues near 0.082.
Filter-and-sort labeling (\cref{fig:histograms:deut}, top subfigure) produces three ordered states with non-overlapping median distributions.
This is an unphysical admixture of three nearly degenerate levels, which have orthogonal eigenvectors that strongly overlap with different interpolating operators.
Using this extra information as encoded in Ritz vector overlaps,
SLRVL produces a result in much closer alignment with physical intuition: 
one state splits off due to spin-orbit coupling, while the other two states remain nearly degenerate, as expected from the lattice $O_h^D$ subductions of the $S$-wave and $D$-wave states.
Inspection of the corresponding overlap factors shows that the state whose energy can be resolved to be slightly lower than the other two states predominantly overlaps with the $S$-wave $|\vec{n}|=2$ dibaryon interpolating operator, as expected on physical grounds.

\subsection{Block Lanczos and GEVP}

Block Lanczos results with SLRVL state identification provide energy estimators whose central values and uncertainties are both consistent with analyses involving multi-state fits to the optimized correlators obtained from GEVP methods, \cref{eq:GEVP_corr}.
\Cref{fig:deut-Lanczos-GEVP-comp} shows the spectra extracted in the $\nn$ and $\deut$ channels, using both block Lanczos and GEVP on the same fixed subsets of Euclidean time ranges and the largest interpolating operator sets ($7D+H$ and $20D+H$, respectively) in each case.
In particular, the figure shows the Euclidean time dependence of GEVP and block Lanczos results for correlated differences 
\begin{equation}
  \overline{\Delta E}_k^{(m)} \equiv  \text{median}_{b} \left[ E_k^{(m,b)NN} - 2 E_0^{(m,b)N} \right],
\end{equation}
where superscripts label energy estimators associated with either $NN$ or single-nucleon correlators.
The uncertainties $\delta \overline{\Delta E}_k^{(m)}$ are computed in a nested bootstrap approach from 68\% empirical bootstrap confidence intervals of 
\begin{equation}
  \overline{\Delta E}_k^{(b,m)} \equiv  \text{median}_{b'} \left[ E_k^{(m,b,b')NN} - 2 E_0^{(m,b,b')N} \right].
\end{equation}

For small times, correlated differences of block-Lanczos energy estimators vary less smoothly than the GEVP principal correlator effective masses.
This non-smoothness
is not present in either $NN$ or single-nucleon energy estimators individually, so it  reflects differences in the rate at which excited-state effects are removed from $NN$ and single-nucleon correlators by block Lanczos.
For large $m$, there is $1\sigma$ consistency between block Lanczos and GEVP results for correlators with the same maximum $t$ for states $k \in \{0,\ldots,6\}$ in the $\nn$ channel and states $k \in \{0,\ldots,20\}$ in the $\deut$ channel.
Interesting differences between block Lanczos and GEVP energy estimators appear for the $k=8$ state in the $\nn$ channel and $k=21$ state in the $\deut$ channel, both of which predominantly overlap with hexaquark interpolating operators. 
Block-Lanczos energy estimators are much more stable with increasing $t$ than GEVP energy estimators for these states. However, it is not clear whether this signals improved genuine convergence or is instead of symptom of finite-statistics stagnation. 

\begin{figure*}[t]
    \includegraphics[width=0.48\textwidth]{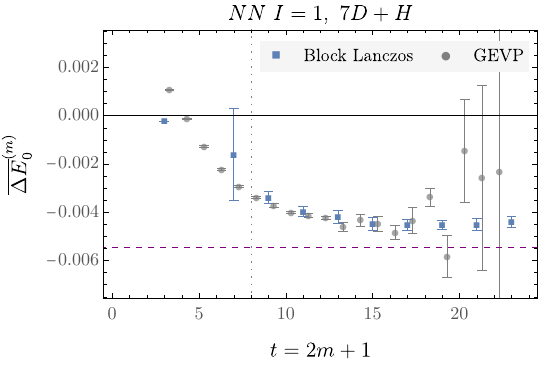}
    \includegraphics[width=0.48\textwidth]{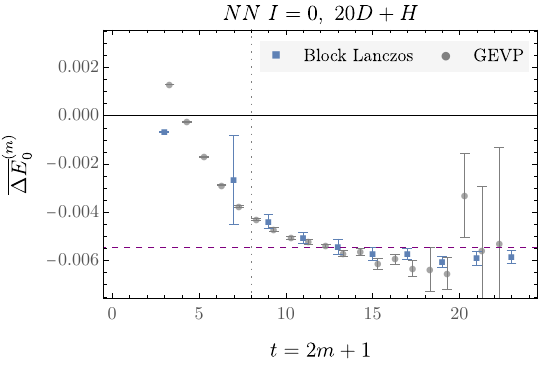} 
    \caption{
    Block-Lanczos results  for $\nn$ (left) and $\deut$ (right) $NN$ ground-state energy shifts $\overline{\Delta E}_0^{(m)}$ are compared to GEVP-optimized effective-mass differences involving correlator matrices with the same maximum $t$. In each case, the largest interpolating operator set is used: $7D+H$ for $\nn$ and $20D+H$ for $\deut$.
    The dashed purple line indicates the threshold between a bound and unbound state at leading order in a large-volume expansion of L{\"u}scher's quantization condition as discussed in Sec. 2.6 of Ref.~\cite{Beane:2010em}. 
    }
    \label{fig:nn-deut-DEMP}
\end{figure*}

\begin{figure*}
    \includegraphics[width=0.4\textwidth]{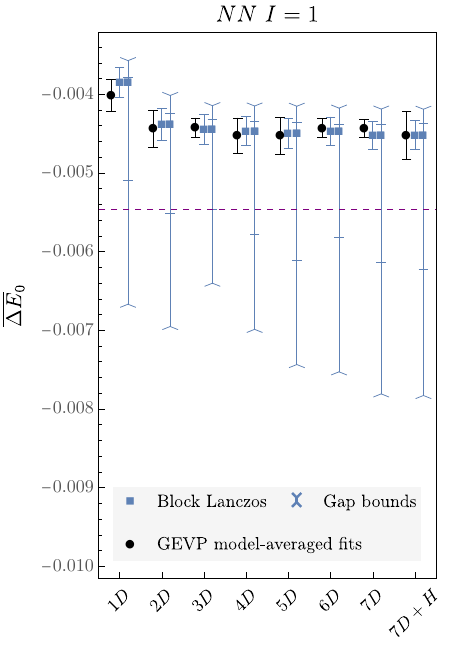}\qquad
    \includegraphics[width=0.4\textwidth]{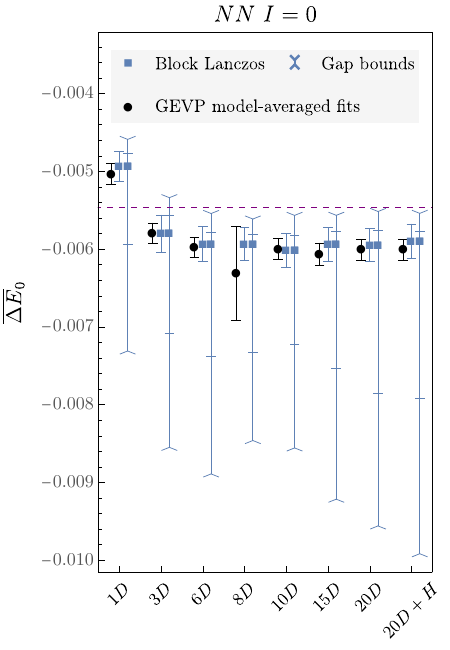} 
    \caption{
      Block-Lanczos results for $\nn$ (left) and $\deut$ (right) $NN$ ground-state energy shifts $\overline{\Delta E}_0$ from the median of iterations $m \in \{8,9,10\}$ are compared to GEVP fit results obtained as detailed in Refs.~\cite{NPLQCD:2020ozd,Amarasinghe:2021lqa,Detmold:2024iwz} using BMA weights~\cite{Jay:2020jkz} with flat priors.
    Haas-Nakatsukasa gap bounds, \cref{eq:Gap_bound}, are also shown.
    The dashed purple line shows the $k\cot\delta = 0$ threshold as in \cref{fig:nn-deut-DEMP}.
    }
    \label{fig:nn-deut-rank}
\end{figure*}

Ground-state energy estimators in particular are consistent at the $1\sigma$-level between block Lanczos and GEVP for $t \gtrsim 9$ in both channels; see \cref{fig:nn-deut-DEMP}.
Both block Lanczos and GEVP energy estimators are intriguingly close to the threshold where the phase shift at this FV energy changes sign.
Under $N$-state-saturation assumptions, this suggests that the inverse scattering length in both channels is unnaturally small \cite{NPLQCD:2020lxg}.

For an asymptotically large volume in the continuum limit,  variational bounds that are below the threshold for $k \cot\delta$ to be negative would be sufficient to conclude that there is a bound state. 
This situation is realized
in the $\deut$ channel at more than $2\sigma$; however, because of the unnaturally small scattering length, it is also possible that sub-leading finite-volume effects are significant  and that a bound state is not present.
Alternatively, the variational bounds that these energy estimators provide may be far from saturated in either or both channels, which would imply that a relatively deep bound state is present.

It is interesting to compare energy estimators obtained by analyzing different interpolating operator sets.
\Cref{fig:nn-deut-rank} compares the medians of $m \in \{8,9,10\}$ block-Lanczos results with multi-state fits to GEVP-optimized correlators for various operator sets.
Agreement at the $1\sigma$ level between block Lanczos and GEVP is found for all sets.
However, differences are clearly visible in both $NN$ channels between energy estimates from the $1D$ interpolating operator sets (the $|\vec{n}|=0$ diagonal correlators discussed in \cref{sec:diag}) and the $jD$ (and $jD+H$) interpolating operator sets with $j > 1$. 
Correlated differences between $1D$ and $7D+H$ block-Lanczos results show 4$\sigma$ deviations of $5.8(1.3) \times 10^{-4}$ in the $nn$ channel, while correlated differences of $1D$ and $20D+H$ block Lanczos results show 5$\sigma$ deviations of 3.8(9) $\times 10^{-4}$ in the $\deut$ channel.
Uncorrelated differences are less clearly resolved but still larger than $2\sigma$ in both channels.
These discrepancies show concretely that caution is required in interpreting statistical uncertainties on energy estimators for systems with small gaps and that the convergence in Euclidean time or Lanczos iteration observed in \cref{sec:diag} is not sufficient to 
guarantee correctness.

Statistically significant differences between $\overline{E}_0^{(m)}$ estimates for different interpolator sets over many iterations could be interpreted in the Lanczos context as stagnation---$1D$ ground-state energy estimates are approximately constant in $m$ but have not converged within their statistical uncertainties to the same result provided by larger sets of interpolating operators.
Even more striking examples of the same phenomenon are observed for other operator sets below.
Here, it is noteworthy that multi-state fits to GEVP principal correlators show the same discrepancies between $1D$ and larger interpolator sets.
The close correspondence between (block) Lanczos and (GEVP principal correlator) multi-state fits suggests that the same finite-statistics effects causing stagnation in the Lanczos context lead to a saturation of fit accuracy when the maximum fit range $t_{\rm max}$ is increased at finite statistics.
This may simply reflect the fact that sufficiently noisy large-$t$ correlators do not provide useful information for any analysis scheme.

\begin{figure*}[t]
    \includegraphics[width=0.48\textwidth]{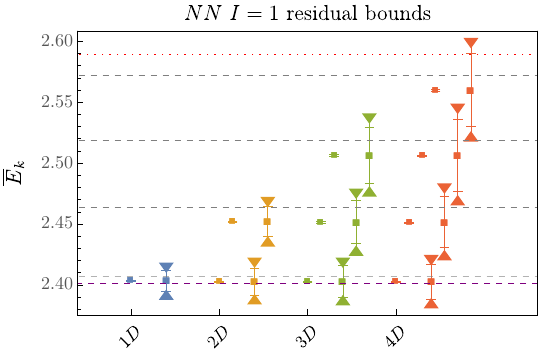}
    \includegraphics[width=0.48\textwidth]{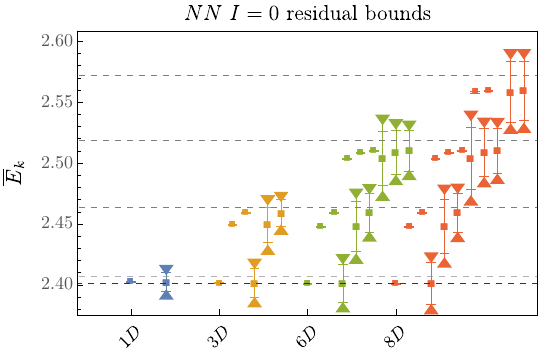}
    \caption{
        Block-Lanczos energy estimators (points with error bars) and residual-bound windows (triangle-bounded intervals) for the interpolating operator sets indicated in both $NN$ channels. Results for each operator set are grouped by color. Dashed gray and dotted red lines correspond to non-interacting $NN$ and $N\Delta$ levels, respectively.
        The dashed purple line shows the $k\cot\delta = 0$ threshold as in \cref{fig:nn-deut-DEMP}.
    }
    \label{fig:nn-bounds}
\end{figure*}

\begin{figure*}[t]
    \includegraphics[width=0.48\textwidth]{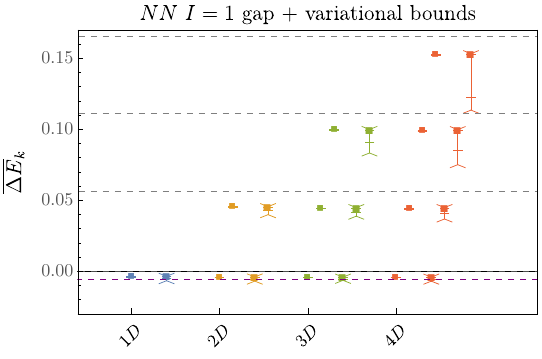} 
    \includegraphics[width=0.48\textwidth]{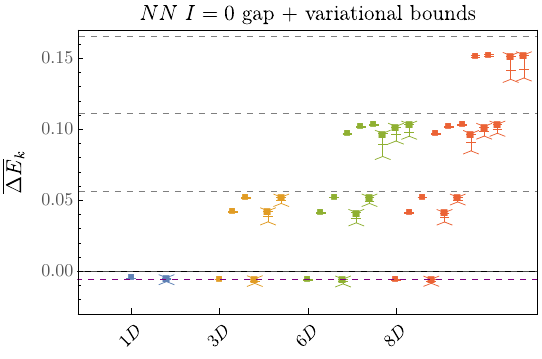}
    \caption{
        Block Lanczos energy-difference estimators  (points with error bars) and combined variational--Haas-Nakatsukasa gap-bound windows (chevron-bounded intervals) for the interpolating operator sets indicated in both $NN$ channels.
        Results for each operator set are grouped by color. Dashed horizontal lines correspond to non-interacting $NN$ levels.
        The dashed purple line shows the $k\cot\delta = 0$ threshold as in \cref{fig:nn-deut-DEMP}.
    \label{fig:nn-gap-bounds}}
\end{figure*}

\subsection{Two-sided bounds}\label{sec:qcd_bounds}

The violations of $N$-state-saturation assumptions seen in the LQCD examples above  motivate the study of two-sided residual and gap bounds for $NN$ systems.

Residual bounds provide the least precise constraints, but only require the assumption of Hermiticity.
At the level of statistics achieved here, residual bounds guarantee that $\nn$ and $\deut$ energy estimators $\overline{E}_0$ have located genuine energy levels with $\alt 1\%$ precision at $1\sigma$ confidence and $\alt 2\%$ precision at $5\sigma$ confidence. 
Clearly, these provide meaningful constraints on the energy spectra of two-nucleon systems that go beyond variational upper bounds.
However, ground-state 
energy shifts in units of the nucleon mass, $\overline{\Delta E}_0 / M_N$, are $O(10^{-3})$ for this volume, and residual bounds are not precise enough  to determine  whether interactions are attractive or repulsive; see \cref{fig:nn-bounds}.
At asymptotically large statistics $B_k^{(m)}$ must converge to a non-zero value, but the statistical precision of this study is still far from this regime.
The $1/\sqrt{N_{\rm cfgs}}$ scaling of both the central values and uncertainties of $\overline{B}_0^{(m)}$ in \cref{sec:diag} persists for larger sets of interpolating operators across the full range of statistics available here.
This indicates that residual-bound constraints could be improved in principle through increased statistics.

\begin{table*}[t]
    \centering
    \begin{ruledtabular}
    \begin{tabular}{ccccccc}
      $k$ & $\overline{\Delta E}_k$ & $\overline{E}_k$  & Residual (68\%)&  Residual (95\%)&  Gap + variational (68\%) &Gap + variational (95\%) \\\hline  
    0 & $-0.0045(2)_{(-25)}^{(+3)}$ 
    & $2.4025(4)_{(-25)}^{(+0)}$ 
      & $[2.39, 2.42]$ &  $[2.39, 2.42]$ 
       & $[2.400, 2.403]$ & $[2.398, 2.403]$ \\
    1 & $0.0442(3)_{(-74)}^{(+6)}$
    & $2.4511(5)_{(-74)}^{(+0)}$  
      & $[2.43, 2.47]$ & $[2.43, 2.48]$ 
      & $[2.444, 2.452]$ & $[2.437, 2.452]$ \\
    2 & $0.0991(4)_{(-241)}^{(+4)}$
    & $2.5060(6)_{(-241)}^{(+0)}$  
      & $[2.47, 2.54]$ & $[2.46, 2.55]$ 
       & $[2.482, 2.507]$ & $[2.455, 2.507]$ \\
    3 & $0.1524(5)_{(-389)}^{(+6)}$
    & $2.5594(7)_{(-393)}^{(+0)}$ 
      & $[2.52, 2.60]$ & $[2.50, 2.62]$ 
       & $[2.520, 2.560]$ & $[2.502, 2.561]$
    \end{tabular}
    \end{ruledtabular}
    \caption{Constraints on $\nn$ two-nucleon energies and energy differences. The second and third columns show energies and energy differences with 68\% statistical uncertainties followed by a ``systematic uncertainty'' that when added to the statistical uncertainty in quadrature produces the widths of combined variational and Haas-Nakatsukasa gap bounds at 68\% confidence. In the case of $\overline{\Delta E}_k$, correlated differences are formed using $NN$ and single-nucleon combined gap and variational bounds as described in the main text. The fourth (fifth) columns shows residual bounds at 68\% (95\%) confidence computed using the $k D$ interpolator set, which in all cases are tighter than bounds obtained from larger interpolator sets. The sixth (seventh) column shows gap bounds at 68\% (95\%) confidence. All gap bounds results are computed using the $4D$ interpolator set with the non-interacting $N\Delta$ energy used to estimate the gap parameter.
    }
    \label{tab:spectrum_summary_1}
\end{table*}

\begin{table*}[t]
    \centering
    \begin{ruledtabular}
    \begin{tabular}{ccccccc}
      $k$ & $\overline{\Delta E}_k$ & $\overline{E}_k$  & Residual (68\%)&  Residual (95\%)&  Gap + variational (68\%) &Gap + variational (95\%) \\\hline  
    0 & $-0.0060(2)_{(-25)}^{(+4)}$ 
    & $2.4009(4)_{(-25)}^{(+0)}$ 
      & $[2.39, 2.41]$ &  $[2.38, 2.42]$ 
      & $[2.398, 2.401]$ & $[2.396, 2.402]$ \\
    1 & $0.0407(3)_{(-54)}^{(+5)}$
    & $2.4477(5)_{(-55)}^{(+0)}$  
      & $[2.43, 2.47]$ & $[2.42, 2.48]$ 
      & $[2.442, 2.448]$ & $[2.437, 2.449]$ \\
    2 & $0.0518(2)_{(-29)}^{(+2)}$
    & $2.4588(4)_{(-30)}^{(+0)}$  
      & $[2.44, 2.47]$ & $[2.44, 2.48]$ 
      & $[2.456, 2.459]$ & $[2.454, 2.460]$ \\
    3 & $0.0962(4)_{(-92)}^{(+4)}$
    & $2.503(5)_{(-94)}^{(+0)}$ 
      & $[2.47, 2.54]$ & $[2.46, 2.55]$ 
      & $[2.494, 2.504]$ & $[2.488, 2.504]$ \\
    4 & $0.1012(4)_{(-50)}^{(+6)}$
    & $2.508(7)_{(-50)}^{(+0)}$ 
      & $[2.48, 2.53]$ & $[2.48, 2.54]$ 
      & $[2.503, 2.509]$ & $[2.501, 2.510]$ \\
    5 & $0.1013(4)_{(-47)}^{(+0)}$
    & $2.510(5)_{(-47)}^{(+0)}$ 
      & $[2.49, 2.53]$ & $[2.48, 2.54]$ 
      & $[2.505, 2.511]$ & $[2.503, 2.511]$ \\
    6 & $0.1511(6)_{(-109)}^{(+5)}$
    & $2.558(7)_{(-107)}^{(+0)}$ 
      & $[2.53, 2.59]$ & $[2.52, 2.60]$ 
      & $[2.547, 2.559]$ & $[2.540, 2.559]$ \\
    7 & $0.1519(5)_{(-102)}^{(+0)}$
    & $2.559(6)_{(-102)}^{(+0)}$ 
      & $[2.53, 2.59]$ & $[2.52, 2.60]$ 
      & $[2.549, 2.559]$ & $[2.543, 2.560]$ \\
    8 & $0.2026(9)_{(-485)}^{(+5)}$
    & $2.6096(10)_{(-488)}^{(+0)}$ 
      & $[2.55, 2.67]$ & $[2.54, 2.69]$ 
      & $[2.571, 2.611]$ & $[2.549, 2.612]$ \\
    9 & $0.2051(10)_{(-527)}^{(+10)}$
    & $2.6121(12)_{(-524)}^{(+0)}$ 
      & $[2.56, 2.67]$ & $[2.54, 2.69]$ 
      & $[2.560, 2.613]$ & $[2.542, 2.615]$
    \end{tabular}
    \end{ruledtabular}
    \caption{Constraints on $\deut$ two-nucleon energies.  Details are as in \cref{tab:spectrum_summary_1}, except that $10D$ is the largest interpolator set included and the non-interacting $\Delta\Delta$ energy is used to estimate the gap parameter. }
    \label{tab:spectrum_summary_0}
\end{table*}

At infinite-statistics, the inclusion of more interpolating operators leads to larger Krylov spaces where Prony-Ritz methods become strictly more accurate~\cite{Abbott:2025yhm}.
However, at finite statistics, adding additional operators that have small overlap with a given energy eigenstate can add noise to the distribution of $\overline{B}_k^{(m)}$ that more than compensates for the small decrease in the central value of $\overline{B}_k^{(m)}$ that must occur at infinite statistics.
Such behavior, which was already noted in $2\times 2$ nucleon correlator matrix applications~\cite{Hackett:2024nbe} and scalar field theory examples in \cref{sec:scalar-matrices}, is observed in both the $\nn$ and $\deut$ channels.
Starting with an interpolating operator that has large overlap with a particular state and then adding more operators leads to approximately monotonic growth in the uncertainty of $\overline{B}_k^{(m)}$, and corresponding growth in residual bound estimates at a given statistical confidence level; see for example Fig.~\ref{fig:nn-bounds}.
If the gap used to compute gap bounds is held fixed as in Davis-Kahan gap bounds, two-sided bound windows at a given level of statistical confidence will also grow linearly with the uncertainty of $\overline{B}_k^{(m)}$.
However, adding operators can also lead to larger gaps between, e.g., the $k=0$ Ritz value and the first energy level without a strongly overlapping Ritz vector, which increases the effective gap in Haas-Nakatsukasa gap bounds.

When considering energy differences, combined variational and gap bounds should be augmented by one further consideration beyond the discussion in \cref{sec:gap_var}---the nucleon ground-state energy estimator $\overline E_0^{N}$ provides a variational upper bound on $M_N$ but not a lower bound.
The nucleon ground-state gap bound does provide such a lower bound.
A genuine upper bound on $\Delta E_k = E_k - 2 M_N$ can be constructed using twice the gap lower bound on $M_N$.
In conjunction with the lower gap bound on $E_k$, this allows two-sided bounds on $\Delta E_k$ to be constructed under  no-missing-states assumptions for both the one- and two-nucleon sectors.

Combined variational and Haas-Nakatsukasa gap bounds on ground-state energy differences for various interpolating operator sets are shown alongside block Lanczos and GEVP energy estimator results in \cref{fig:nn-gap-bounds}.
The precision of these gap bounds is insensitive to the size of the interpolating operator set for relatively small set sizes, $r$.
In practice,  this insensitivity arises from competition between increases in the statistical uncertainty of $B_0^{(m)}$ and  increases in estimated gap parameters $\hat{G}_0^{(m)}$ as $r$ increases. 
Once $r$ is large enough that $N\Delta$ or $\Delta\Delta$ thresholds dictate the size of $\hat{G}_0^{(m)}$ and the gap parameter no longer increases with $r$, further increases in $r$ lead to less precise gap bound constraints.

The $1\sigma$ Haas-Nakatsukasa gap bounds on the energy for the $k=0$ state achieve relative uncertainty of $6\times 10^{-4}$ for $1D$, $2D$, and $3D$ interpolator sets in the $\nn$ channel.
Larger interpolator sets involve gap parameters set by the $N\Delta$ threshold and significantly larger gap-bound widths. 
These bounds are approximately an order of magnitude more precise than the corresponding residual bounds, although they are still a factor of 3--5 less precise than the corresponding statistical uncertainties on $\overline{\Delta E}_0^{(m)}$.
Results for the lowest gap bound in each $NN$ channel are reported in \cref{tab:results}.
This hierarchy appears to be fairly generic across low-energy levels in the $NN$ system, and reflects the relative strength of the assumptions (or lack thereof) built into residual bounds, gap bounds, and energy estimators.
Further results for energy estimators and bounds are tabulated in \cref{tab:spectrum_summary_1,tab:spectrum_summary_0}.

Critically, unlike statistical uncertainties on Lanczos estimators or BMA fits to GEVP-optimized correlators, gap bounds are conservative enough such that  $1D$ and larger interpolating operator sets are consistent at 1$\sigma$ in both channels. 
This consistency between all interpolator sets is found both for gap bounds on energies and also for combined variational and gap bounds on correlated energy differences.
In particular, consistency of gap bounds obtained from all interpolator sets holds despite the fact that $NN$ interactions shift LQCD energy levels slightly below their non-interacting counterparts and therefore cause $\hat{G}_0^{(m)}$ to slightly underestimate the true gap parameter $G_0^{(m)}$.

\begin{figure}
  \includegraphics[width=0.48\textwidth]{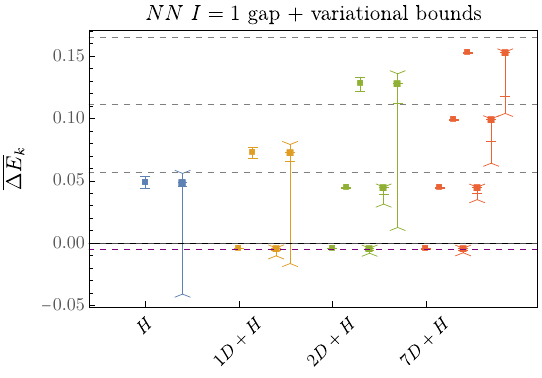}\\\vspace{10pt}
\includegraphics[width=0.48\textwidth]{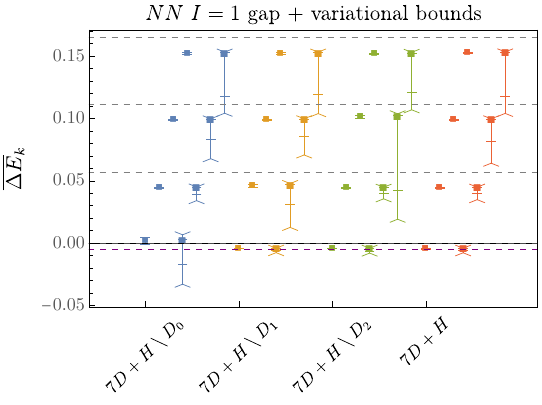}
    \caption{
        Block-Lanczos FV energy-difference estimators (points with error bars) and combined variational and Haas-Nakatsukasa gap-bound windows (chevron-bounded intervals) for various $\nn$ interpolating-operator sets containing a hexaquark operator are shown.
        Results for each operator set are grouped by color. Dashed horizontal lines correspond to non-interacting $NN$ levels.
        The dashed purple line shows the $k\cot\delta = 0$ threshold, as in \cref{fig:nn-deut-DEMP}.
        In both panels, the results from the full 7D+H operator set are shown for comparison.
    \label{fig:nn-H-bounds}
  }
\end{figure}

Analyses of interpolating operator sets that include hexaquark operators but not all of the available dibaryon operators provide even more striking examples in which $N$-state-saturation assumptions are violated but gap bounds lead to valid constraints.
\Cref{fig:nn-H-bounds} compares Lanczos energy estimator and gap bound results for interpolating operator sets with  a hexaquark operator and various combinations of dibaryon operators, which lead to significant inconsistencies if $N$-state saturation is assumed in all cases. 
For example, the set $1D+H$ leads to an energy estimator $\overline{E}_1$ that disagrees with all $7D+H$ energy estimators by at minimum a $6\sigma$ correlated difference of $0.026(4)$. 
The gap bounds associated with these discrepant energy levels, which predominantly overlap with the hexaquark operator, are significantly larger than those associated with other energy levels and are consistent within $1\sigma$ with energy estimators from larger interpolating operator sets.
For all operator sets in \cref{fig:nn-H-bounds}, the $1\sigma$ two-sided windows provided by combined variational and Haas-Nakatsukasa gap bounds are consistent with an energy-level estimate from the largest interpolating operator set.
This includes adversarially chosen interpolator sets analogous to the ones studied in Ref.~\cite{Amarasinghe:2021lqa} where individual dibaryon operators are omitted and $N$-state-saturation assumptions fail catastrophically.

\section{Phase-shift analysis}
\label{sec:phase-shift}

\begin{figure*}[t]
    \includegraphics[width=0.48\textwidth]{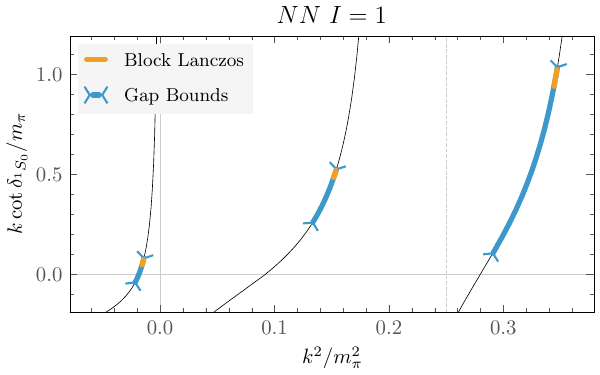}\qquad
    \includegraphics[width=0.48\textwidth]{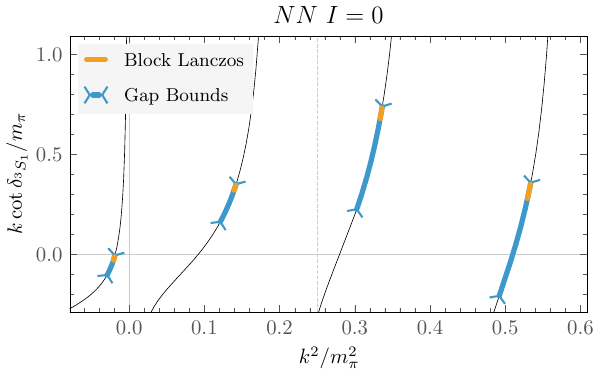}
    \caption{
      The $\nn$ (left) and $\deut$ (right) $s$-wave phase shifts, with orange bands representing the statistical uncertainties, and chevron-bounded blue bands represent the uncertainties from the gap bounds. The black lines are the L\"uscher zeta functions, and the dashed vertical line shows the position of the right-hand $t$-channel cut ($k^2/m^2_\pi = 1/4$).
      }
    \label{fig:phase-shifts}
\end{figure*}

Neglecting lattice artifacts, the  assumptions underlying gap bounds, namely that there are no additional unexpected eigenstates present in a particular range of the spectrum, are consistent with the assumptions underlying the use of FV quantization conditions \cite{Luscher:1986pf,Luscher:1990ux}, namely that there is a tower of energy levels close to threshold that asymptotically approach the threshold with an approximately known $1/L^2$ scaling, as well as energy levels that approach bound-state poles if any are present.
Under this assumption, two-nucleon phase shifts obtained using both the statistical uncertainties and gap bound windows discussed above are shown in \cref{fig:phase-shifts}, using the energies from the interpolator sets $3D$ for $\nn$ and $10D$ for $\deut$. 
Assuming that only the lowest partial wave contributes in the FV quantization condition for each irrep (${}^1S_0$ for $A^+_1$ in the $\nn$ channel, and ${}^3S_1$ for $T^+_1$ in the $\deut$ channel), the following relation is used for both isospin channels,
\begin{equation}
\label{eq:Luscher}
    k \cot\delta(k) = \frac{2}{\sqrt{\pi}L}\mathcal{Z}_{0,0}(1; \tilde{k}^2)\ ,
\end{equation}
with $\tilde{k}=kL/(2\pi)$ (where $k$ is the relative scattering momentum defined from the extracted energy $E=2\sqrt{M_N^2+k^2}$), and $\mathcal{Z}_{0,0}(1;\tilde{k}^2)$ being the zeta function~\cite{Luscher:1986pf,Luscher:1990ux}. 
The numerical values of the extracted phase shifts are shown in~\cref{tab:phase_shift}. Only the phase shifts corresponding to energies below the $N\Delta$ and $\Delta\Delta$ thresholds
are shown.\footnote{The $k=8$ state with $I=0$ is not shown because its gap bounds do not span over only a single branch of the zeta function.}
Residual bounds are not propagated through \cref{eq:Luscher} because they span multiple branches of the zeta functions and overlap with each other  and thus do not provide constraints on the phase shifts.
While gap-bound uncertainties on $\overline{E}_0$ are $\sim 6$ times larger than statistical uncertainties, one- and two-nucleon residual-norm-squares are less correlated than the corresponding energy estimators, resulting in gap-bound uncertainties on $\Delta \overline{E}_0$ that are $\sim 12$ times larger than statistical uncertainties.
However, the gap bound uncertainties are still sufficiently precise to meaningfully constrain phase shifts.

\begin{table}[h]
\renewcommand{\arraystretch}{1.4}
\centering
\begin{ruledtabular}
    \centering
    \begin{tabular}{l l}
    \multicolumn{2}{c}{$\nn$ channel, ${}^1S_0$} \\
    \multicolumn{1}{c}{$k^2/m^2_{\pi}$} & \multicolumn{1}{c}{$k \cot\delta/m_\pi$} \\ \hline 
    $-0.0151_{(-6)(-66)}^{(+6)(+10)}$	            &	$0.061_{(-13)(-99)}^{(+14)(+21)}$	\\
    $\phantom{-}0.153_{(-1)(-20)}^{(+1)(+1)}$	&	$0.50_{(-2)(-24)}^{(+2)(+3)}$	\\
    $\phantom{-}0.346_{(-1)(-55)}^{(+1)(+2)}$	&	$0.98_{(-4)(-88)}^{(+4)(+5)}$	\\ \hline \hline
    \multicolumn{2}{c}{$\deut$ channel, ${}^3S_1$} \\
    \multicolumn{1}{c}{$k^2/m^2_{\pi}$} & \multicolumn{1}{c}{$k \cot\delta/m_\pi$} \\ \hline 
    $-0.0202_{(-6)(-89)}^{(+8)(+12)}$	            &	$-0.021_{(-8)(-81)}^{(+10)(+16)}$	\\
    $\phantom{-}0.141_{(-1)(-19)}^{(+1)(+2)}$	&	$\phantom{-}0.33_{(-1)(-17)}^{(+1)(+2)}$	\\
    $\phantom{-}0.335_{(-1)(-32)}^{(+1)(+2)}$	&	$\phantom{-}0.71_{(-3)(-48)}^{(+3)(+3)}$	\\
    $\phantom{-}0.531_{(-2)(-39)}^{(+2)(+3)}$	&	$\phantom{-}0.31_{(-3)(-52)}^{(+4)(+5)}$	\\
    \end{tabular}
    \end{ruledtabular}
    \caption{Relative momentum $k^2$ and $s$-wave phase shifts $k\cot\delta$ in units of $m_\pi$, for $\nn$ and $\deut$ systems. The first uncertainties are statistical. When added to them in quadrature, the second uncertainties produce the widths of combined variational and gap bounds at 68\% confidence. Results are shown for energies in \cref{tab:spectrum_summary_1,tab:spectrum_summary_0} that are below the right-hand $t$-channel cut and predominantly overlap with $s$-wave interpolating operators.
    }
    \label{tab:phase_shift}
\end{table}

In principle, these phase shifts can be fitted to an effective range expansion (below the right-hand $t$-channel cut) to extract the scattering parameters.
However, due to the small number of points available, as well as the severe truncation on the quantization conditions, the systematic uncertainties associated with that fit would be large and difficult to quantify.
If such a fit is nonetheless attempted, the scattering length is found to be consistent with zero at $1\sigma$ in the $I=0$ and $2\sigma$ in the $I=1$ channel using block Lanczos statistical uncertainties under an $N$-state-saturation assumption. The scattering lengths are also consistent with zero at $1\sigma$ in both channels using gap bounds with the no-missing-states assumption. Consequently, the sign of the scattering length is ambiguous and the presence of a bound state cannot be conclusively ruled in or out in either channel.

On the other hand, spectroscopy results provide precise constraints on the magnitudes of the $NN$ scattering lengths (estimated from the $k\cot\delta$ value of the ground-state energy) if discretization effects and residual FV effects are neglected.
Gap bounds constrain $|a^{-1}_{^{1}S_0/^{3}S_1}|\alt m_\pi/10$ at $68\%$ confidence in both channels under no-missing-states assumptions.
These bounds indicate that both the $\nn$  and $\deut$ NN systems are fine-tuned \cite{NPLQCD:2020lxg}.
Under $N$-state-saturation assumptions,  which are seen above to be invalid for at least some interpolator sets, block Lanczos statistical uncertainties provide a similar constraint in the $\nn$ channel and a three-times tighter constraint in the $\deut$ channel.

\section{Asymmetric correlation function analysis}\label{sec:asymm}

As discussed in the introduction, asymmetric correlation functions involving a hexaquark operator at the source and various dibaryon operators at the sink have been used in previous studies of $NN$ systems at various unphysical values of the quark masses~\cite{Beane:2009py,Yamazaki:2012hi,NPLQCD:2011naw,NPLQCD:2012mex,Yamazaki:2015asa,Berkowitz:2015eaa} and led to the conclusions that there were bound states in these systems.
The oblique Lanczos algorithm can also be applied directly to asymmetric correlators, rather than to the matrices containing them. Consequently, it can be used to test the previous results in a context where rigorous bounds are available.
As will be discussed in \cref{sec:scalar-asymmetric}, this requires modifications to Lanczos analysis at both infinite and finite statistics.

\subsection{Oblique Lanczos methods}\label{sec:scalar-asymmetric}

Lanczos analysis of asymmetric correlators requires some modifications relative to the symmetric case because $L$ and $R$ quantities are distinct, even at infinite statistics.
The required formalism is introduced below, with $C_{LL}(t) = \left< \mathcal{O}_L(t) \mathcal{O}_L^\dagger(0) \right>$ and $C_{RR}(t) = \left< \mathcal{O}_R(t) \mathcal{O}_R^\dagger(0) \right>$ denoting the symmetric correlators constructed from the same interpolating operators appearing in a generic asymmetric correlator $C_{LR}(t)= \left< \mathcal{O}_L(t) \mathcal{O}_R^\dagger(0) \right>$.
Notably, calculation of residual bounds for Ritz values obtained from analysis of $C_{LR}(t)$ requires calculation of either or both of $C_{LL}(t)$ and $C_{RR}(t)$.
Gap bounds do not apply to asymmetric correlator Ritz values, as already noted in \cref{sec:gap} and discussed further below.

In an asymmetric correlator analysis, the residual-norm-squares $B^{R(m)}_k$ and $B^{L(m)}_k$, as defined in \cref{eq:Bdef}, are distinct quantities whose values may differ significantly.
Computing $B^{R(m)}_k$($B^{L(m)}_k$) requires data from the corresponding symmetric correlator $C_{RR}$ ($C_{LL}(t)$); see \cref{eq:Bexplicit} and  Ref.~\cite{Abbott:2025yhm}. 
A unique value for $B^{(m)}_k$ can be obtained by setting it to whichever of $B^{R(m)}_k$ and $B^{L(m)}_k$ has the smaller absolute value.

Hermitian-subspace and ZCW filtering were designed for the symmetric correlator (matrix) case and require some modification and reinterpretation in the asymmetric case.
For real-valued correlators, some subset of Ritz values $\lambda^{(m)}_k$ and  amplitudes $Z^{R(m)*}_k Z^{L(m)}_k$ will be real (but not necessarily positive) to working precision.
Since $T$ has real eigenvalues, it is reasonable to 
filter 
to this subset of real Ritz values.
However, because $\ket{y^{R(m)}_k} \neq \ket{y^{L(m)}_k}$ in general, asymmetric correlators do not admit an exact Hermitian subspace.

The same numerical algorithm for the ZCW test can be applied as in the symmetric case, i.e., removing states with small overlap products:
\begin{equation}
    \left| \frac{Z^{R(m)*}_k Z^{L(m)}_k}{C_{LR}(2m_0)} \right| 
    = \left| \frac{\braket{\psi^L | y^{R(m)}_k} \braket{y^{L(m)}_k | \psi^R}}{\braket{\psi^L | \psi^R}} \right|
    \leq \varepsilon_\mathrm{ZCW} ~ .
    \label{eq:scalar-ZCW}
\end{equation}
With $\ket{y^{R(m)}_k} \neq \ket{y^{L(m)}_k}$, this condition cannot be interpreted as a constraint on Hilbert-space states in the same way as the symmetric case.
In particular, the constraints that the $L$ and $R$ Ritz vectors are individually non-spurious would involve constraints on $|Z_k^L|^2$ and $|Z_k^R|^2$.
However, since \cref{eq:scalar-ZCW} involves the product of overlaps $[Z_k^R]^* Z_k^L$ that explicitly enters the Prony-Ritz decomposition for an asymmetric correlator, \cref{eq:prony_ritz}, the same logic underlying the ZCW test applies in the asymmetric case. These contributions to the Prony-Ritz decomposition smaller than a threshold set by statistical precision are more likely to be noise artifacts than genuine signals of states making very small correlator contributions.

The SLRVL scheme introduced above would also require generalization to treat the asymmetric case. 
For simplicity, filter-and-sort state identification is used in the analyses of this section as the focus is on ground state extractions for scalar correlators, for which filter-and-sort is reliable.

Further differences between vectors obtained from symmetric and asymmetric correlator analyses can be studied by computing the angles between $L$ and $R$ Krylov and Ritz vectors.
For asymmetric correlators, right and left Krylov vectors are distinct even 
at infinite statistics; that is $\ket{k^R_t} \neq \ket{k^L_t}$ for all finite $t$.
The similarity of the $L$ and $R$ Krylov vectors can be examined by computing the angle between them as a function of $t$,
\begin{equation} 
\label{eq:krylov-angle}
\begin{split}
  \cos(\theta_{LR}^{k}(t)) = \frac{ \braket{ k_{m}^L | k_{m}^R } }{ \left|\ket{k_{m}^L}\right| \ \left|\ket{k_{m}^R}\right| 
  } = \frac{C_{LR}(t)}{ \sqrt{ C_{LL}(t) C_{RR}(t) }},
  \end{split}
\end{equation}
where $m = t/2-m_0$.
For a symmetric correlator, $\cos(\theta_{LR}^{k}(t)) = 1$ holds by construction.
For an asymmetric correlator, this quantity differs from unity for finite $t$ to the extent that excited-state effects differ between source and sink interpolating operators.

Ritz vectors are constructed from superpositions of Krylov vectors as  $\ket{y^{R(m)}_k} = \sum_t \ket{k^{R}_t} P^{R(m)}_{tk}$ and $\bra{y^{L(m)}_k} = \sum_t P^{L(m)}_{kt} \bra{k^{L}_t}$.
Similar to \cref{eq:krylov-angle} above, the angle between the $k$th left- and right-Ritz vectors can therefore be computed as
\begin{equation}\label{eq:ritz-angle}
  \begin{split}
    &\cos \theta^{y_k(m)}_{LR} 
    \equiv \frac{\braket{y^{L(m)}_k | y^{R(m)}_k}}{\left|\ket{y^{L(m)}_k}\right|\  \left|\ket{y^{R(m)}_k}\right |}
    \\
    &\qquad = \left[
    \sum_{st} P^{L(m)}_{ks} C_{LL}(s+t+2m_0) P^{L(m)*}_{kt}
    \right.
    \\ &\qquad\qquad  \times
    \left.
    \sum_{st} P^{R(m)*}_{sk} C_{RR}(s+t+2m_0) P^{R(m)}_{tk}
    \right]^{-1/2},
\end{split}
\end{equation}
where the oblique Lanczos bi-orthonormality condition $\braket{y^{L(m)}_k | y^{R(m)}_k} = 1$ has been used to simplify the second line.
The ground-state $LR$ Ritz angle appears in discussions below and is denoted $\theta^{y(m)}_{LR} \equiv \theta^{y_0(m)}_{LR} $.
For $m=1$, it coincides with the $LR$ Krylov angle, $\theta^{y(1)}_{LR} = \theta_{LR}^{k}(1+2m_0)$.

The ground-state $LR$ Ritz angle provides a diagnostic for whether Ritz values can be interpreted as (approximately) symmetric matrix elements of $T$ or whether they must be interpreted as off-diagonal transition matrix elements.
When $\cos \theta^{y(m)}_{LR} \neq 1$, the $R/L$ states are distinct and therefore, because $\braket{y_0^{L(m)} | y_0^{R(m)}} = 1$ by construction\footnote{Bi-orthonormality holds for $R/L$ eigenvectors defined as they appear in the eigendecomposition and thus for the Ritz vectors as they appear in the correlator spectral decomposition.}, $\ket{y^{R(m)}_0}$ and $\ket{y^{L(m)}_0}$ cannot be simultaneously unit normalized.
A transition matrix element of $T$ defined between unit-normalized $R/L$ Ritz vectors 
differs from the Ritz value by this Ritz angle factor,
\begin{equation}\begin{aligned}
  \frac{ \braket{ y_0^{L(m)} | T | y_0^{R(m)} } }{ \left| \ket{y_0^{L(m)}} \right| ~ \left| \ket{y_0^{R(m)}}\right| } 
  &= \frac{ \braket{ y_0^{L(m)} | T | y_0^{R(m)} } }{\braket{ y_0^{L(m)} | y_0^{R(m)}}  } \cos \theta^{y(m)}_{LR}
  \\
  &= \lambda^{(m)}_0 \cos \theta^{y(m)}_{LR}.
\end{aligned}\end{equation}
This normalization difference is not important for spectroscopy, which is concerned with the estimation of $T$ eigenvalues.
However, this difference means that estimators for matrix elements of generic operators constructed from Ritz coefficients computed from asymmetric two-point correlators without explicitly accounting for Ritz-vector normalization are incorrect.
Both $L$ and $R$ symmetric correlators are required to obtain correct values for matrix elements in asymmetric correlator Lanczos analyses unless $\cos \theta^{y_k(m)}_{LR}=1$ for all states $k$ involved.
Analogous state normalization ambiguities are also an issue for standard matrix element analyses.

A different $R/L$ ambiguity that is directly relevant to spectroscopy is whether to consider energy estimators derived from Ritz values, or instead from \emph{symmetric Rayleigh quotients} (SRQs) constructed from diagonal matrix elements in left- or right-Ritz states, 
\begin{equation}
   \eta_k^{R/L(m)} \equiv \frac{ \braket{y_k^{R/L(m)} | T | y_k^{R/L(m)} } }{ \braket{y_k^{R/L(m)} |  y_k^{R/L(m)} } }.
\end{equation}
The $\eta_k^{R/L(m)}$ both coincide with $\lambda_k^{(m)}$ in symmetric Lanczos applications, but for oblique Lanczos they lead to distinct energy estimators when $\cos \theta^{y_k(m)}_{LR} \neq 1$.
As with the $B^{R/L(m)}_k$, these SRQs can be computed using the corresponding diagonal correlators and Ritz coefficients from the asymmetric analysis as
\begin{equation}\begin{aligned}
  \eta_k^{R(m)} 
  &= \frac{ \sum_{t,t'} [P_{tk}^{R(m)}]^* C_{RR}(t+t'+1) P_{t'k}^{R(m)} }{ \sum_{t,t'} [P_{tk}^{R(m)}]^* C_{RR}(t+t') P_{t'k}^{R(m)} } ~ ,
  \\
  \eta_k^{L(m)} 
  &= \frac{ \sum_{t,t'} P_{kt}^{L(m)} C_{LL}(t+t'+1) [P_{kt'}^{L(m)}]^*}  { \sum_{t,t'} P_{kt}^{L(m)} C_{LL}(t+t') [P_{kt'}^{L(m)}]^* }
  ~.
\end{aligned}\end{equation}
The SRQs are estimators for $T$ eigenvalues that may be used to construct energy estimators,
\begin{equation}
    E^{R/L(m)}_k = - \ln \eta_k^{R/L(m)} ~ ,
\end{equation}
just as with Ritz values.

The optimality properties of Ritz values in symmetric Lanczos follow from the fact that they are SRQs of Hermitian operators~\cite{Abbott:2025yhm}.
Ritz values of asymmetric correlators are not SRQs of $T$ and therefore do not provide ``optimal'' approximations to transfer-matrix eigenvalues in the Rayleigh-Ritz (or maxmin~\cite{Abbott:2025yhm}) sense.
SRQs constructed from asymmetric correlator Ritz vectors are not optimal in the same sense as symmetric Ritz values, but still provide variational bounds on true energy eigenvalues.
They can further be used to construct gap bounds, which specifically constrain the position of true eigenvalues relative to SRQs as
\begin{equation}
   \min_{\lambda \in \{\lambda_n\}} \left| \eta^{R/L(m)}_k - \lambda \right| \leq  \frac{ B^{R/L(m)}_k   }{ g_{k}^{R/L(m)} }.
    \label{eq:asymm_gap_bound}
\end{equation}
where the gap parameter $g_k^{R/L(m)}$ may be defined analogously to \cref{eq:gkdef},
\begin{equation}\label{eq:asymm_gkdef}
   g_k^{R/L(m)} \equiv \min_{\lambda_{n'} \in \{\lambda_n \neq \lambda\}} \left|  \lambda_{n'} - \eta_k^{R/L(m)} \right|,
\end{equation}
Note that since $\eta_k^{R(m)} \neq \eta_k^{L(m)} \neq \lambda_k^{(m)}$ for asymmetric correlators, there are distinct gap bounds associated with $E_k^{R(m)}$ and $E_k^{L(m)}$, and neither is centered around $E_k^{(m)}$.

\subsection{Complex harmonic oscillator}

\begin{figure}[]
    \includegraphics[width=0.48\textwidth]{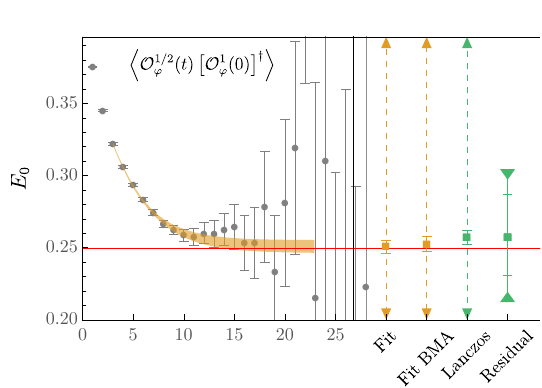}
    \includegraphics[width=0.48\textwidth]{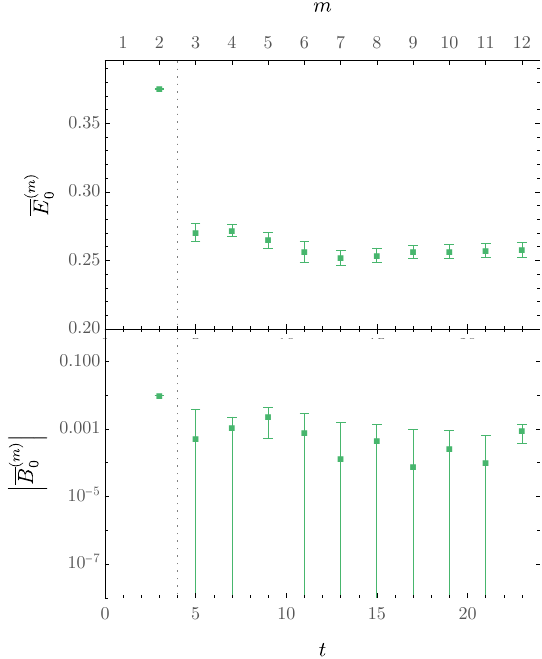}
    \caption{
      Upper: Effective masses, energy estimators, and two-sided bounds analogous to  \cref{fig:complex-scalar} for the scalar field theory asymmetric correlator. Arrows emphasize that statistical uncertainties provide ``zero-sided bounds'' because they can be shifted in either direction by uncontrolled excited-state contamination. 
      Lower: Iteration dependence of Lanczos energy-estimators and residual-norm-square estimators for the asymmetric correlator.
    }
    \label{fig:complex-scalar-od}
\end{figure}

For comparisons to LQCD results below, it is interesting to consider the asymmetric correlation function appearing in the off-diagonal entry, $\left< \mathcal{O}_{\varphi}^1(t) [\mathcal{O}_{\varphi}^{1/2}(0)]^\dagger \right>$, of the complex harmonic oscillator correlator matrix discussed in \cref{sec:scalar-matrices}.
The effective mass for this asymmetric correlator is shown in \cref{fig:complex-scalar-od}, along with multi-state fit results, Lanczos energy estimates (obtained using the asymmetric filtering scheme discussed above and filter-and-sort state identification), and residual bounds.
Statistical uncertainties in this asymmetric correlator are typically larger than those arising from the symmetric correlator built from $\mathcal{O}_{\varphi}^{1/2}$ but smaller than those arising from that constructed from $\mathcal{O}_{\varphi}^1$.
Residual-bound widths for the asymmetric case are also intermediate between those of the $\mathcal{O}_{\varphi}^{1/2}$ and $\mathcal{O}_{\varphi}^1$ symmetric correlators.
For comparison with LQCD results below, note that $\overline{B}_0^{(m)}$ is consistent with zero at $1\sigma$ for almost all iterations with $m \gtrsim 2$, which qualitatively resembles the symmetric correlator results above.

\begin{figure}[]
    \includegraphics[width=0.48\textwidth]{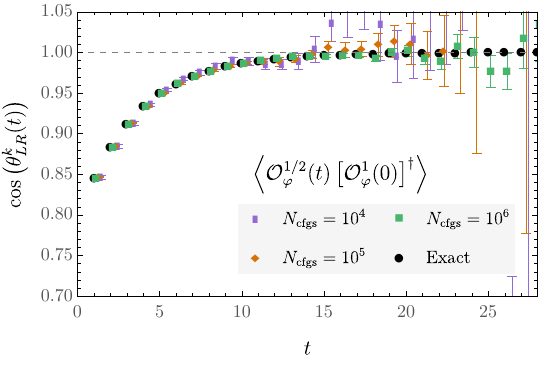}
    \includegraphics[width=0.48\textwidth]{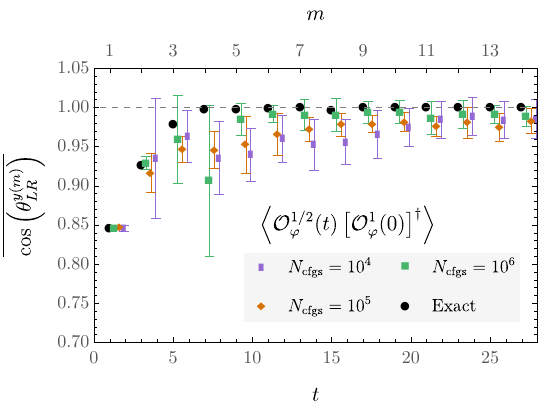}
    \caption{
      Angles between left- and right-Krylov vectors (top) and left- and right-Ritz vectors (bottom) in the scalar field theory case.
    }
    \label{fig:complex-scalar-od-angle}
\end{figure}

Results for the Krylov and Ritz angles (Eqs. \eqref{eq:krylov-angle} and \eqref{eq:ritz-angle}, respectively) for the $\mathcal{O}_{\varphi}^{1/2} [\mathcal{O}_{\varphi}^1]^\dagger $ correlator are shown in \cref{fig:complex-scalar-od-angle}.
For $t \gtrsim 15$, $\cos(\theta_{LR}^{k}(t))$ is consistent with unity within $1\sigma$ uncertainties for finite-statistics results, and within one part in $10^4$ for exact results.
For $m \geq 2$, exact results for $\cos \theta^{y(m)}_{LR}$ approach one more rapidly than $\cos(\theta_{LR}^{k}(2m-1+2m_0))$.
This means that, as expected from the improved convergence of Lanczos over power iteration, the $R/L$ Ritz vectors become parallel more rapidly than the $R/L$ Krylov vectors. 
However, at finite statistics, this relationship is less clear.

In this scalar field theory model,  $\cos(\theta^k_{LR}(t)) \approx 1$ within statistical uncertainties for $t \gtrsim 15$ suggests that the normalization ambiguities discussed in \cref{sec:scalar-asymmetric} are not a practical concern.
Note, however, that $\cos(\theta^k_{LR}(t)) \approx 1$ only holds in this example for $t \delta_1 \gtrsim 4$.

\subsection{Residual bounds}

\begin{figure*}
    \includegraphics[width=0.48\textwidth]{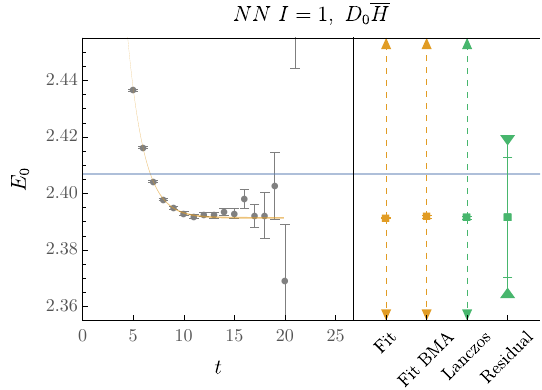}\quad
    \includegraphics[width=0.48\textwidth]{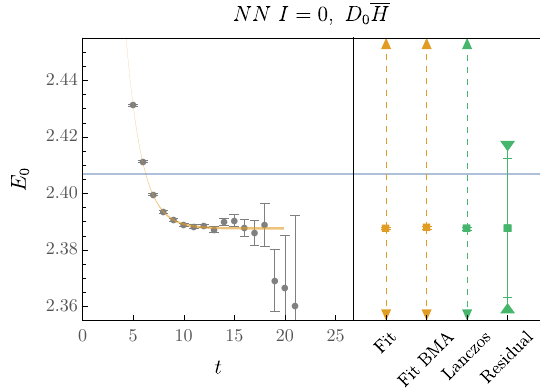}\\\vspace{15pt} 
    \includegraphics[width=0.48\textwidth]{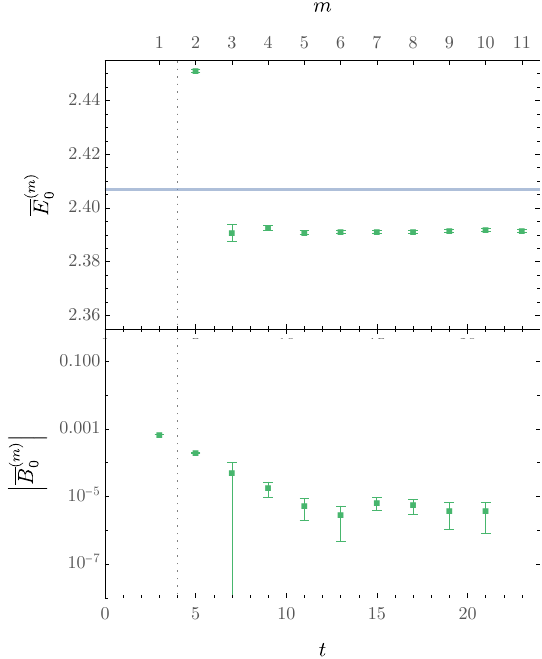}\quad
    \includegraphics[width=0.48\textwidth]{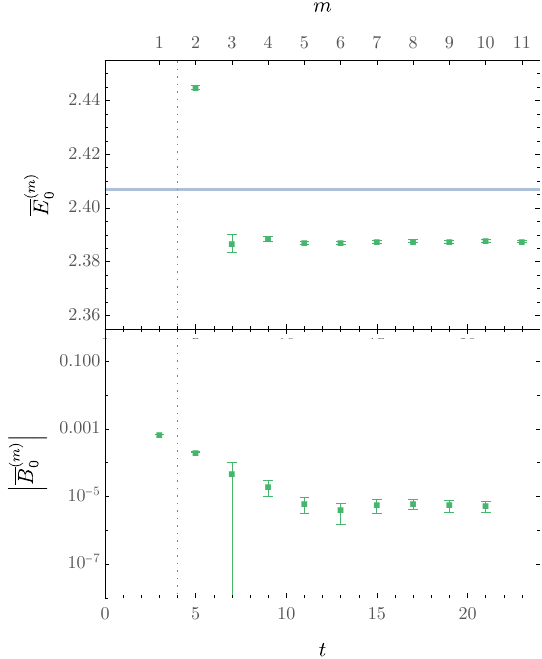}
    \caption{
      Upper: Effective masses, multi-state fit results, Lanczos energy estimators, and two-sided bounds using $\left< D_0 \overline{H}\right>$ correlators for the $\nn$ (left) and $\deut$ (right) channels. Lower: Iteration dependence of Lanczos energy estimators and residual-norm-square estimators. Details are analogous to \cref{fig:nn-Lanczos-EMP}.
    }
    \label{fig:asym:od-results}
\end{figure*}

\begin{figure*}
    \includegraphics[width=0.48\textwidth]{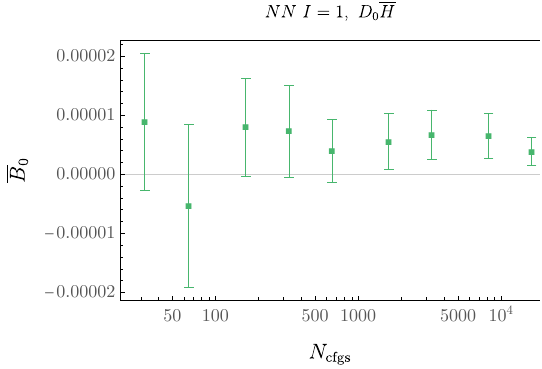}\qquad
    \includegraphics[width=0.48\textwidth]{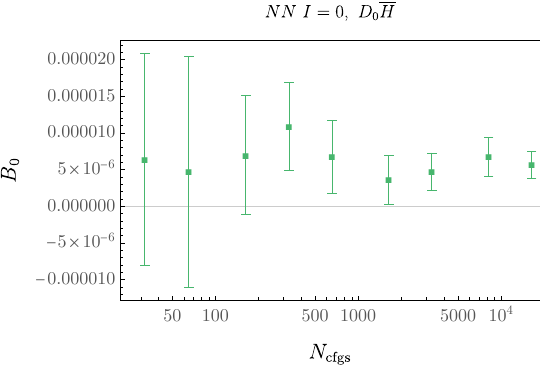}\qquad
    \caption{
      Convergence of Lanczos ground-state residual-norm-square estimators from $D_0\overline{H}$ correlator results, constructed from a median of iterations $m \in \{8,9,10\}$, as a function of $N_{\rm cfgs}$. 
    }
    \label{fig:nn-DH-stats}
\end{figure*}

\begin{figure*}
    \includegraphics[width=0.48\textwidth]{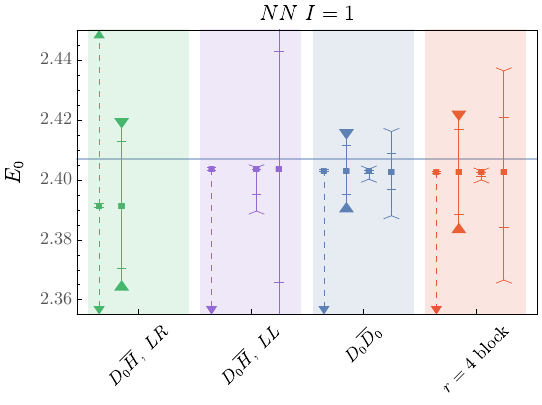} 
    \quad
    \includegraphics[width=0.48\textwidth]{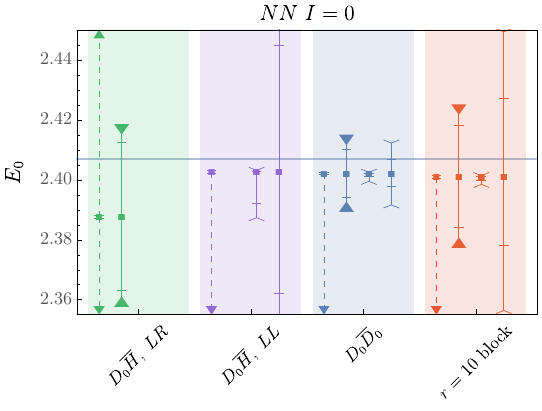} 
    \caption{
        Lanczos $\nn$ (left) and $\deut$ (right) ground-state energy  estimates and corresponding two-sided bounds on true energies. 
        In each panel, the green region indicates bounds on the asymmetric Ritz value $\braket{y^L_0 | T | y^R_0}$ derived from $D_0\overline{H}$ asymmetric correlators.
        The purple region indicates bounds on the $LL$ symmetric Rayleigh quotient $\braket{y^L_0|T|y^L_0}/\braket{y^L_0|y^L_0}$ derived using $D_0\overline{H}$ and  the $D_0 \overline{D_0}$ symmetric correlator.
        The blue and red regions indicate results from symmetric analyses (scalar and block, respectively). 
        Solid lines with inwards-facing arrows indicate residual bounds, while lines with inwards-facing chevrons indicate gap bounds for two scenarios---Haas-Nakatsukasa gap bounds computed as in \cref{sec:mind_the_gap} (left-most gap-bound in each set) and Davis-Kahan gap bounds computed assuming the central value of the ${D_0\overline{H}}$ Ritz value has converged to a genuine energy level (right-most gap-bound in each set).
        Dashed lines with arrows extending to the bottom of the plot range indicate a variational bound extending to $-\infty$; dashed lines extending over the entire plot range indicate that asymmetric Ritz values are not bounds.
        The pale horizontal line indicates the combined variational and gap bound on the non-interacting threshold $2M_N$.
    }
    \label{fig:asym:nn-Lanczos-bounds-comp}
\end{figure*}

Returning to LQCD, \Cref{fig:asym:od-results} shows the effective mass of the $D_0\overline{H}$ zero-momentum dibaryon--hexaquark correlation functions in the $\nn$ and $\deut$ channels, along with fits and Lanczos energy-estimators.
As already noted, residual-norm-squares can be computed for asymmetric LQCD correlators using the corresponding symmetric correlator data; the resulting $B$ values are as presented in \cref{fig:asym:od-results} and their scaling with $N_{\rm cfgs}$ is shown in \cref{fig:nn-DH-stats}.
These can be used to construct residual-bound windows for asymmetric Lanczos energy-estimators.

Residual-bound results are shown in \cref{fig:asym:nn-Lanczos-bounds-comp} and compared with residual bounds computed from the various symmetric analyses of \cref{sec:lqcd-lanczos} (results from other bounds that are also shown are discussed below).
The residual bounds computed from the asymmetric $D_0\overline{H}$ correlator require one of the corresponding diagonal correlators as input. 
Residual bounds computed using the $D_0 \overline{D}_0$ correlator as additional input are more precise than those involving the $H\overline{H}$ correlator, but they are still less precise than the residual bounds for the $D_0 \overline{D}_0$ Ritz value.
The $D_0\overline{H}$ residual-bound windows are broad enough to encompass the $D_0\overline{D}_0$ energy estimators in both channels. Importantly, this means that residual bounds do not provide evidence for a second sub-threshold energy level in either channel.

As shown in \cref{fig:asym:od-results}, analysis of the residual-norm-squares provides statistical evidence for non-zero $\overline{B}_0^{(m)}$ at large $m$ for $D_0 \overline{H}$ correlators.
Further, there is no significant decrease in $\overline{B}_0^{(m)}$ for $m \gtrsim 5$.
Scaling with $N_{\rm cfgs}$ shown in \cref{fig:nn-DH-stats} further suggests that $\overline{B}_0^{(m)}$ is converging towards a positive non-zero value.
Non-zero residual-norm-squares indicate that the accuracy of energy estimates is limited by finite-$m$ truncation effects rather than by noise. Similarly, lack of reduction of $B$ as $m$ increases points towards stagnation.
Taken together, this suggests that asymmetric correlator results are exhibiting the emergence of a finite-$m$ stagnation effect that would be present even at infinite statistics, rather than stagnation simply due to finite statistics stalling the convergence of the Lanczos process.\footnote{One may wonder whether the apparent lack of $\overline{B}_0$ converging to zero arises from an issue with the oblique Lanczos formalism and is common to all asymmetric correlator analyses. Comparison with \cref{fig:complex-scalar-od} shows that it is not.}

\subsection{$L$ and $R$ Krylov- and Ritz-vector angles}

\begin{figure*}
  \includegraphics[width=0.48\textwidth]{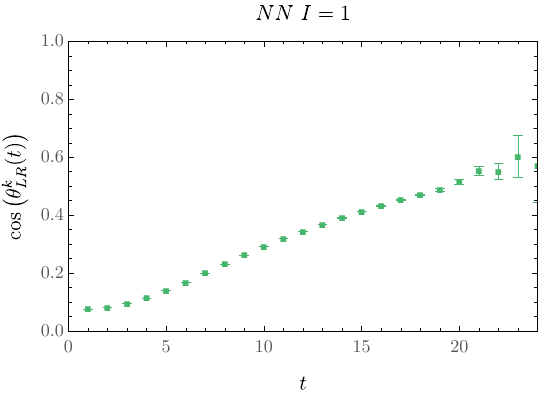}\qquad
    \includegraphics[width=0.48\textwidth]{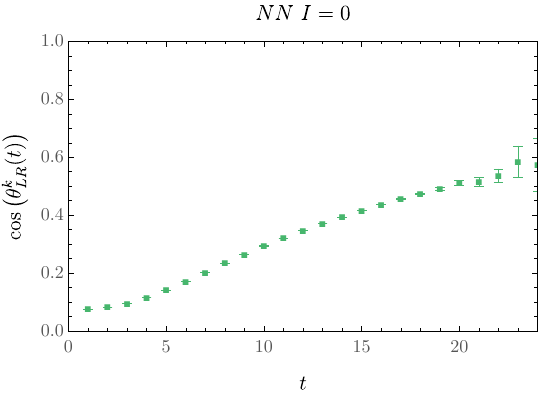}
    \caption{
        Cosine of the angle between $L$ and $R$ Krylov vectors corresponding to the $D_0\overline H$ correlator in the $\nn$ and $\deut$ channels, as defined in \cref{eq:krylov-angle}.
    }
    \label{fig:asym:krylov-angles}
\end{figure*}

\begin{figure*}
  \includegraphics[width=0.48\textwidth]{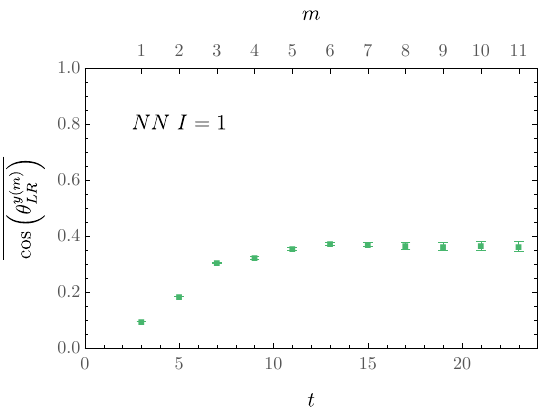}\qquad
    \includegraphics[width=0.48\textwidth]{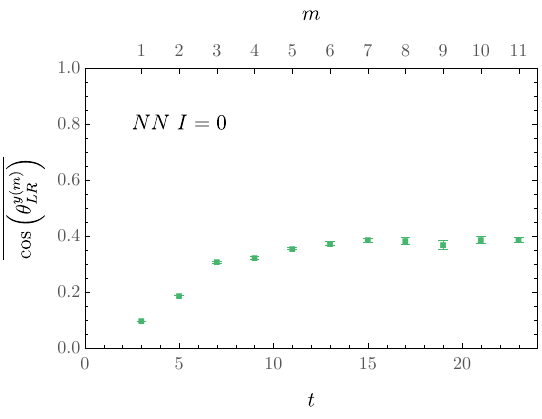}
    \caption{
        Cosine of the angle between $L$ and $R$ Ritz vectors corresponding to the $D_0\overline H$ correlator in the $\nn$ and $\deut$ channels, as defined in \cref{eq:ritz-angle}.
    }
    \label{fig:asym:ritz-angles}
\end{figure*}

Further insight into the ground-state Ritz vectors produced by oblique Lanczos for asymmetric correlators is provided by the Ritz and Krylov $LR$ overlap angles, $\theta_{LR}^k(t)$ and $\theta^{y(m)}_{LR}$, as introduced in \cref{sec:scalar-asymmetric}.
Estimates for the $\nn$ and $\deut$ channel Krylov angles are shown in \cref{fig:asym:krylov-angles}.
Notably, these angles can be resolved from zero with high statistical precision.
The results indicate that, in practice, the $L$ and $R$ Krylov vectors are distinct for all available $t$.
Moreover, $\cos \theta_{LR}^k(t)$ grows steadily with $t$ and does not reach a plateau near unity over the full range of Euclidean times available.
This change indicates that the overlaps of the ground-state left- and right-eigenvector approximations (i.e., the Krylov vectors) continue to evolve significantly over the full range of Euclidean time.
In this situation, even if the effective mass or other energy estimators have plateaued to a stable value for an asymmetric correlator, it is not because a good approximation to the ground-state energy eigenvector has been constructed.
Rather, two different approximations have conspired to create an apparently stable $T$ transition-matrix element result. 

The apparent stability of the energy and residual-norm-square estimators in \cref{fig:asym:od-results}, combined with the behavior of $\theta_{LR}^k(t)$, is in fact signaling the stagnation of oblique Lanczos.
This can be diagnosed by examining the ground-state Ritz-vector angles $\cos \theta^{y(m)}_{LR}$ defined in \cref{eq:ritz-angle}.
As shown in \cref{fig:asym:ritz-angles}, with increasing $m$, $\cos \theta^{y(m)}_{LR}$ asymptote to $\sim 0.35$ in both channels.
This behavior is qualitatively different from that seen in the scalar field theory example.
This indicates that the $R/L$ Ritz vectors stabilize to \emph{distinct} states, rather than each converging to a common ground state.
The individual identities of $\ket{y^{R/L(m)}_0}$ can be inferred from the analysis of symmetric Rayleigh quotients discussed below. 

The observation that the $R$ and $L$ Ritz vectors are distinct
has several immediate consequences.
First, the Ritz values are {transition matrix elements} of $T$, rather than symmetric estimators of a $T$ eigenvalue as would be obtained if $\cos \theta^{y(m)}_{LR} \approx 1$.
Second, it follows that, as with asymmetric effective masses, the asymmetric Ritz values are not variational bounds---this would require $\cos \theta^{y(m)}_{LR} \approx 1$. 

As discussed in \cref{sec:adversary} below, $\cos \theta^{y(m)}_{LR}\slashed{\approx} 1$ does not forbid the possibility that asymmetric correlators provide accurately converged energy estimators. 
However, it does require that this convergence arises from the orthogonality of $R$ and $L$ excited states, rather than from the convergence of the $R$ and $L$ Ritz vectors to a common ground state. 

\subsection{Symmetric Rayleigh quotients}\label{sec:srq}

\begin{figure*}[t]
    \includegraphics[width=0.48\textwidth]{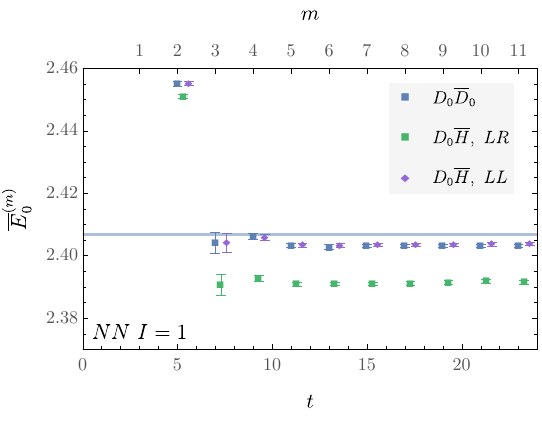}
    \includegraphics[width=0.48\textwidth]{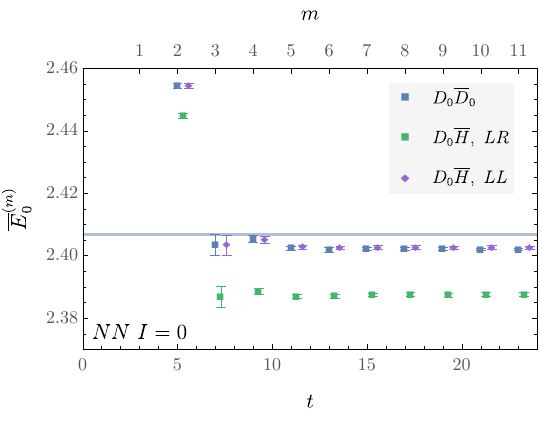} \\
    \includegraphics[width=0.48\textwidth]{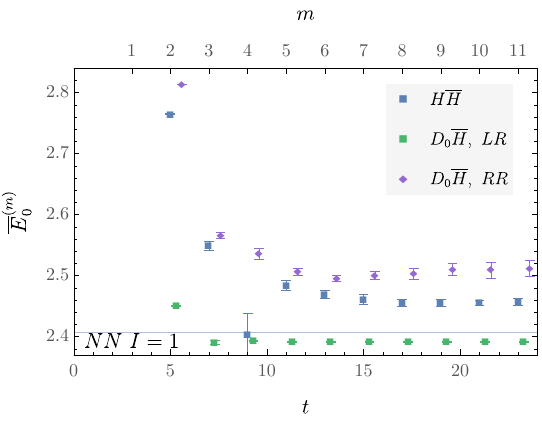}
    \includegraphics[width=0.48\textwidth]{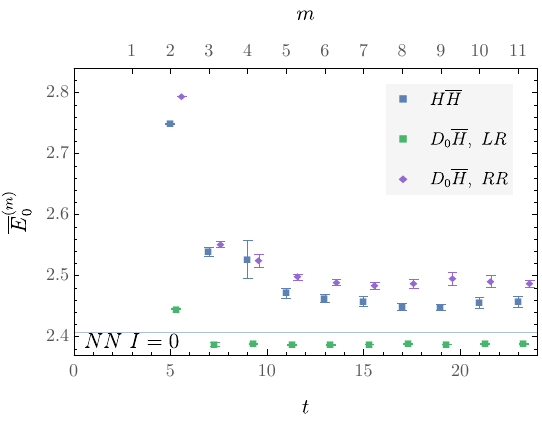}
    \caption{
      Upper: Comparison of Lanczos energy estimators derived from $D_0 \overline{H}$ Ritz values with those derived from SRQs involving $T^t \ket{D_0}$ states and $ D_0 \overline{D}_0$ Lanczos energy estimators.
      Lower: Comparison of the same Lanczos energy estimators derived from $D_0 \overline{H}$ Ritz values with those derived from SRQs involving $T^t \ket{H}$ states and $ H \overline{H}$ Lanczos energy estimators.
    }
    \label{fig:asym:DH-Lanczos-EMP-RQ}
\end{figure*}

\Cref{fig:asym:DH-Lanczos-EMP-RQ} show SRQ estimators for the $\nn$ and $\deut$ channels.
Also shown for comparison are energy estimates from the asymmetric Ritz values, as well as the Ritz values from analyses of the corresponding diagonal correlators.
Both $R$ and $L$ SRQs differ from the asymmetric Ritz values at high significance.
However, $\overline{E}^{L(m)}_0$, which is constructed from $T^t \ket{D_0}$ states, is consistent at the $1\sigma$ level with the Ritz values derived from the corresponding $D_0 \overline{D}_0$ diagonal correlator.
Similarly, the other SRQ, $\overline{E}^{R(m)}_0$, constructed from $T^t \ket{H}$ states, is comparable to the $H \overline{H}$ energy estimator.
Differences between $\overline{E}_0^{(m)}$, $\overline{E}_0^{R(m)}$, and $\overline{E}_0^{L(m)}$ should not be surprising given the behavior of $\cos \theta^{y(m)}_{LR}$ in \cref{fig:asym:ritz-angles} above: $\ket{y^{R(m)}_0}$ and $\ket{y^{L(m)}_0}$ are distinct states which have not separately converged to a common ground state estimate.
The SRQs do, however, provide additional information about the separate identities of the $R$ and $L$ states: the left Ritz vector $\ket{y^{L(m)}_k}$ is approximately the same state as yielded by a symmetric analysis of the ${D_0 \overline{D_0}}$ correlator, and similar for $\ket{y^{R(m)}_k}$ and the ${H\overline{H}}$ correlator.
In a vector-space interpretation of asymmetric correlator data, the asymmetric Ritz values are again seen to correspond to transition matrix elements.

Qualitatively similar results are obtained when analyzing $D_1 \overline{H}$ correlators: the asymmetric Ritz-value-based estimator $\overline{E}_0^{(m)}$ is significantly below the $D_1 \overline{D}_1$  energy estimator (which is consistent with $\overline{E}_1^{(m)}$ from $2D$ block Lanczos), while $\overline{E}_0^{L(m)}$ is consistent with the $D_1 \overline{D}_1$ ground-state energy estimator.
Asymmetric correlators constructed from different dibaryon operators such as $D_1 \overline{D}_0$ can also be analyzed, but they have no real Ritz values.
Effective masses, and the real parts of these complex-valued Ritz values, for $D_1 \overline{D}_0$ correlators are significantly further below threshold than those of $D_0\overline{H}$ correlators~\cite{Amarasinghe:2021lqa}. However, all estimators based on the corresponding SRQs are above threshold for  $D_1 \overline{D}_0$ correlators.

As already noted above, SRQs provide variational bounds on true energy levels in the same way as symmetric Ritz values.
Moreover, they can be used to construct gap bounds using \cref{eq:asymm_gap_bound},
where, importantly, $B^{R(m)}_k$ must be used to construct the bound for $\eta^{R(m)}_k$, and similarly for the left version.
The resulting gap bounds are shown in \cref{fig:asym:nn-Lanczos-bounds-comp}.
In this case, gap bounds arising from two definitions of the gaps going into the bounds are shown, using either (1) the gap definitions in \cref{sec:bounds}, or (2) alternatively, the gap to the energy extracted from the asymmetric $D_0\overline{H}$ correlator. In the first case, it is assumed that the counting of the true eigenvalues up to the $N\Delta$ or $\Delta\Delta$ threshold is the same as that of the non-interacting two-nucleon energies. Under this assumption, the gap bound determined from the left SRQ arising from the $D_0\overline{H}$ correlator indicates that there is a true energy just below threshold that is at a significantly higher energy than the energy found from fits to the $D_0\overline{H}$ correlator alone in both the $\nn$ and $\deut$ NN channels. However, the second less constraining choice for the gap leads to gap bounds on the same SRQ that are broader (even larger than the SRQ residual bounds) and are consistent with the energy found from fits to the $D_0\overline{H}$ correlator alone in both channels. 

The results above lead to the conclusion that there are only two scenarios that are consistent with LQCD data for this volume. The first scenario is that there is only one energy level near threshold (in particular, much closer than the first non-interacting excitation energy): the no-missing-states assumptions in \cref{sec:strategies} are valid in this scenario and the relevant energy difference is constrained by gap bounds at 68\% confidence to be within $[-0.007, -0.004]$ in the $\nn$ channel and $[-0.009, -0.006]$ in the $\deut$ channel. The second scenario is that there are multiple energy levels near threshold: in this case, the gap bounds are much larger and the multiplicity and locations of these levels is much less constrained. In both scenarios, the presence of other states outside the gap bound windows is not constrained. The next section discusses ways in which both scenarios could be realized and demonstrates the difficulty of establishing a preference for one scenario over the other based on LQCD data without additional physics assumptions.

\section{Scenarios with multiple states near threshold}\label{sec:adversary}

The roles of different assumptions in spectroscopy are particularly important when there are closely spaced states or other sharp spectral features.
In this section, constrained fits to the LQCD data and adversarial examples are used to demonstrate how various constraints can fail when the relevant assumptions are not valid.  These examples highlight why physically motivated assumptions are required to draw well-quantified conclusions from spectroscopy analysis of LQCD data.
EFT$(\slashed{\pi}$)~\cite{Kaplan:1996nv,Bedaque:1998km,Beane:2000fi} is also used to demonstrate that such near degeneracies can arise straightforwardly in two-nucleon systems.

\subsection{The $\varepsilon \gg \Omega$ model}
\label{app:epsilon}

The $\varepsilon$ model was contrived in Ref.~\cite{Amarasinghe:2021lqa} as an adversarial scenario whose existence demonstrates that assumptions about the spectrum are required to exclude the existence of states with lower energies than those obtained in GEVP analyses.
The key mechanism appearing in this scenario is (approximate) orthogonality between the excited states that overlap with the two different interpolating operators.
In the original version, this orthogonality was taken to be exact for simplicity.
A generalized parameterization of the $\varepsilon$ model where this orthogonality is only approximate is given by a three-state model where two operators, labeled $H$ and $D$, have overlaps  of the form
\begin{equation}
  \begin{split}
    Z_D &= (\varepsilon, \sqrt{1 - \varepsilon^2 - \Omega^2}, \Omega), \\
    Z_H &= (\varepsilon, \Omega, \sqrt{1 - \varepsilon^2 - \Omega^2}),
    \label{eq:toy-model-2}
  \end{split}
\end{equation}
where it is essential that $\Omega \ll \varepsilon^2 < 1$.\footnote{Ref.~\cite{BaSc:2025yhy} noted that this model behaves differently at the $\Omega = 0$ point and in the regime $\varepsilon < \Omega < 1$. This work explores the neighborhood of small $\Omega\ll\varepsilon^2$ where the model behaves similarly to the $\Omega = 0$ point. 

Note also that the original $\varepsilon$-overlap model in Ref.~\cite{Amarasinghe:2021lqa} considered $\varepsilon \ll 1$ and $\Omega=0$. The condition $\varepsilon \ll 1$ is not satisfied in the numerical fits to LQCD data below, which prefer $\varepsilon \sim $ 0.6, motivating the more general model considered here.
}
In this model, asymmetric $D \overline{H}$ correlators can have almost perfect ground-state overlap with excited-state effects suppressed by $O(\Omega / \varepsilon^2)$.
Within the same model, both diagonal correlators overlap predominantly with excited states rather the ground state until $t \gtrsim \ln(\varepsilon^2 / (1 - \varepsilon^2)) / (E_1 - E_0)$.
The approximate orthogonality of excited states associated with the two operators, encoded by the smallness of $\Omega / \varepsilon^2$, leads to the  possibility that asymmetric correlators can asymptote to the ground state faster than either of the corresponding diagonal correlators, as illustrated in the constrained fits to the data below.

Following the suggestion in Ref.~\cite{BaSc:2025yhy} to include three interpolating operators and more than three energy levels, one can generalize the model in Eq.~(\ref{eq:toy-model-2}) to one involving $d+h+1$ states with overlaps
\begin{equation}
    \begin{split}
        Z_H &= (\varepsilon_H, \underset{d}{\underbrace{\Omega ,  \dots, \Omega}}, Z_H^{(1)},\dots, Z_H^{(h)}), \\
        Z_{D_1} &= (\varepsilon_{D_1}, Z_{D_1}^{(1)},\dots, Z_{D_1}^{(d)}, \underset{h}{\underbrace{\Omega ,  \dots, \Omega}}), \\
        Z_{D_2} &= (\varepsilon_{D_2}, Z_{D_2}^{(1)}, \dots, Z_{D_2}^{(d)}, Z_{D_2}^{(d+1)}, \ldots, Z_{D_2}^{(d+h)}) ~ .
    \end{split}
    \label{eq:toy-model-3}
\end{equation}
In the regime $\Omega \ll \varepsilon_a \varepsilon_b < Z_a^{(n)}Z_b^{(n)}$ $\forall\ a,b \in~\{H, D_1, D_2\}$, this generalized $\varepsilon$ model shares the same qualitative features as \cref{eq:toy-model-2}. In particular, each diagonal correlator overlaps predominantly with an excited state rather than the ground state until $t \gtrsim \min_{a,n}\ln(\varepsilon_a^2 / [Z_a^{(n)}]^2 ) / (E_n - E_0)$. Meanwhile, the asymmetric correlator constructed from $H$ and $D_1$ has large ground-state overlap, with excited-state effects suppressed by $O(\Omega Z^{(n)}_{D_1}/\varepsilon_H \varepsilon_{D_1})$.

\begin{figure}
    \centering
    \includegraphics[width=\linewidth]{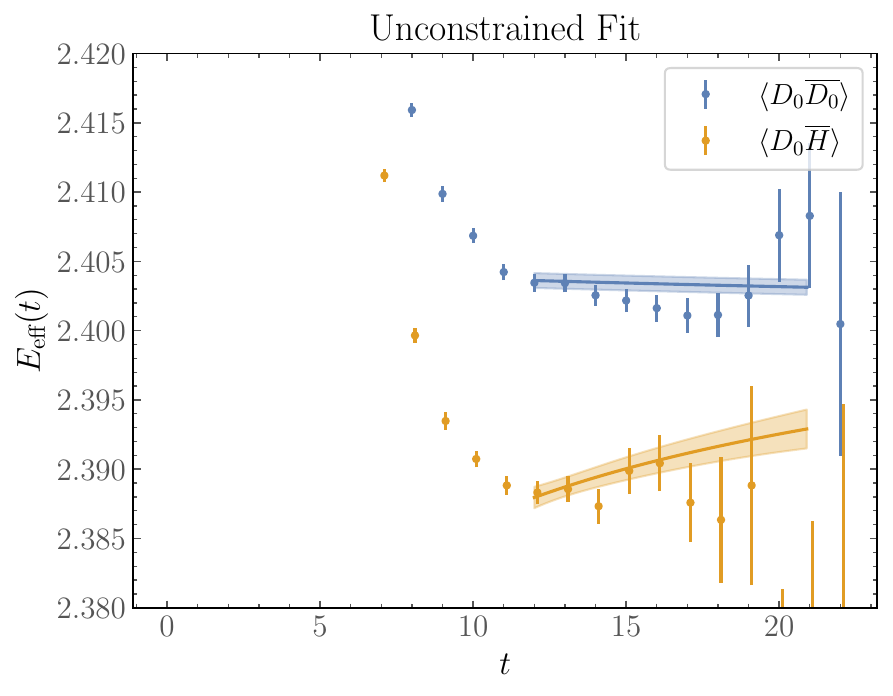}
    \includegraphics[width=\linewidth]{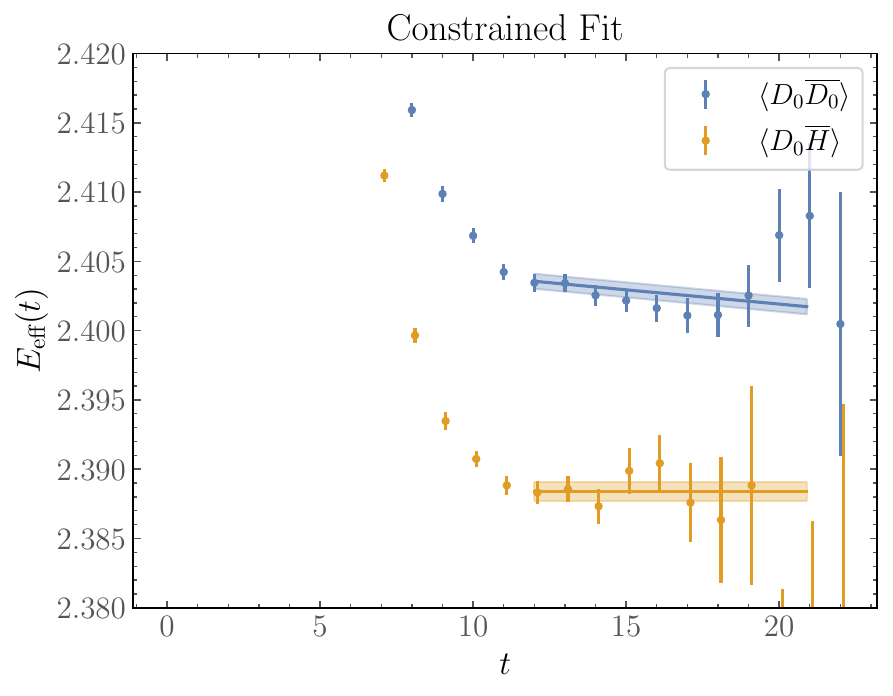}
    \caption{Effective masses of the $D_0\overline{D}_0$ and $D_0\overline{H}$ correlators and results from fits to the $3\times 3$ correlation matrix built from interpolating operators corresponding to the hexaquark and $D_0$ and $D_1$ dibaryon interpolating operators for the deuteron.  The fit in the upper panel is unconstrained and finds no bound state; the fit in the lower panel is a constrained fit to the model in Eq.~(\ref{eq:toy-model-3}) and has a bound state.  
    \label{fig:3-correlator-fits}
    }
\end{figure}

Constrained fits can be used to inquire how well the $3\times 3$ matrix of  $\deut$ $NN$ correlators with the $2D+H$ interpolating operator set can be described by this model for some choice of energies $E_0, E_1, \ldots, E_{d+h}$ and overlaps  $\{\epsilon_a,Z_a^{(n)}\}$.  For concreteness, fits are performed to \cref{eq:toy-model-3} with $\Omega=0$ and all other parameters left unconstrained. This constrained fit is then compared to a $N$-state fit with all overlap factors, and the value of $N$, unconstrained.
The best fits of the $2D+H$ correlator matrix data to each scenario, as measured by the AIC~\cite{AkaikeAIC}, are 
\begin{itemize}
    \item Unconstrained fit: an unconstrained 4-state fit starting with $t_\text{min} = 12$,
    \item Constrained fit: a 5-state fit starting with $t_\text{min} = 12$ constrained by Eq.~(\ref{eq:toy-model-3}) with $d=1$, $h=3$.
\end{itemize}
These fits are shown in Fig.~\ref{fig:3-correlator-fits}, along with the  $D_0\overline{D}_0$ and $D_0\overline{H}$ effective masses. Both provide good descriptions of the full $3\times3$ correlator matrix over the fitted time range.
The unconstrained and constrained fits provide mutually inconsistent descriptions of the data, with $E_0^{(\text{Unconstrained})} = 2.4016(23)$ and $E_0^{(\text{Constrained})} = 2.3884(7)$, respectively. The other parameters of the fits are shown in Tabs.~\ref{tab:fit-1} and \ref{tab:fit-2}.
With the zeroed-out overlap factors not counted as free parameters for the purpose of computing the AIC, the constrained fit is somewhat preferred ($\Delta_\text{AIC}\equiv \text{AIC}({\rm Unconstrained})-\text{AIC}({\rm Constrained}) = 24.4 - 28.3  = -3.9$).  If the zeroed-out overlap factors are counted as free parameters (on the basis of the model being considered too contrived), the unconstrained fit is preferred by a similar margin, $\Delta_{\text{AIC}'}\equiv \text{AIC}'({\rm Unconstrained})-\text{AIC}'({\rm Constrained}) = 24.4 - 20.3 = 4.1$.  However, the preference in either direction is too weak to exclude either scenario.\footnote{Note that many other forms of Bayesian and frequentist hypothesis tests that look for a preference between curvature in diagonal and asymmetric correlators can be constructed. Other examples that we have explored are similarly inconclusive.}

It is  important to note that if the $\varepsilon$-model scenario is realized, it does not provide a straightforward starting point from which the structure of approximate energy eigenstates can be studied by computing matrix elements of external currents and other generic operators.
In this scenario, the presence of the operator matrix element could spoil the $\varepsilon$-model orthogonality mechanism in three-point functions and quantifying the excited-state contamination present in operator matrix elements will be challenging.

\begin{table}[]
    \centering
    \begin{ruledtabular}
       \begin{tabular}{cS[table-format=1.2]S[table-format=1.2]S[table-format=1.2]S[table-format=1.2]@{}r} 
        $n$ & {$E_n$} & {$Z^{(n)}_{D_1}$} & {$Z^{(n)}_{D_2}$} & {$Z^{(n)}_H$} \\ \hline
        0 & 2.4016(23) & 0.955(87) & 0.150(38) & 0.207(65) \\
        1 & 2.433(48) & 0.29(23) & -0.4(1.1) & -0.25(36) \\
        2 & 2.457(17) & -0.06(25) & 0.90(48) & 0.16(42) \\
        3 & 2.633(46) & -0.007(34) & -0.010(73) & 0.9329(90)
    \end{tabular}
    \end{ruledtabular}
    
    \caption{Fit parameters for the unconstrained fit, with correlators normalized so that the squared coefficients sum to 1 for each of the diagonal correlators.  The fit has a $\chi^2/\text{dof} = 63.6/44 = 1.44$ and an AIC of 24.4. 
    }
    \label{tab:fit-1}
\end{table}

\begin{table}[]
    \centering
    \begin{ruledtabular}
    \begin{tabular}{cS[table-format=1.2]S[table-format=1.2]S[table-format=1.2]S[table-format=1.2]@{}r}
        $n$ & {$E_n$} & {$Z^{(n)}_{D_1}$} & {$Z^{(n)}_{D_2}$} & {$Z^{(n)}_H$} \\ \hline
        0 & 2.38840(69) & 0.622(17) & 0.216(13) & 0.184(28) \\
        1 & 2.4171(15) & 0.783(13) & -0.1924(40) & 0 \\
        2 & 2.456(39) & 0 & -0.7(3.4) & -0.32(17) \\
        3 & 2.464(41) & 0 & 0.7(3.3) & 0.01(1.75) \\
        4 & 2.650(64) & 0 & -0.015(36) & -0.929(28)
    \end{tabular}
    \end{ruledtabular}   
    \caption{Fit parameters for the constrained fit, constrained such that $Z^{(n)}_{D_1} Z^{(n)}_H = 0$ except for the ground state, with correlators normalized so that the squared coefficients sum to 1 for each of the diagonal correlators.  The fit has a $\chi^2/\text{dof} = 59.7/44 = 1.36$ and an AIC of 28.3, where parameters constrained to be 0 are not included as degrees of freedom. 
    }
    \label{tab:fit-2}
\end{table}

\subsection{A small-gap adversarial model}\label{sec:gap_attack}

\begin{table*}[t]
\begin{ruledtabular}
\begin{tabular}{rrrrrrrrrr}
\multicolumn{10}{c}{\textbf{Scattering scenario}} \\\hline
$k$ & $E^S_k$ & $Z^S_{k,D_0}$ & $Z^S_{k,D_1}$ & $Z^S_{k,D_2}$ & $Z^S_{k,D_3}$ & $Z^S_{k,D_4}$ & $Z^S_{k,D_5}$ & $Z^S_{k,D_6}$ & $Z^S_{k,H}$
\\ \hline
 0 &  $2.402$ & $19.683$ & $  1.880$ & $  0.535$ & $  0.079$ & $ -0.034$ & $-0.128$ & $-0.134$ & $ 1.590$ \\
 1 &  $2.451$ & $ 2.662$ & $-17.463$ & $ -1.574$ & $ -0.140$ & $  0.064$ & $ 0.261$ & $ 0.265$ & $-2.700$ \\
 2 &  $2.506$ & $ 0.729$ & $  2.541$ & $-15.806$ & $ -0.333$ & $  0.127$ & $ 0.289$ & $ 0.319$ & $-2.754$ \\
 3 &  $2.560$ & $ 0.335$ & $  0.458$ & $  0.759$ & $-14.572$ & $  0.020$ & $ 0.372$ & $ 0.263$ & $-1.938$ \\
 4 &  $2.614$ & $ 0.307$ & $  0.149$ & $  0.128$ & $  0.098$ & $-12.009$ & $-0.045$ & $ 0.063$ & $-1.214$ \\
 5 &  $2.667$ & $ 0.396$ & $  0.119$ & $  0.200$ & $ -0.037$ & $  0.215$ & $-9.845$ & $ 0.072$ & $-1.859$ \\
 6 &  $2.717$ & $ 0.505$ & $  0.209$ & $  0.141$ & $  0.048$ & $  0.184$ & $ 0.009$ & $-9.140$ & $-0.723$ \\
 7 &  $2.902$ & $ 2.740$ & $  1.289$ & $  0.682$ & $  0.241$ & $  0.295$ & $ 0.472$ & $ 0.414$ & $ 0.408$ \\
 8 &  $2.987$ & $13.591$ & $  5.596$ & $  1.296$ & $  0.273$ & $ -0.110$ & $-0.288$ & $ 0.069$ & $ 2.371$ \\
 9 &  $3.014$ & $ 7.541$ & $-10.610$ & $ -0.791$ & $  0.290$ & $  0.379$ & $ 0.456$ & $ 0.056$ & $-0.420$ \\
10 &  $3.051$ & $ 3.533$ & $ -1.493$ & $  6.270$ & $ -0.357$ & $  0.112$ & $ 0.044$ & $-0.191$ & $ 0.150$ \\
11 &  $3.084$ & $ 2.806$ & $ -0.892$ & $  0.659$ & $ -7.673$ & $  0.536$ & $-0.342$ & $-0.469$ & $-0.026$ \\
12 &  $3.053$ & $ 3.417$ & $ -0.158$ & $  0.036$ & $ -0.530$ & $ -6.234$ & $-0.230$ & $ 0.090$ & $-0.078$ \\
13 &  $3.129$ & $ 2.458$ & $ -0.327$ & $  0.400$ & $ -0.575$ & $  0.065$ & $-8.287$ & $-0.103$ & $-0.240$ \\
14 &  $3.182$ & $ 2.293$ & $  0.027$ & $  0.242$ & $ -0.422$ & $  0.630$ & $ 0.386$ & $-7.714$ & $-0.195$ \\
15 &  $3.428$ & $ 2.261$ & $  0.484$ & $  0.406$ & $ -0.239$ & $  0.450$ & $ 0.017$ & $ 0.331$ & $ 0.585$ \\
16 &  $3.797$ & $ 8.620$ & $  2.784$ & $  0.750$ & $ -0.098$ & $  0.620$ & $ 0.031$ & $-0.000$ & $ 1.160$ \\
17 &  $3.821$ & $ 3.800$ & $ -4.210$ & $  0.237$ & $ -0.120$ & $  0.080$ & $-0.007$ & $ 0.174$ & $ 0.105$ \\
18 &  $3.848$ & $ 2.805$ & $ -0.996$ & $  1.012$ & $  0.078$ & $ -0.245$ & $-0.012$ & $-0.053$ & $ 0.252$ \\
19 &  $3.870$ & $ 2.535$ & $ -0.252$ & $ -0.058$ & $  1.390$ & $ -0.308$ & $-0.099$ & $-0.031$ & $ 0.096$ \\
20 &  $3.878$ & $ 2.524$ & $ -0.245$ & $ -0.244$ & $  0.154$ & $ -0.605$ & $ 0.058$ & $ 0.147$ & $ 0.136$ \\
21 &  $3.941$ & $ 1.902$ & $ -0.145$ & $  0.163$ & $ -0.273$ & $  0.182$ & $-2.657$ & $ 0.063$ & $ 0.068$ \\
22 &  $4.000$ & $ 1.663$ & $ -0.019$ & $  0.018$ & $ -0.052$ & $  0.258$ & $ 0.038$ & $ 0.435$ & $-0.048$ \\
23 &  $4.219$ & $ 1.096$ & $  0.181$ & $  0.170$ & $ -0.101$ & $  0.263$ & $ 0.129$ & $ 0.226$ & $ 0.306$ \\
\end{tabular}
\end{ruledtabular}

\vspace{1em}

\begin{ruledtabular}
\begin{tabular}{rrrrrrrrrr}
\multicolumn{10}{c}{\textbf{Bound scenario}} \\\hline
$k$ & $E^B_k$ & $Z^B_{k,D_0}$ & $Z^B_{k,D_1}$ & $Z^B_{k,D_2}$ & $Z^B_{k,D_3}$ & $Z^B_{k,D_4}$ & $Z^B_{k,D_5}$ & $Z^B_{k,D_6}$ & $Z^B_{k,H}$
\\ \hline
$0\phantom{'}$  &   2.392 &  8.466 &  0.646 &  0.173 &  0.029 & $-0.012$ & $-0.050$ & $-0.048$ & 0.573 \\
$0'$ &   2.4045 & 17.769 &  1.766 &  0.506 &  0.073 & $-0.032$ & $-0.118$ & $-0.125$ & 1.483 \\
$\geq 1$ & \multicolumn{9}{c}{(Same as Scattering)} \\
\end{tabular}
\end{ruledtabular}
\caption{
Energies and overlaps that
define the scattering and bound scenarios of the small-gap adversarial exercise. 
Parameters for the scattering scenario are taken from the central values\footnote{While the model involves 24 states, those for $k\ge7$ are primarily included to keep the spectrum high-rank and to match the synthetic correlator data as closely as possible to the LQCD data.} of a block Lanczos analysis of the $7D+H$ interpolator set in the $\nn$ channel.
Parameters for the bound scenario are identical to those of the scattering scenario for $k \geq 1$, with construction of the $k=0$ and $0'$ states detailed in the main text.
}
\label{tab:small-gap-adversarial-params}
\end{table*}

\begin{figure*}
    \includegraphics[width=0.45\linewidth]{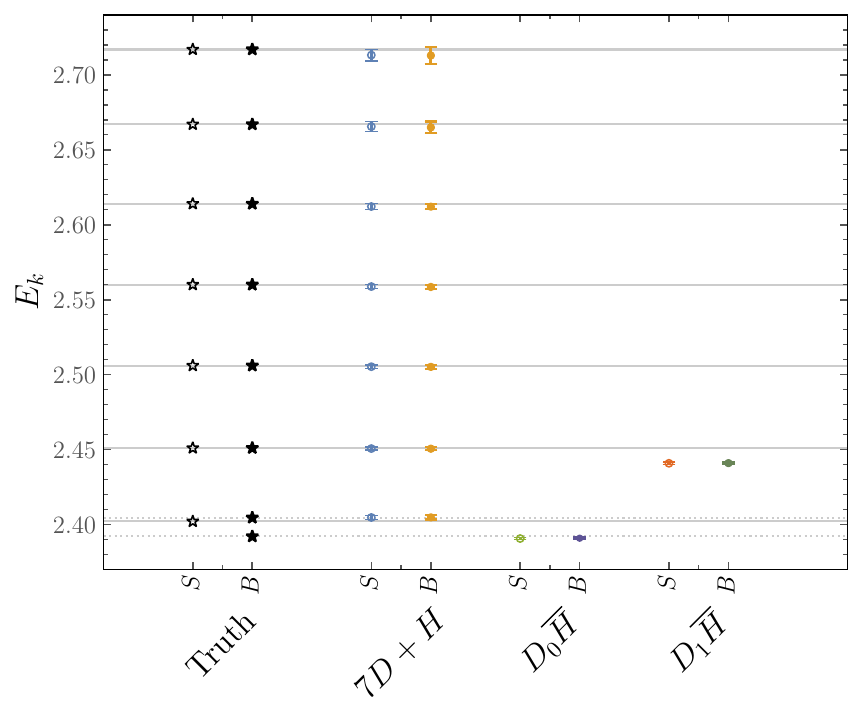}
    \includegraphics[width=0.45\linewidth]{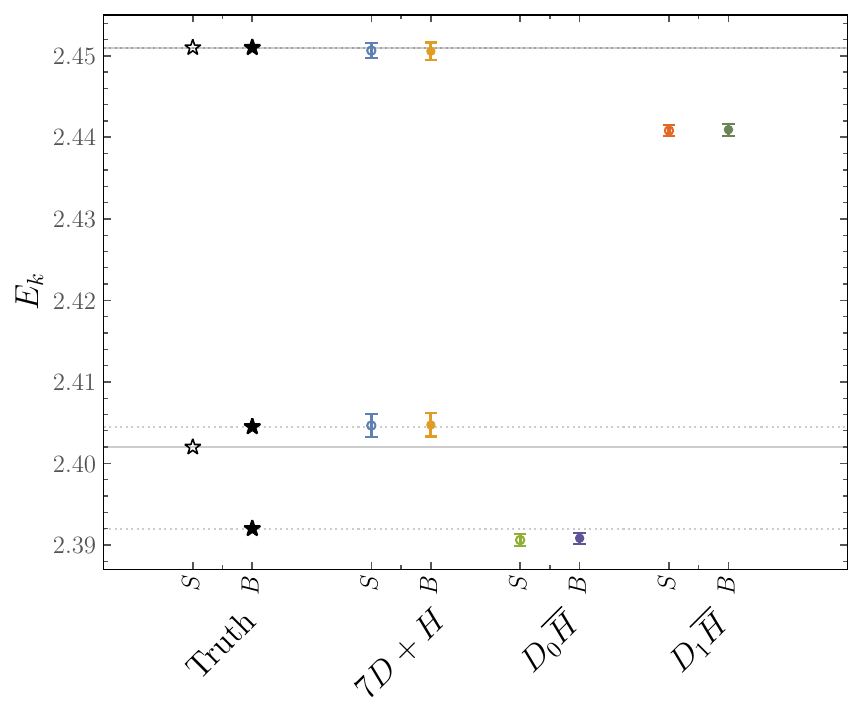}
    \caption{
        Results of the small-gap adversarial scenario, as defined in the text.
        Starred points labeled ``Truth'' are the true spectra in the scattering and bound scenarios.
        Other series are the energies extracted from analysis of synthetic data generated under each scenario, with $7D+H$ indicating a symmetric block Lanczos analysis of the $8 \times 8$ correlator matrix, and with $D_0 \overline{H}$ and $D_1 \overline{H}$ indicating asymmetric Lanczos analyses of the respective off-diagonal correlators.
        The error bars show the statistical uncertainty only.
        The right panel is a zoom on the lowest-lying states in the left panel.
        Solid lines correspond to the true energies in the scattering scenario, while dashed lines correspond to the true energies in the bound scenario.
    }
    \label{fig:gap-attack}
    \label{fig:gap-attack-zoom}
\end{figure*}

\begin{figure}
    \includegraphics[scale = 0.6]{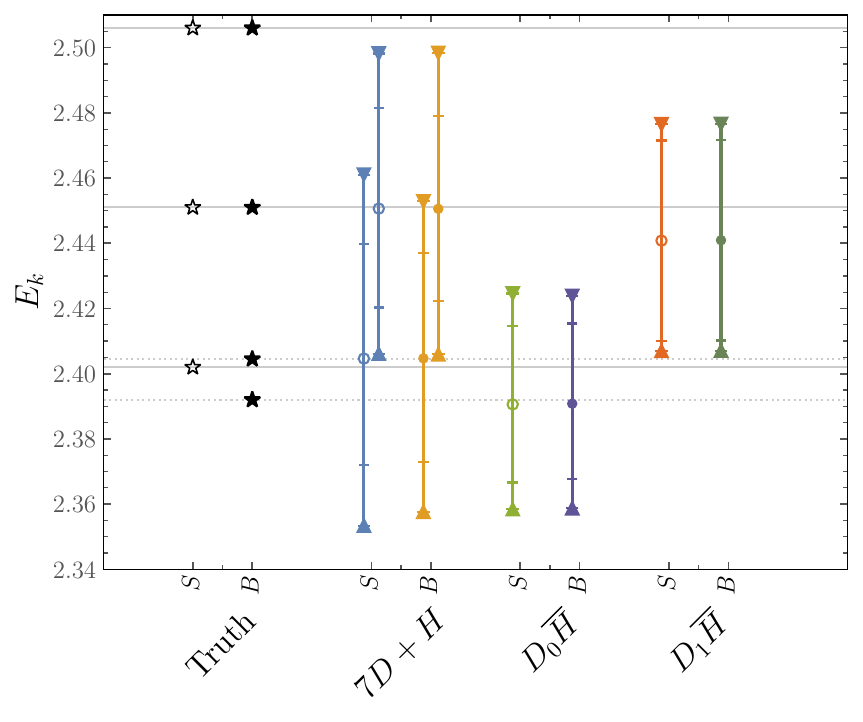}
    \caption{
        As in \cref{fig:gap-attack-zoom}, but with residual bounds shown as uncertainties.
        For the asymmetric analyses, the narrower $B^L$ bound corresponding to the dibaryon interpolator is shown.
    }
    \label{fig:gap-attack-res-bounds}
\end{figure}

In this section, a simple mechanism of near degeneracies in the spectrum is explored. A spectral model is matched to the symmetric Lanczos analysis of LQCD data. This is then used to produce a second model by splitting one of the states into a near-degenerate pair. Synthetic data generated from these models is then confronted against both symmetric and asymmetric correlator analysis techniques. At achievable statistical precision, all available analysis techniques fail to distinguish between these models.

The ``scattering'' (S) example is constructed using the central values of the block Lanczos analysis of the $7D+H$ correlator matrix for $\nn$, as presented in \cref{sec:lqcd-lanczos}.
The energies and overlap factors ($E_k^S$ and $Z_{k,a}^S$ for $k\in\{0,\ldots,23\}$ and $a\in\{D_0,\ldots, D_6, H\}$) are presented in \cref{tab:small-gap-adversarial-params}.

The ``bound'' (B) example is constructed from the scattering model by separating the $k=0$ ground state into two close-by states: one slightly below ($k=0$), and the other ($k=0'$) slightly above, the scattering model ground state.
All other states are kept the same. For the modified energies, $E^B_{0} = 2.392$ is chosen corresponding to the value extracted from analysis of the $D_0\overline{H}$ correlator alone, while $E^B_{0'}=2.4045$ was chosen by hand. 
The overlaps for the two low-lying states in the bound model are taken from the overlaps $Z^S_{ka}$ from the scattering example, and divided between the two states as
\begin{equation}\begin{aligned}
    Z^B_{0a}  &= \sqrt{p_a} Z^S_{0a} \\
    Z^B_{0'a} &= \sqrt{1-p_a} Z^S_{0a}
\end{aligned}\end{equation}
where 
\begin{equation}
    p_a = \{0.185, 0.118, 0.105, 0.136, 0.128, 0.153, 0.131, 0.13\} ~ .
\end{equation}
are coefficients drawn randomly from $[0.1,0.3]$ to avoid fine tuning and produce overlaps with the lowest state $k=0$ that are slightly smaller than those for the next-to-lowest state $k=0'$.
Using different values for each interpolator ensures that there is linearly independent information available about each state; note especially that the overlaps with the $D_0$ and $H$ interpolating operators are in different ratios.

A common noise model is used for both the scattering and bound examples and is tuned to match the SNR of the LQCD data. Specifically, synthetic noise is applied multiplicatively to the noiseless correlator matrix after it has been constructed from the exact parameters in \cref{tab:small-gap-adversarial-params}:
\begin{equation}
    C_{n,ab}(t) = (1 + \eta_{n,ab}(t)) \sum_k Z_{ka} Z_{kb} e^{-E_k t}
\end{equation}
where $\eta_{n,ab}(t)$ is the $n$th noise draw.\footnote{Note that noise draws are independent for each inner bootstrap used in a nested bootstrap setup in Lanczos analyses. This neglects correlations between different inner ensembles.}
The $\eta_{n,ab}(t)$ are sampled to reproduce the relative noise levels observed in the LQCD data.
Specifically, the $\eta_{n,ab}(t)$ are sampled from a multivariate Gaussian with zero means and a covariance matrix
\begin{equation}
    \Sigma_{abt,a'b't'} = \frac{
        \mathrm{Cov}[C_{ab}(t), C_{a'b'}(t')]
    }{
        \braket{C_{ab}(t)} \braket{C_{a'b'}(t')}
    }
\end{equation}
constructed using bootstrap estimators on the LQCD data.

Block Lanczos with SLRVL state-labeling inside a nested-bootstrap scheme, as in \cref{sec:lqcd-lanczos}, is used to analyze the resulting data sets.\footnote{Some modifications are necessary due to the simplicity of the noise model.
Specifically, despite matching the SNR levels with the original data, the synthetic data admits only one symmetric Lanczos iteration.
This is insufficient to enable a SLRVL analysis exactly as employed in the analysis of \cref{sec:lqcd-lanczos}. Instead, a fixed $m_\mathrm{last}=3$ is used and pre-processed using ZCW filtering to remove the small fraction of  label states that are not Hermitian at this iteration.
A more realistic noise model would admit construction of synthetic data that does not require this special treatment. It is not expected that this complication changes the qualitative conclusions of the exercise.}
Asymmetric Lanczos analyses of the off diagonal $D_0 \overline{H}$ and $D_1 \overline{H}$ correlators are also performed.
These are expected to primarily target the ground and first excited state, respectively.
In each asymmetric case, filtering for real eigenvalues and amplitudes and using the ZCW test is used as in \cref{sec:asymm} with $\epsilon^\mathrm{ZCW} = 10^{-3}$, and only the ground state is retained.

\Cref{fig:gap-attack} compares the three resulting analyses with the true spectra in each scenario, restricting to the lowest-lying 7 states that are expected to be well-resolved from the $7D+H$ interpolating operator set.
The problem is immediately apparent: analyses of data from the scattering and bound scenarios give effectively identical outputs in both value and statistical uncertainty under all three types of analysis.
Focusing on the low-energy region, it is also notable that none of the analyses are correct within statistical uncertainties.
In the block analysis of the scattering scenario, the extracted ground state is $\approx 2.1\sigma$ higher than truth. 
In the block analysis of the bound scenario, the extracted ground state aligns well with the true first excited state, but the true ground state is missed entirely and the degeneracy is entirely missed.
In both the scattering and bound scenarios, the asymmetric analyses of the $D_0 \overline{H}$ correlator find low-lying energies in the neighborhood of the true ground state of the bound scenario.
In the scattering scenario, this is simply incorrect, while in the bound scenario, the estimate is still $\approx 1.7\sigma$ lower than the true eigenvalue (this is allowed since the asymmetric estimator is not a variational bound).
The asymmetric analysis of the $D_1 \overline{H}$ correlator is many standard deviations below the true eigenvalue in both scenarios.

\Cref{fig:gap-attack-res-bounds} shows the residual bounds for the lowest two states in each scenario, and for each analysis approach.
All bounds are satisfied, but they are too broad to discriminate between the two scenarios, as most of the bounds contain multiple true eigenvalues.
As with the energy estimators, the bounds derived from each analysis method are nearly  identical between the two scenarios. Gap bounds are not shown for these examples as these require information on the spectrum that is being deliberately occluded in this discussion.

The results of this exercise demonstrate explicitly that spectroscopy on noisy correlators of finite extent is an ill-posed inverse problem: qualitatively different underlying spectra may produce statistically indistinguishable correlation function data.
In this scenario, any analysis of the data cannot distinguish between any of the true underlying spectra that produce the given correlators.\footnote{Note that when the exercise is repeated at levels of simulated statistics that are significantly higher than those of the LQCD data, the issues described below eventually resolve.}
A particular choice of analysis methods will preferentially return one kind of spectrum but this is, in essence, an implicit regularization of the inverse problem and should not be over-interpreted. 
Further discussion of paths forward is given in \cref{sec:discussion}.

\subsection{Physics interpretation}
\label{sec:EFT}

As demonstrated by the LQCD and synthetic data examples above, unexpected small gaps and near degeneracies are especially problematic for existing analysis methods.
Absent a data-driven diagnostic, it is important to consider how likely it is that such fine tunings appear in physical systems.
Well-motivated assumptions should be backed by physics arguments: would it be physically reasonable for the system to exhibit near degeneracies, or can one argue that it should not?
As argued below, \textit{a priori}, there is no  reason to exclude small gaps in $NN$ systems.

Indeed, it is straightforward to provide a EFT$(\slashed{\pi})$ description of the $\varepsilon$ model that shows that it is a genuine possibility for the spectrum of a hadronic theory.
The relevant Lagrangian is
\begin{equation}
\begin{split}
\mathcal{L}&=N^\dagger \bigg(i D_t -\frac{\vec{D}^2}{2M_N}\bigg)N 
+ d^*\bigg(iD_t -\frac{\vec{D}^2}{4M_N}-\Delta\bigg)d 
\\
&-g_{dNN} (d^* NN+\text{h.c})-\frac{1}{2}C (N^\dagger N)^2+\dots,
\end{split}
\end{equation}
where $N$ and $d$ are nucleon and dibaryon fields, respectively~\cite{Kaplan:1996nv,Bedaque:1998km,Beane:2000fi}, $D_t$ and $\vec{D}$ are temporal and spatial derivatives, and $\Delta$, $g_{dNN}$ and $C$ are couplings.
The appropriate values of the parameters\footnote{Note that $C$ and  $\Delta$ are renormalization-group scheme and scale dependent. These values correspond to the scheme used in Ref.~\cite{Detmold:2023lwn}.} to describe the LQCD data in the $\varepsilon$-model scenario  are 
$C \approx 0.1$ GeV $\cdot$ fm$^3$, 
$\Delta \approx 15$ MeV, and $g_{dNN}^2 \lesssim 0.1$ GeV$^{-1}$.
With these parameters,  EFT$(\slashed{\pi})$ can describe a two-nucleon FV spectrum in which there is both a $\sim15$ MeV bound state and a scattering state appearing $\sim5$ MeV below threshold for this volume.\footnote{These parameters are sufficient to ensure that the $NN$ scattering amplitude includes a pole at $E \approx \Delta$ while 
keeping the effective potential $C + g_{dNN}^2/(E - \Delta)$ dominated by the $(N^\dagger N)^2$ contact operator rather than by Yukawa interactions for $E - \Delta \gtrsim 10$ MeV. This in turn allows a scattering state pole to be described using a similar value of $C$ to that in Ref.~\cite{Detmold:2023lwn}, while at the same time introducing an additional bound state pole with binding energy $\approx \Delta$.}
For comparison, the (very) low-energy ${}^1S_0$ scattering amplitude at the physical quark masses, where there is no bound state, can be described in an effective theory including pions as well as dibaryon fields with corresponding $NN$ contact operators and dibaryon parameters $C \approx -0.2$ GeV $\cdot$ fm$^3$, $\Delta \approx 15$ MeV, and $g_{dNN}^2 \approx 2$~GeV$^{-1}$~\cite{Kaplan:1996nv}. These values are of the same magnitudes as those needed to describe the LQCD data.

Since $NN$ scattering is finely tuned in nature, and the scattering lengths bounded in \cref{sec:phase-shift} are also fine-tuned, it does not seem that the EFT$(\slashed{\pi})$ parameters needed to describe the $\varepsilon$-model scenario are unrealistic for QCD at unphysical quark masses. 
Similar matching to EFT$(\slashed{\pi})$ or to chiral EFT can likely also accommodate both the bound and scattering scenarios from the small-gap exercise of \cref{sec:gap_attack}, although determining the appropriate values for EFT counter-terms would require more sophisticated analysis similar to that performed in Ref.~\cite{Detmold:2023lwn}.

The central point that can be learned from the $\varepsilon$-model description of the LQCD data and the small-gap adversarial example is that physical assumptions are \textit{required} to draw conclusions about the nature of the two-nucleon spectrum from LQCD calculations with present algorithms and computing resources where $\delta_1 t \gg 1$ is not achievable.
This conclusion holds independently of whether 
these scenarios provide
a realistic description of QCD or any other theory.
At the physical quark masses, enough is known experimentally about two-nucleon systems that it could be very reasonable to make no-missing-states or $N$-state-saturation assumptions in a variety of phenomenological applications.
For unphysically large quark masses, where less is known about the spectrum \emph{a priori}, it is especially important to be explicit about what input assumptions lead to particular conclusions about the spectrum.
These exercises also 
demonstrate
that caution must be applied to both symmetric and asymmetric correlator analyses.

\section{Discussion and conclusion}
\label{sec:discussion}

\begin{table}[t]
    \centering
    \begin{ruledtabular}
    \begin{tabular}{ccc}
    Quantity  & Constraint & Reason/Assumption \\ \hline
    $E_0$ & $\leq 1.2037$  & Variational bound \\
    $E$ & $\in[1.196,1.211]$  & Residual bound \\
    $E$ & $\in[1.2032,1.2038]$  & Gap bound \\
    $E_0$ & $\in[1.2033,1.2037]$  & $N$-state saturation \\
    \end{tabular}
    \end{ruledtabular}
    \caption{Constraints on single-nucleon energies. Bounds and intervals are presented at 68\% statistical confidence.
    Quantities denoted $E$ without a definite state index emphasize that they constrain \emph{some} energy without knowledge or assumption of its position relative to the ground state.
    }
    \label{tab:bounds_summary_N}
\end{table}

A primary strength of lattice QCD calculations is the ability to quantify systematic uncertainties.
Discretization artifacts, finite-volume effects, and other systematic uncertainties are typically quantified by identifying a small parameter and performing an expansion in that parameter whose terms can be fit to LQCD results using theoretical guidance or neglected if they are expected to be sufficiently small.
For single-hadron systems, excited-state effects fall into the same pattern, where $e^{-\delta_1 t} = e^{-(E_1-E_0)t}$ is the small parameter.
However, for multi-hadron systems with signal-to-noise problems, energy gaps are small and computationally accessible imaginary times are limited such that statistical control is only achieved for $\delta_1 t \lesssim 1$ and excited-state contributions are not effectively suppressed by imaginary-time evolution. State-of-the-art analyses of multi-baryon spectroscopy have therefore been forced to adopt \textit{$N$-state-saturation assumptions} implicitly.
In addition to assuming the qualitative structure of the spectrum, this amounts to making assumptions about overlaps between particular interpolating operators and LQCD eigenstates. 
In certain situations, such assumptions may be justified, however many examples exist, both for symmetric and asymmetric correlation functions, in which they fail~\cite{Dudek:2012xn,Lang:2012db,Kiratidis:2015vpa,Wilson:2015dqa,Amarasinghe:2021lqa}. 
In this work, bounds on multi-hadron energy levels that require 
weaker assumptions are introduced and investigated.
In addition to  variational bounds, new two-sided bounds that can be computed within a Lanczos framework  are studied.
\textit{Residual bounds} only require the assumption of transfer-matrix Hermiticity to provide two-sided windows where one or more LQCD eigenvalue must be present.
\textit{Gap bounds} importantly require a \textit{no-missing-states assumption}---that there are the same number of Lanczos energy estimators and genuine energy eigenvalues within a specified window---however, they do not require any assumptions about operator overlaps.
All of these bounds are valid for finite imaginary-time data, even when the gap is not small ($\delta_1 t \slashed{\gg} 1$).
Their constraining power instead arises from the smallness of the Lanczos residual-norm-square $B$, 
with residual bounds scaling
as $O(\sqrt{B})$ and gap bounds scaling as $O(B)$.
In addition to these bounds, Lanczos methods provide a novel way to assess asymmetric correlation functions through the construction of \textit{Krylov-} and \textit{Ritz-vector overlap angles} that query whether asymmetric correlators provide matrix elements of the transfer matrix in a definite state.

Here, these bounds and angles are investigated, both analytically and numerically, in a solvable scalar field theory model in $0+1$D whose correlation functions behave in a statistically similar way to LQCD $NN$ correlators. While Lanczos energy estimators and BMA fits to GEVP principal correlators are found to underestimate the true uncertainties on extracted energies, the three types of bounds provide reliable uncertainties in this system where the true energies are known. Since residual bounds apply even in pathological cases, these bounds significantly overestimate the uncertainties.  Gap bounds, however, are found to be well-calibrated in this model, although generalizations to other contexts should be pursued with caution. The Ritz and Krylov angles indicate that the asymmetric correlators that are studied in this model have interpretations as diagonal transfer matrix elements at sufficiently large Euclidean times.

These new constraints, and state-of-the-art analysis techniques, are used to investigate a high statistics LQCD calculation of two-nucleon energies in a single finite volume at unphysically heavy quark masses and at a single lattice spacing. 
The results for energy estimators and the bounds discussed above can be used to draw various conclusions about the LQCD spectrum under different assumptions.
These conclusions are summarized in \cref{tab:bounds_summary_N,tab:bounds_summary_0,tab:bounds_summary_1,tab:bounds_summary_0_diff,tab:bounds_summary_1_diff}.
The $\gg 5\sigma$ inconsistencies between the  constraints on $\Delta E_0^{I=0,1}$ coming from GEVP/block Lanczos under $N$-state-saturation assumptions and those from $D_0\overline{H}$ asymmetric correlators under the same class of  assumptions (differences of 0.0134(4) and 0.0111(4) for $I=0$ and $I=1$, respectively) mean that these assumptions cannot be true in both cases. The lowest energy residual bounds derived from GEVP/block Lanczos, as well as those built from $D_0\overline{H}$ asymmetric correlators and their symmetric counterparts, are consistent with each other and with both classes of energy estimators. Gap bounds can only be computed in the symmetric case, or through the construction of symmetric Rayleigh quotients that provide energy estimators consistent with those obtained in the symmetric case.
Gap bounds are found to be significantly more constraining that residual bounds, although dependent on the choice of the spectral gap, see \cref{tab:bounds_summary_0,tab:bounds_summary_1}.

Quantitative constraints on $NN$ scattering amplitudes can be obtained from FV energies through L{\"u}scher's quantization condition \cite{Luscher:1986pf,Luscher:1990ux} or using effective field theory \cite{Barnea:2013uqa,Eliyahu:2019nkz,Sun:2022frr,Detmold:2023lwn,Detmold:2021oro}. Variational bounds are one sided, and therefore provide overlapping phase-shift constraints that are not useful. A key advantage of the Lanczos-based bounds is that they provide finite intervals in which an eigenvalue of the LQCD transfer matrix exists. It is therefore possible to use such bounds to constrain scattering phase shifts rigorously. While the Lanczos approach will miss states in the spectrum with small overlaps onto the interpolating operators that are used, the phase-shift constraints constructed from the states that are determined remain valid (it is also noteworthy that this task does not require knowledge of which state an eigenvalue corresponds to).
In the current study with more than 20 million quark propagators, residual bounds provide percent-level constraints on energy levels but are not precise enough to  provide useful constraints on $NN$ phase shifts. Consequently, some physical assumptions are required to produce meaningful constraints on $NN$ scattering amplitudes from LQCD using existing spectroscopic techniques at achievable statistics.
Under no-missing-states assumptions, gap bounds provide significantly tighter constraints on the FV two-nucleon energies. These lead to phenomenologically relevant constraints on scattering phase shifts and to the conclusion that both $\nn$ and $\deut$ $NN$ channels are attractive and have fine-tuned scattering lengths, as discussed in \cref{sec:phase-shift}.

\begin{table}[t]
    \centering
    \begin{ruledtabular}
    \begin{tabular}{cccc}
    Quantity  & Constraint & Reason/Assumption & $\{\mathcal{O} \}$ \\ \hline
    $E_{3}$ & $\leq 2.560$ & Variational bounds & $4D$ \\
    $E_{0}$ & $\leq 2.404$ & Variational bounds & $1D$ \\
    $E_{0}$ & $\leq 2.403$ & Variational bounds & $4D$ \\
    $E$ & $\in[2.39,2.42]$ & Residual bound & $1D$ \\
    $E$ & $\in[2.38,2.42]$ & Residual bound & $4D$ \\
    $E$ & $\in[2.36,2.42]$ & Residual bound & $D_0\overline{H}$ \\
      $E$ & $\in[2.400,2.406]$  & $NN$ $k=1$ gap  bound & $1D$  \\
      $E$ & $\in[2.400,2.405]$  & $N\Delta$ gap  bound & $4D$ \\
      $E$ & $\in[2.39,2.42]$  & $NN$ $k=1$ gap  bound & $D_0\overline{H}$ \\
      $E$ & $\in[2.39,2.41]$  & $E_{\text{asymm}}$ gap  bound & $1D$  \\
      $E$ & $\in[2.37,2.43]$  & $E_{\text{asymm}}$ gap  bound & $4D$ \\
      $E$ & $\in[2.34,2.47]$  & $E_{\text{asymm}}$ gap  bound & $D_0\overline{H}$ \\
      $E_0$ & $\in[2.403,2.404]$  & $N$-state saturation & $1D$ \\
      $E_0$ & $\in[2.402,2.403]$  & $N$-state saturation & $4D$ \\
      $E_0$ & $\in[2.391,2.392]$  & $N$-state saturation & $D_0\overline{H}$ 
    \end{tabular}
    \end{ruledtabular}
    \caption{Constraints on $\nn$ two-nucleon energies. Bounds and intervals are presented at 68\% confidence. The $E_{\text{asymm}}$ gap bounds arises from taking $\Egap$ to be the energy extracted from analysis of the $D_0\overline{H}$ asymmetric correlators alone under $N$-state saturation. Quantities denoted $E$ without a definite state index emphasize that they constrain \emph{some} energy without knowledge or assumption of its position relative to the ground state.
    For a visual comparison of some of these constraints, see \cref{fig:asym:nn-Lanczos-bounds-comp}. 
    }
    \label{tab:bounds_summary_1}
\end{table}

Constraining nucleon-nucleon and other multi-baryon scattering amplitudes from LQCD remains challenging, but the new tools introduced here suggest paths towards robust excited-state systematic uncertainty quantification under explicit and quantifiable assumptions. One major question to address is the nature of the ground-states of the $NN$ system at unphysically heavy quark masses. 
Previous works that have concluded the existence \cite{NPLQCD:2013bqy,Wagman:2017tmp,Orginos:2015aya,NPLQCD:2020ozd,NPLQCD:2020lxg,Yamazaki:2012hi,Yamazaki:2015asa,Berkowitz:2015eaa}  or non-existence \cite{BaSc:2025yhy} of bound states in these channels from symmetric- or asymmetric-correlator energy-estimators have necessarily only established such results under an assumption of $N$-state saturation.
In principle, one-sided variational bounds  can demonstrate the existence of an infinite-volume bound state in one or both $NN$ channels. However, to date they have not done so. By their nature, residual and gap bounds provide evidence for the existence of states in particular energy windows, and could also demonstrate the existence of a given state; however, they cannot exclude states they do not find. In the LQCD studies in the current work, none of these bounds provide positive evidence for $\nn$ $NN$ bound states. 
The $\deut$ ground state is too close to the bound-state threshold to reliably assess its nature with results from a single lattice volume.

\begin{table}[t]
    \centering
    \begin{ruledtabular}
    \begin{tabular}{cccc}
    Quantity  & Constraint & Reason/Assumption & $\{\mathcal{O} \}$ \\ \hline
    $E_{9}$ & $\leq 2.613$ & Variational bounds & $10D$ \\
    $E_{0}$ & $\leq 2.402$ & Variational bounds & $1D$ \\
    $E_{0}$ & $\leq 2.401$ & Variational bounds & $10D$ \\
    $E$ & $\in[2.39,2.41]$ & Residual bound & $1D$ \\
    $E$ & $\in[2.38,2.43]$ & Residual bound & $10D$ \\
    $E$ & $\in[2.36,2.42]$ & Residual bound & $D_0\overline{H}$ \\
    $E$ & $\in[2.399,2.405]$  & $NN$ $k=1$ gap  bound & $1D$ \\
    $E$ & $\in[2.398,2.404]$  & $\Delta\Delta$ gap  bound & $10D$ \\
    $E$ & $\in[2.39, 2.42]$  & $NN$ $k=1$ gap  bound & $D_0\overline{H}$ \\
      $E$ & $\in[2.40,2.41]$  & $E_{\text{asymm}}$ gap  bound & $1D$  \\
      $E$ & $\in[2.37,2.44]$  & $E_{\text{asymm}}$ gap  bound & $4D$ \\
      $E$ & $\in[2.34,2.46]$  & $E_{\text{asymm}}$ gap  bound & $D_0\overline{H}$ \\
      $E_0$ & $\in[2.402,2.402]$  & $N$-state saturation & $1D$ \\
      $E_0$ & $\in[2.400,2.401]$  & $N$-state saturation & $10D$ \\
      $E_0$ & $\in[2.387,2.388]$  & $N$-state saturation & $D_0\overline{H}$
    \end{tabular}
    \end{ruledtabular}
    \caption{Constraints on $\deut$ two-nucleon energies. Bounds and intervals are presented at 68\% confidence. Details are as in \cref{tab:bounds_summary_1}. 
    }
    \label{tab:bounds_summary_0}
\end{table}

An oblique Lanczos analysis of the asymmetric LQCD correlators casts doubt on their physical interpretation and ability to provide positive evidence for a low-lying state.
In the current study, the Krylov- and Ritz-vector overlap angles for $D_0\overline{H}$ asymmetric correlators change significantly over the available range of Euclidean times. This means that if the energy extracted from the $D_0\overline{H}$ asymmetric correlator alone is correct, it can only be so through approximate orthogonality of excited-state overlaps of the two different interpolating operators, as discussed in \cref{app:epsilon}. 
However, scenarios in which there are multiple states near threshold are still consistent with the LQCD data.
As shown in the small-gap adversarial example of \cref{sec:gap_attack}, 
it may not be possible to assess the validity of $N$-state assumptions from spectroscopy analyses alone.
Without making some assumptions about the LQCD spectrum, the natures of the ground states in the $\nn$ and $\deut$ $NN$ channels remain unresolved. 

The conclusions of the present analysis of a very high statistics dataset at relatively large quark masses and in a relatively small volume can be interpreted in a negative fashion, indicating that more phenomenologically relevant applications of QCD to multi-hadron systems are, for the most part, hopeless.
We advocate for a more nuanced interpretation, in which there is an acceptance of the current role of physics-based constraints in such calculations.
Taking this point of view, it is important to ask what assumptions  are reasonable. 
One might ask whether the near-degeneracy required for shallow bound states in the $NN$ system is realistic.  Given the $NN$ system in nature is somewhat fine-tuned, this is a particularly relevant question in these channels. Effective field theory provides a  framework in which to phrase this question. Indeed, estimating gap bounds is somewhat akin to  postulating the degrees of freedom in an EFT. As discussed in \cref{sec:EFT}, EFT$(\slashed{\pi}$) is able to describe both bound and unbound scenarios for the $NN$ system at unphysically large quark masses.

\begin{table}[t]
    \centering
    \begin{ruledtabular}
    \begin{tabular}{cccc}
    Quantity  & Constraint & Reason/Assumption & $\{\mathcal{O} \}$ \\ \hline
    $\Delta E_0$ & $\leq -0.0036$ & Variational + $N$ gap & $1D$ \\
    $\Delta E_0$ & $\leq -0.0042$ & Variational + $N$ gap & $4D$ \\
    $\Delta E$ & $\in[-0.030,0.023]$ & Residual bounds & $1D$ \\
    $\Delta E$ & $\in[-0.037,0.028]$ & Residual bounds & $4D$ \\
    $\Delta E$ & $\in[-0.056,0.025]$ & Residual bounds & $D_0\overline{H}$ \\
    $\Delta E_0$ & $\in[-0.017,-0.0036]$ & Res + var + $N$ gap & $1D$ \\
    $\Delta E_0$ & $\in[-0.024,-0.0042]$ & Res + var + $N$ gap & $4D$ \\
      $\Delta E_0$ & $\in[-0.0067,-0.0036]$  & $NN$ $k=1$ gap  + var & $1D$  \\
      $\Delta E_0$ & $\in[-0.0070,-0.0042]$  & $N\Delta$ gap  + var & $4D$ \\
            $\Delta E_0$ & $\in[-0.017,-0.0030]$  & $NN$ $k=1$ gap  + var & $D_0\overline{H}$ \\
      $\Delta E_0$ & $\in[-0.015,-0.0036]$  & $E_{\text{asymm}}$ gap  + var & $1D$  \\
      $\Delta E_0$ & $\in[-0.033,-0.0042]$  & $E_{\text{asymm}}$ gap  + var & $4D$ \\
      $\Delta E_0$ & $\in[-0.067,-0.0030]$  & $E_{\text{asymm}}$ gap  + var & $D_0\overline{H}$ \\
      $\Delta E_0$ & $\in[-0.0040,-0.0037]$  & $N$-state saturation & $1D$ \\
      $\Delta E_0$ & $\in[-0.0047,-0.0043]$  & $N$-state saturation & $4D$ \\
      $\Delta E_0$ & $\in[-0.016,-0.015]$  & $N$-state saturation & $D_0\overline{H}$ 
    \end{tabular}
    \end{ruledtabular}
    \caption{Constraints on $\nn$ two-nucleon energy differences. Bounds and intervals are presented at 68\% confidence. 
    Rows with ``Residual bounds'' involve differences of $NN$ residual lower (upper) bounds and twice $N$ residual upper (lower) bounds. 
    Rows with ``Variational + $N$ gap'' involve differences between $NN$ variational upper bounds and twice $N$ gap lower bounds.
    Rows with ``Res + var + $N$ gap'' involve the same upper bounds as well as differences of $NN$ residual lower bounds and twice $N$ variational upper bounds. 
    Rows with ``gap + var'' show analogous constraints using $NN$ gap lower bounds.
    Other details are as in \cref{tab:bounds_summary_1}.
    }
    \label{tab:bounds_summary_1_diff}
\end{table}

The type of assumptions that should be used also clearly depends on the context, physics applications, and precision goals of a given calculation.
For example, at the physical values of the quark masses in the continuum limit, there is significant prior knowledge from experiment that can be informative. However, away from the physical point, and in theories other than QCD, such prior knowledge is less easy to acquire. Different levels of assumptions are also appropriate when making qualitative statements than when making definitive statements that attempt to rule out entire scenarios.

As with other systematics in LQCD calculations, assumptions used in the analysis of correlators should be tested in all possible ways, and extensive consistency checks are critical.
For example, in the symmetric Lanczos framework, Ritz vectors offer a means to compute matrix elements of operators in approximations of energy eigenstates \cite{Hackett:2024xnx}. Using operators such as the vector and axial currents, matrix element calculations can be used to probe the structure of the states in question and thereby test the assumptions being made in correlator analysis.  

\begin{table}[t]
    \centering
    \begin{ruledtabular}
    \begin{tabular}{cccc}
    Quantity  & Constraint & Reason/Assumption & $\{\mathcal{O} \}$ \\ \hline
    $\Delta E_0$ & $\leq -0.0046$ & Variational + $N$ gap & $1D$ \\
    $\Delta E_0$ & $\leq -0.0056$ & Variational + $N$ gap & $10D$ \\
    $\Delta E$ & $\in[-0.030,0.020]$ & Residual bounds & $1D$ \\
    $\Delta E$ & $\in[-0.043,0.032]$ & Residual bounds & $10D$ \\
    $\Delta E$ & $\in[-0.062,0.024]$ & Residual bounds & $D_0\overline{H}$ \\
    $\Delta E_0$ & $\in[-0.017,-0.0046]$ & Res + var + $N$ gap & $1D$ \\
    $\Delta E_0$ & $\in[-0.031,-0.0056]$ & Res + var + $N$ gap & $10D$ \\
    $\Delta E_0$ & $\in[-0.0073,-0.0046]$  & $NN$ $k=1$ gap  + var & $1D$ \\
    $\Delta E_0$ & $\in[-0.0086,-0.0056]$  & $\Delta\Delta$ gap + var & $10D$ \\
    $\Delta E_0$ & $\in[-0.020,-0.0040]$  & $NN$ $k=1$ gap + var & $D_0\overline{H}$ \\
    $\Delta E_0$ & $\in[-0.011,-0.0046]$  & $E_{\text{asymm}}$ gap  bound & $1D$  \\
      $\Delta E_0$ & $\in[-0.042,-0.0056]$  & $E_{\text{asymm}}$ gap  bound & $10D$ \\
      $\Delta E_0$ & $\in[-0.062,-0.0040]$  & $E_{\text{asymm}}$ gap  bound & $D_0\overline{H}$ \\
      $\Delta E_0$ & $\in[-0.0051,-0.0047]$  & $N$-state saturation & $1D$ \\
      $\Delta E_0$ & $\in[-0.0062,-0.0058]$  & $N$-state saturation & $10D$ \\
      $\Delta E_0$ & $\in[-0.020,-0.019]$  & $N$-state saturation & $D_0\overline{H}$
    \end{tabular}
    \end{ruledtabular}
    \caption{Constraints on $\deut$ two-nucleon energy differences. Bounds and intervals are presented at 68\% confidence. Details are as in \cref{tab:bounds_summary_1_diff}. 
    }
    \label{tab:bounds_summary_0_diff}
\end{table}

The lack of conclusive results in this analysis should also be seen as motivation to develop improved methods. New approaches may exist that dramatically reduce the sizes of the residual bounds in the Lanczos approach. Methods may exist that allow for \textit{systematic} improvement of spectroscopy calculations by introducing new interpolating operators in a way that provably converges to full coverage of the low-energy sector of the LQCD Hilbert space. The current study highlights the importance of pursuing such directions, as it  demonstrates that increasing statistics alone is unlikely to resolve many questions in spectroscopy. 

\begin{acknowledgements}
We are grateful to R.~Brice{\~n}o, Z.~Davoudi, Y.~Fu, W.~Jay, C.~Morningstar, A.~Nicholson, A.~Parre{\~n}o, F.~Romero-L\'opez, M.~J.~Savage, and A.~Walker-Loud for helpful discussions.
The software packages chroma \cite{Edwards_2005}, qdp-jit \cite{Winter:2014dka}, quda \cite{Clark_2010}, QPhiX \cite{qphix}, and GLU \cite{glu} 
were used to generate gauge field configurations and compute quark propagators.
WD, PES and RJP are supported in part by the U.S. Department of Energy, Office of Science under grant Contract Number DE-SC0011090 and by the SciDAC5 award DE-SC0023116.
PES is additionally supported by the U.S. DOE Early Career Award DE-SC0021006 and by Simons Foundation grant 994314 (Simons Collaboration on Confinement and QCD Strings).
AVG is supported by the National Science Foundation under Grant No.~PHY-240227.
MI is partially supported by the Quantum Science Center (QSC), a National Quantum Information Science Research Center of the U.S. Department of Energy.
This document was prepared using the resources of the Fermi National Accelerator Laboratory (Fermilab), a U.S. Department of Energy, Office of Science, Office of High Energy Physics HEP User Facility. Fermilab is managed by Fermi Forward Discovery Group, LLC, acting under Contract No.~89243024CSC000002.

This research used resources of the Oak Ridge Leadership Computing Facility at the Oak Ridge National Laboratory, which is supported by the Office of Science of the U.S. Department of Energy under Contract number DE-AC05-00OR22725
and the resources of the National Energy Research Scientific Computing Center (NERSC), a Department of Energy Office of Science User Facility using NERSC award NP-ERCAPm747. 
We acknowledge EuroHPC JU for awarding the project ID EHPC-REG-2023R03-082 access to LUMI at CSC, Finland. 
The research reported in this work made use of computing facilities of the USQCD Collaboration, which are funded by the Office of Science of the U.S. Department of Energy.
\end{acknowledgements}

\appendix

\crefalias{section}{appendix}

\section{$B$ distributions and correlations}\label{app:lanczos}

\begin{figure}[t!]
    \includegraphics[width=0.48\textwidth]{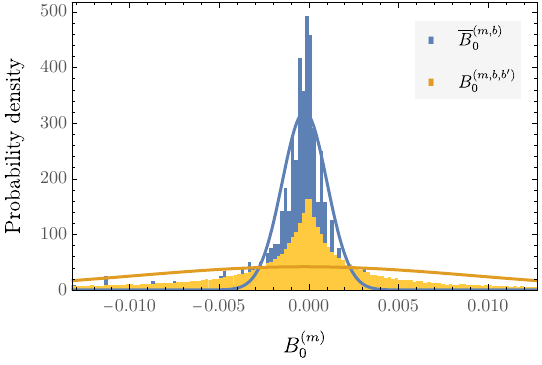}
    \includegraphics[width=0.48\textwidth]{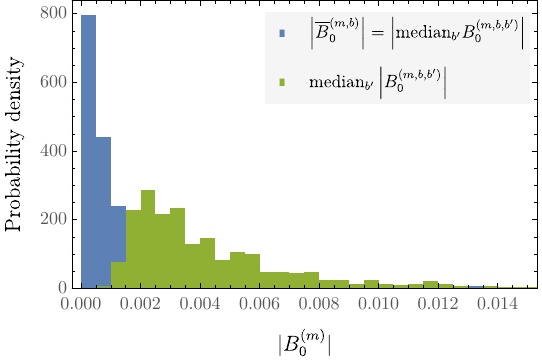}
    \caption{
      Top: Probability densities for various estimators of the residual-norm-square defined in \cref{eq:Bdef} for the ground-state Ritz value of the scalar field theory correlator $C_\varphi^{11}$ defined in \cref{eq:Cphi}.  
      Distributions of inner bootstrap samples $B_0^{(m,b,b')}$ are compared with those of bootstrap-median estimators $\overline{B}_0^{(m,b)}$. Gaussian distributions with the same mean and variance as each distribution are shown for comparison in curves of each color (note that the $B_0^{(m,b,b')}$ distribution is very heavy-tailed on scales larger than this plot range) and provide a qualitatively better description for the bootstrap-median estimator distribution than for the nested bootstrap distribution. Bottom: The corresponding outer-bootstrap distribution of the absolute values $|\overline{B}_0^{(m,b)}|$ is also compared with the distribution obtained from the median of $|B_0^{(m,b,b')}|$ . In all cases distributions are shown for the one-dimensional complex scalar correlator described in the main text with $m \in \{13,14,15\}$ and $N_{\rm cfg} = 10^6$. 
    }
    \label{fig:complex-scalar-hist}
\end{figure}

\Cref{fig:complex-scalar-hist} shows (nested) bootstrap distributions of $B_0^{(m,b,b')}$ and corresponding bootstrap-median estimators $\overline{B}_0^{(m,b)}$ for an ensemble of $N_{\rm cfg} = 10^6$ scalar correlators as defined in \cref{eq:Cphi} for $k=k'=1$.
Results are shown for the ground state estimator based on the $m \in \{13,14,15\}$ iterations.
In this example, the distribution of $B_0^{(m,b,b')}$ has a mean that is much smaller than its width and is approximately symmetric around zero.  
Moreover, the distribution of $B_0^{(m,b,b')}$ has much broader tails than a Gaussian distribution.

As with estimators for Ritz values, outlier-robust estimators must be used to obtain results whose statistical distributions can be easily interpreted. 
Taking a median over inner bootstraps yields a $\overline{B}_0^{(m,b)}$ distribution that is qualitatively more similar to a Gaussian than the corresponding $B_0^{(m,b,b')}$ distribution.
Nevertheless, some heavy tails are still visible, and 
the Kolmogorov-Smirnov  and Shapiro-Wilk tests provide clear evidence of non-Gaussianity.
Statistical uncertainties and confidence intervals in the main text are therefore estimated based on empirical bootstrap confidence intervals that do not assume Gaussianity.

The fact that $B_0^{(m,b,b')}$ and $\overline{B}_0^{(m,b)}$ both have approximately symmetric distributions centered around zero for large $m$ complicates the construction of two-sided bounds at finite statistics.
Forming gap bounds directly with $\overline{B}_0^{(m,b)}=\text{median}_{b'} B_0^{(m,b,b')}$ has the disadvantage that ``upper'' and ``lower'' gap bounds are effectively reversed on bootstrap samples with $\overline{B}_0^{(m,b)} < 0$, which occur frequently.
Positive-width gap and residual bounds that can be interpreted straightforwardly at the bootstrap level can instead be obtained using $|\overline{B}_0^{(m,b)}|=|\text{median}_{b'} {B}_0^{(m,b,b')}|$, which is the choice made in the main text. 
This definition has the advantage that it leads to relatively tight bounds in comparison with, e.g., those obtained by interchanging the order of taking medians and absolute values and forming bounds from $\text{median}_{b'} |{B}_0^{(m,b,b')}|$.
Comparisons of finite-statistics residual- and gap-bound results with exact finite-$m$ truncation errors in \cref{sec:scalar-calibration} demonstrate that bounds formed using $|\overline{B}_0^{(m,b)}|$ are not overly aggressive in these scalar field theory examples.
Further, no apparent inconsistencies are visible in two-sided bounds formed using $|\overline{B}_0^{(m,b)}|$ for LQCD results with different interpolators.
These consistency checks suggest that finite-statistics effects in bounds formed using $|\overline{B}_0^{(m,b)}|$ are not pathological in these examples.
However, significant differences in the magnitudes of $|\overline{B}_0^{(m,b)}|$ and $\text{median}_{b'} |{B}_0^{(m,b,b')}|$ at finite statistics motivate further detailed studies of other examples, and development of more robust estimators for residual-norm-squares.

\begin{figure*}[]
    \includegraphics[width=0.48\textwidth]{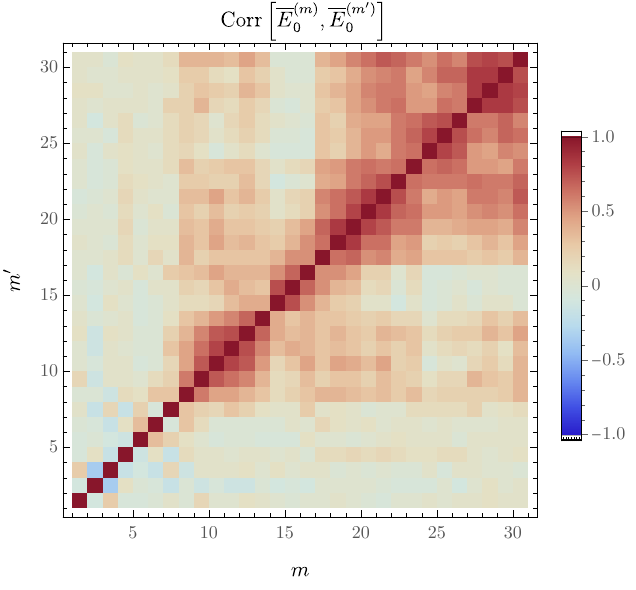}
    \includegraphics[width=0.48\textwidth]{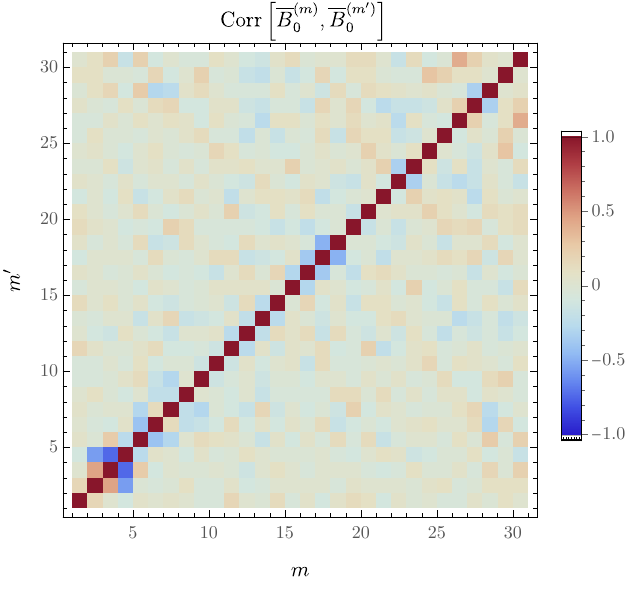}
    \caption{
      Autocorrelations as functions of iteration $m$ for complex-scalar bootstrap-median energy and residual-norm-square estimators, $\overline{E}_0^{(m,b)}$ and $\overline{B}_0^{(m,b)}$. 
    }
    \label{fig:complex-scalar-corr}
\end{figure*}

\begin{figure*}
    \includegraphics[width=0.48\textwidth]{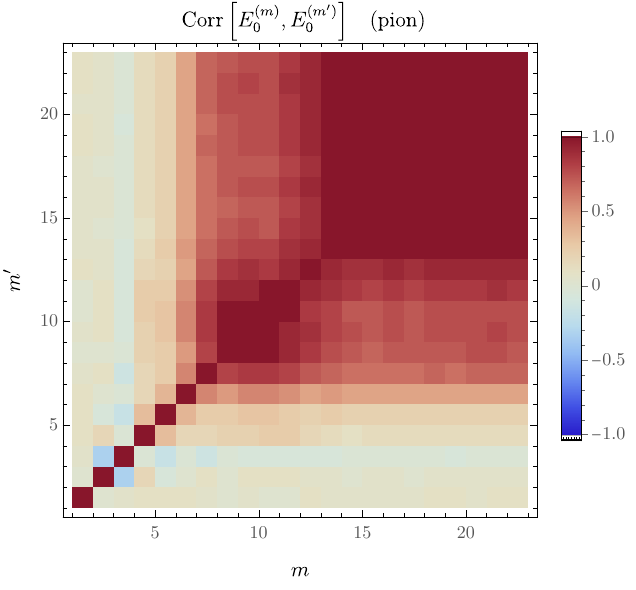}
    \includegraphics[width=0.48\textwidth]{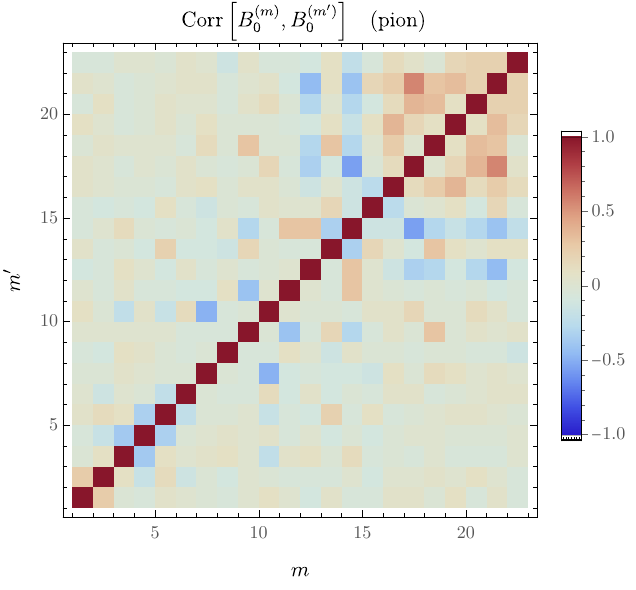}
    \caption{    
        Left: correlations between bootstrap-median $\overline{E}_0^{(m)}$ estimators for different $m$ for the pion. Right: analogous correlations for $\overline{B}_0^{(m)}$.     
        }
    \label{fig:pion-Lanczos-correlations}
\end{figure*}
\begin{figure*}
    \includegraphics[width=0.48\textwidth]{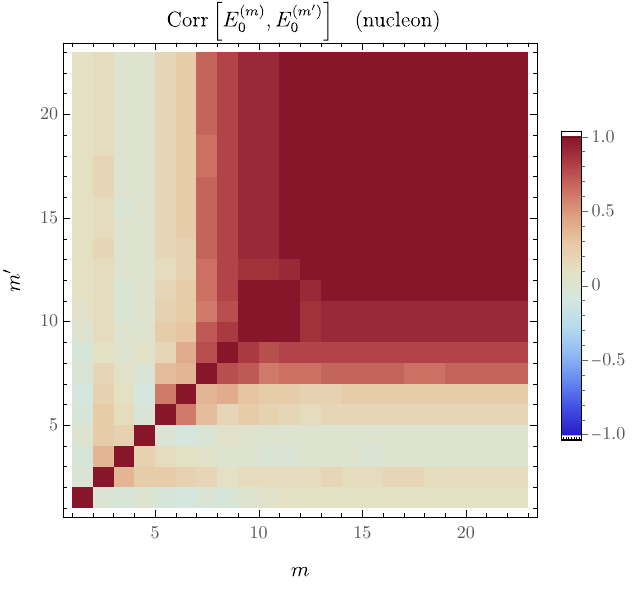}
    \includegraphics[width=0.48\textwidth]{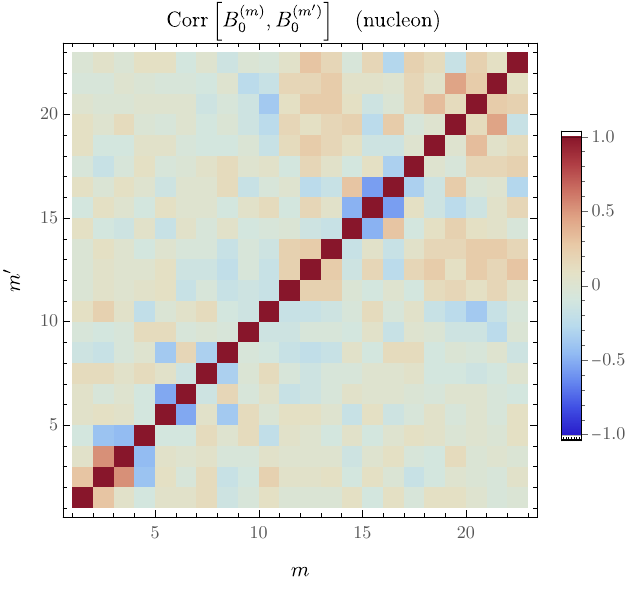}
    \caption{    
        Left: correlations between bootstrap-median $\overline{E}_0^{(m)}$ estimators for different $m$ for the nucleon. Right: analogous correlations for $\overline{B}_0^{(m)}$. 
    }
    \label{fig:nuc-Lanczos-correlations}
\end{figure*}
\begin{figure*}
    \includegraphics[width=0.48\textwidth]{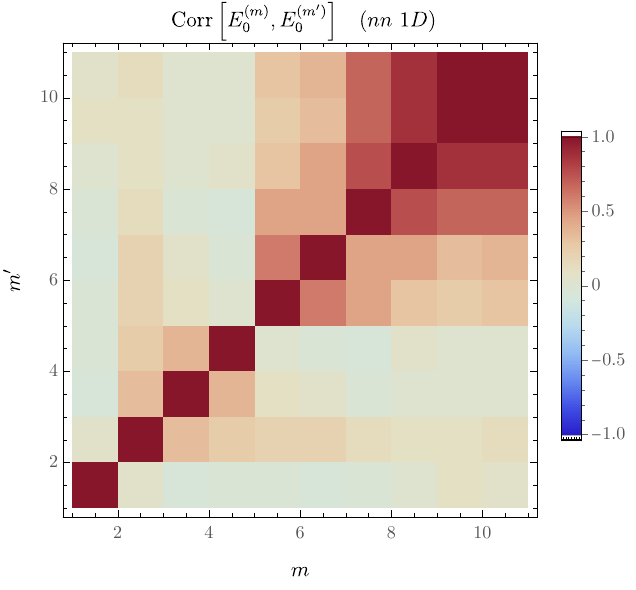}
    \includegraphics[width=0.48\textwidth]{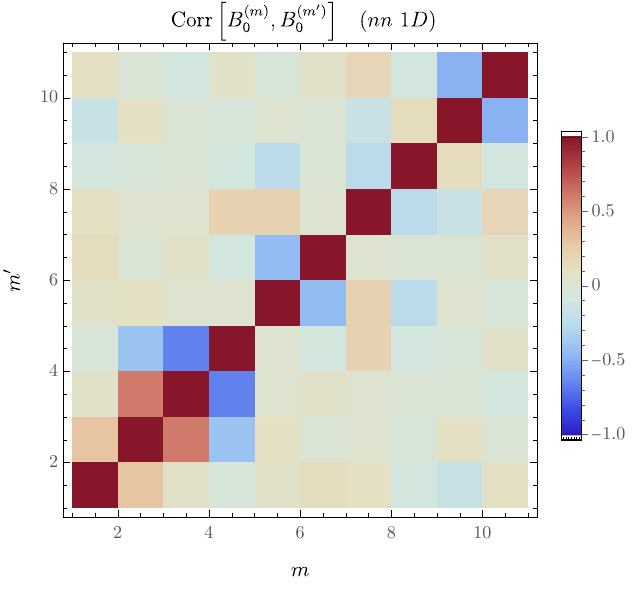}
    \caption{    
        Left: correlations between bootstrap-median $\overline{E}_0^{(m)}$ estimators for different $m$ for two-nucleon $D_0\overline{D}_0$ correlators in the $\nn$ channel. Right: analogous correlations for $\overline{B}_0^{(m)}$. Note that $NN$ correlator results are only available for $m \leq 11$.  
    }
    \label{fig:nn-Lanczos-correlations}
\end{figure*}
\begin{figure*}
    \includegraphics[width=0.48\textwidth]{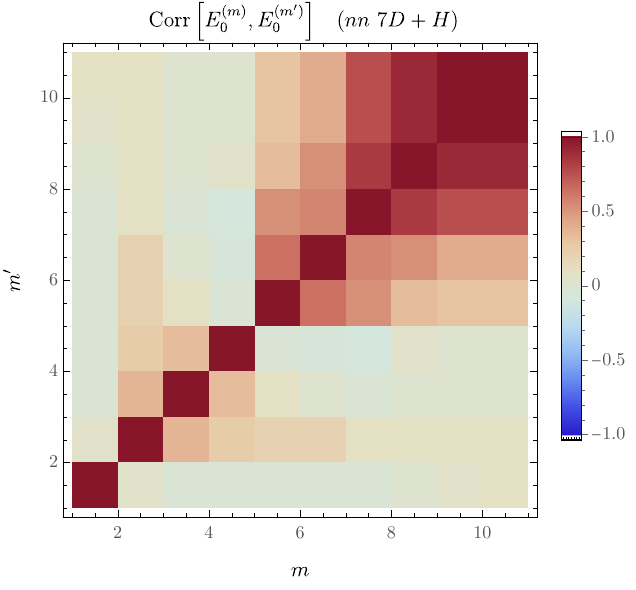}
    \includegraphics[width=0.48\textwidth]{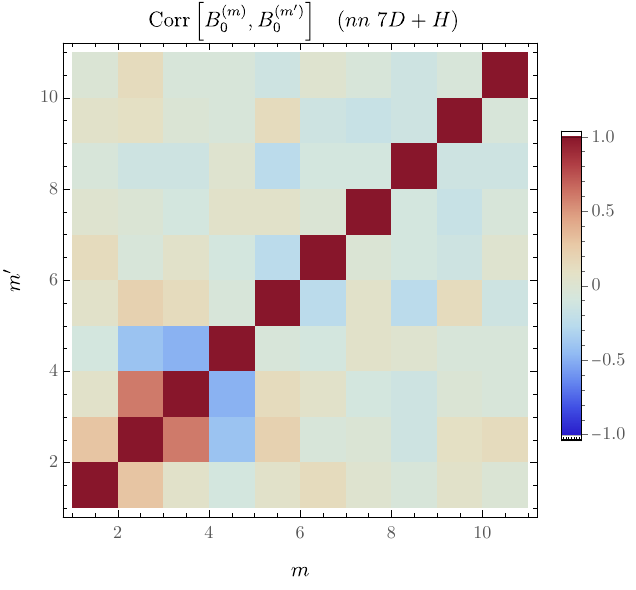}
    \caption{    
        Left: correlations between bootstrap-median $\overline{E}_0^{(m)}$ estimators for different $m$ for two-nucleon $7D+H$ block Lanczos in the $\nn$ channel. Right: analogous correlations for $\overline{B}_0^{(m)}$. 
    }
    \label{fig:nn-block-Lanczos-correlations}
\end{figure*}

Large correlations in $\overline{E}_0^{(m)}$ for large $m$ have been previously observed in LQCD studies~\cite{Wagman:2024rid,Hackett:2024xnx,Ostmeyer:2024qgu,Chakraborty:2024scw,Hackett:2024nbe}.
Similar correlations are seen here in scalar field theory, pion, nucleon, and $NN$ correlator results shown in \cref{fig:complex-scalar-corr,fig:pion-Lanczos-correlations,fig:nuc-Lanczos-correlations,fig:nn-Lanczos-correlations}.
Block Lanczos results show similar features of large correlations emerging at large $m$; see for example \cref{fig:nn-block-Lanczos-correlations}.
However, residual-norm-square estimators $\overline{B}_0^{(m)}$ have a different correlation structure.
Some relatively large correlations of $\gtrsim 0.5$ are seen in large-$m$ results for $\overline{B}_0^{(m)}$ in the pion, see \cref{fig:pion-Lanczos-correlations}, but correlations are generally $\lesssim 0.2$ for nucleon and two-nucleon $\overline{B}_0^{(m)}$ as shown in \cref{fig:nuc-Lanczos-correlations,fig:nn-Lanczos-correlations}.
Significant large-$m$ correlations are also not seen in block-Lanczos results, see e.g.~$r=8$ results for the $\nn$ channel in \cref{fig:nn-block-Lanczos-correlations}.
Consequently, median-averaging over a window of several large-$m$ iterations, as discussed in \cref{sec:B}, can lead to significant reductions in the statistical uncertainty of $\overline{B}_0^{(m)}$.

\bibliography{bib.bib}

\end{document}